%% file: ms.tex
\newif\ifarxiv
\arxivtrue
\ifarxiv
\RequirePackage{snapshot}
\documentclass[
    reprint,
    superscriptaddress,
    amsmath,amssymb,
    aps,
    pra,
]{revtex4-2}
\renewcommand{\doi}[1]{\href{http://doi.org/#1}{doi:#1}}
\else
\documentclass[journal=jacsat,suppinfo,bookmarksdepth=subparagraph]{achemso}
\newcommand{\doi}[1]{\href{http://doi.org/#1}{doi:#1}}
\fi
\newif\ifshowlabels
\showlabelsfalse
\newif\iffigdraft
\figdraftfalse
\newif\ifgentoc
\newif\ifeditmode
\editmodefalse
\gentocfalse
\newif\ifshowparagraphinline
\showparagraphinlinefalse
\ifeditmode\else
\showparagraphinlinefalse
\fi
\ifarxiv
\usepackage[sectionbib]{bibunits}
\defaultbibliographystyle{new}
\defaultbibliography{references}
\else
\fi
\usepackage{layouts}
\usepackage[acronyms]{glossaries}
\usepackage[procnames]{listings}
\usepackage{pygments} 
\usepackage{amsfonts}
\usepackage{blindtext}
\let\oldblindtext\blindtext
\renewcommand{\blindtext}{\textcolor{gray}{\oldblindtext}}
\usepackage{cmap}
\usepackage{datetime}
\usepackage{environ}
\usepackage{etoolbox}
\usepackage[utf8]{inputenc}
\DeclareUnicodeCharacter{03BD}{\ensuremath{\nu}}
\DeclareUnicodeCharacter{2212}{--}
\DeclareUnicodeCharacter{03BC}{\textmu}
\DeclareUnicodeCharacter{03B2}{\ensuremath{\beta}}
\DeclareUnicodeCharacter{03B1}{\ensuremath{\alpha}}
\DeclareUnicodeCharacter{00B0}{\ensuremath{^\circ}}
\DeclareUnicodeCharacter{00B1}{\ensuremath{\pm}}
\DeclareUnicodeCharacter{2192}{\ensuremath{\rightarrow}}
\DeclareUnicodeCharacter{03C3}{\ensuremath{\sigma}}
\DeclareUnicodeCharacter{03C4}{\ensuremath{\tau}}
\DeclareUnicodeCharacter{2009}{ }
\DeclareUnicodeCharacter{2080}{\ensuremath{_0}}
\DeclareUnicodeCharacter{2082}{\ensuremath{_2}}
\DeclareUnicodeCharacter{2084}{\ensuremath{_4}}
\DeclareUnicodeCharacter{1F449}{\ensuremath{\rightarrow}} 
\DeclareUnicodeCharacter{1F448}{\ensuremath{\leftarrow}} 
\iffigdraft
    \usepackage[draft]{graphicx}
\else
    \usepackage{graphicx}
\fi
\usepackage{hyphenat}
\usepackage{ifxetex}
\usepackage{mathrsfs}
\usepackage{calc}

\newcommand{\ie}{\textit{i.e.\@}\xspace}
\newcommand{\vs}{\textit{vs.\@}\xspace}
\newcommand{\eg}{\textit{e.g.\@}\xspace}
\newcommand{\etc}{\textit{etc.\@}\xspace}
\newcommand{\etal}{\textit{et.~al.\@}\xspace}
\newcommand{\via}{\textit{via}\xspace}

\ifeditmode
\usepackage{pdfcomment}
\newcommand{\sepcomment}{\\ \vspace*{3ex}}
\else
\newcommand{\pdfcomment}[2][]{}
\newcommand{\sepcomment}{}
\fi
\usepackage{hyperref}
\hypersetup{colorlinks=true, pdfstartview=FitV, linkcolor=linkcolor, citecolor=dbluecolor, urlcolor=dgreencolor, bookmarksdepth=subparagraph}
\usepackage{bookmark}
\usepackage{sidecap}
\usepackage{soul}
\usepackage{textcomp}
\usepackage{url}
\usepackage{xcolor}
\usepackage{xspace}
\usepackage{multirow}
\usepackage{array}
\newcolumntype{R}{>{\vspace{2ex}\displaystyle}{r}}
\newcolumntype{L}{>{\displaystyle}{l}}
\definecolor{SUgrey}{HTML}{6F777D}
\definecolor{SUorange}{HTML}{D44500}
\definecolor{dbluecolor}{rgb}{.01,.02,0.29}
\definecolor{dgraycolor}{rgb}{0.50,0.50,0.50}
\definecolor{dgreencolor}{rgb}{0.0,0.4,0}
\definecolor{linkcolor}{cmyk}{0,0.7,0.5,0.5}
\usepackage{accsupp}    
\newcommand{\noncopynumber}[1]{%
    \BeginAccSupp{method=escape,ActualText={}}%
    #1%
    \EndAccSupp{}%
}

\usepackage[compact]{titlesec}
\ifshowparagraphinline
\titleformat{\paragraph}[runin]{\color{gray}\normalfont\bfseries\footnotesize}{}{3pt}{\hspace{0.75em}\ul{\footnotesize\thesubsection\theparagraph)\;}}[:]
\else
\renewcommand{\paragraph}[1]{\par\phantomsection\addcontentsline{toc}{paragraph}{#1}}
\fi
\makeatletter
\@ifundefined{linelabel}{%
\newcommand{\linelabel}[1]{}}{}
\makeatother
\makeglossaries
\renewcommand{\glossarysection}[2][]{}
\include{acronyms}
\renewcommand{\thesection}{\Roman{section}}
\renewcommand{\thesubsection}{\thesection.\arabic{subsection}}

\makeatletter
\renewcommand{\p@subsection}{}
\renewcommand{\p@subsubsection}{}
\makeatother
\newcounter{subfigure}[figure]
\newcounter{subfigurenonumber}
\newcounter{tempfigure}
\renewcommand\thesubfigurenonumber{(\alph{subfigurenonumber})}
\newcommand{\subfig}[2]{%
    \setcounter{tempfigure}{\value{figure}}%
    \addtocounter{tempfigure}{1}%
    \refstepcounter{subfigure}%
    \setcounter{subfigurenonumber}{\value{subfigure}}%
    \expandafter\edef\csname ref#2\endcsname{\thesubfigurenonumber}
    \label{#1}%
    }
\newcommand{\titleblock}{
\newcommand\SUaffil{\affiliation{Department of Chemistry, Syracuse University, Syracuse, NY 13210, USA}}
\author{Alec A. Beaton}
\author{Alexandria Guinness}
\author{John M. Franck}
\SUaffil
\email{jmfranck@syr.edu}
\title{Rapidly Screening the Correlation Between
the Rotational Mobility and the Hydrogen Bonding Strength of Confined Water}
\date{Not for Final Submission Purposes: \today, \currenttime}
\date{\today}
}
\ifarxiv\else
    \titleblock
\fi
\newif\ifpoormancref
\poormancreffalse
\ifpoormancref
\usepackage[poorman]{cleveref}
\else
\usepackage{cleveref}
\fi
\crefname{equation}{Eq.}{Eqs.}
\crefname{table}{Table}{Tables}
\crefname{figure}{Fig.}{Figs.}
\crefname{section}{Sec.}{Sec.}
\crefname{subfigure}{Fig.}{Figs.}
\crefname{lstlisting}{listing}{listings}
\Crefname{lstlisting}{Listing}{Listings}

\usepackage{xr}
\ifshowlabels
\usepackage{refcheck}
\makeatletter
\newcommand{\refcheckize}[1]{%
  \expandafter\let\csname @@\string#1\endcsname#1%
  \expandafter\DeclareRobustCommand\csname relax\string#1\endcsname[1]{%
    \csname @@\string#1\endcsname{##1}\wrtusdrf{##1}}%
  \expandafter\let\expandafter#1\csname relax\string#1\endcsname
}
\def\@setmarginlbl{%
    \if@show@ref
        \if@labelled
            \set@fbox@par
            \if@unsdlbl
                \makebox[0pt][l]{\zero@height{$\,$\rotatebox{90}{\scalebox{0.6}{\mark@size
                {\bfseries\upshape?}\underline{\last@lbl}{k\bfseries\upshape?}}}}}%
            \else
                \makebox[0pt][l]{\zero@height{$\,$\rotatebox{90}{\scalebox{0.6}{\fbox{{\mark@size\last@lbl}}}}}}%
            \fi
        \else
            \if@show@unl@bld
                \makebox[0pt][l]{\zero@height{$\,$\rotatebox{90}{\scalebox{0.6}{\unl@bld@mark}}}}%
            \fi\fi
        \fi
        \global\@labelledfalse
    }
\def\@setnmmarginlbl{%
    \if@show@ref
        \set@fbox@par
        \if@unsdlbl
            \hbox to \textwidth{\makebox[0pt][r]{\rotatebox{90}{\scalebox{0.6}{\mark@size{\bfseries
                            \upshape?}$\langle$\last@lbl$\rangle${\bfseries
            \upshape?}}}$\,$}\hfill}%
        \else
            \hbox to \textwidth{\makebox[0pt][r]{\rotatebox{90}{\scalebox{0.6}{\mark@size$\langle$%
            \last@lbl$\rangle$}}$\,$}\hfill}%
        \fi
    \fi
    \global\@labelledfalse
}
\def\@bibitem@proceed@#1{%
    \@ifundefined{cit@#1}{\@warning@rc@{Unused bibitem `#1'}%
        \if@show@cite
            \gdef\@biblabel{\makebox[0pt][r]{\zero@height{\rotatebox{90}{\scalebox{0.6}{{\mark@size{\bfseries\upshape?}}%
                \underline{\@verbatim@{#1}}{\mark@size{\bfseries\upshape?}}}}$\,$}}%
            \@@@biblabel@@}%
        \fi
    }{%
        \if@show@cite
            \set@fbox@par
            \gdef\@biblabel{\makebox[0pt][r]{\zero@height{\rotatebox{90}{\scalebox{0.6}{\fbox{TESTTESTTEST\@verbatim@{#1}}}}$\,$}}\@@@biblabel@@}%
        \fi
}}%
\makeatother
\refcheckize{\cref}
\refcheckize{\Cref}
\fi
\begin{document}
\newlength\myfigwidth
\setlength{\myfigwidth}{3.5in}
\ifarxiv
\titleblock
\fi

\begin{abstract}
\input{abstract}
\end{abstract}

\input{figures}
\ifarxiv\begin{bibunit}\fi
\maketitle
\ifarxiv\glsresetall\fi
\input{content}
\appendix
\ifarxiv
\putbib
\end{bibunit}
\pagebreak
\onecolumngrid
\begin{center}
\makeatletter
\textbf{\large Supplemental Materials for:\\
\@title}
\makeatother
\end{center}
\twocolumngrid
\begin{bibunit}
\setcounter{equation}{0}
\setcounter{figure}{0}
\setcounter{table}{0}
\setcounter{page}{1}
\makeatletter
\renewcommand{\theequation}{S\arabic{equation}}
\renewcommand{\thefigure}{S\arabic{figure}}
\renewcommand{\thesection}{S\arabic{section}}
\renewcommand{\bibnumfmt}[1]{[S#1]}
\renewcommand{\citenumfont}[1]{S#1}
\glsresetall
\let\maincref\cref
\input{supp_content}
\putbib
\end{bibunit}
\else
\bibliography{references}
\fi
\end{document}

%% file: acronyms.tex
\newacronym{oec}{OEC}{oxygen-evolving complex}
\newacronym{psii}{PSII}{Photosystem II}
\newacronym{lna}{LNA}{Low Noise Amplifier}
\newacronym{odnp}{ODNP}{Overhauser Dynamic Nuclear Polarization}
\newacronym{nmr}{NMR}{Nuclear Magnetic Resonance}
\newacronym{esr}{ESR}{Electron Spin Resonance}
\newacronym{deer}{DEER}{Double Electron-Electron Resonance}
\newacronym{pds}{PDS}{Pulse Dipolar Spectroscopy}
\newacronym{psd}{PSD}{Power Spectral Density}
\newacronym{eseem}{ESEEM}{Electron Spin Echo Envelope Modulation}
\newacronym{sdsl}{SDSL}{Site-Directed Spin-Labeling}
\newacronym{pre}{PRE}{Paramagnetic Relaxation Enhancement}
\newacronym{awg}{AWG}{Arbitrary Waveform Generation}
\newacronym{br}{BR}{bacteriorhodopsin}
\newacronym{pr}{PR}{proteorhodopsin}
\newacronym{md}{MD}{molecular dynamics}
\newacronym{mtsl}{MTSL}{\textit{S}-(1-oxyl-2,2,5,5-tetramethyl-2,5-dihydro-1H-pyrrol-3-yl)methyl methanesulfonothioate)}
\newacronym{ednmr}{EDNMR}{electrically detected \gls{nmr}}
\newacronym{mmm}{MMM}{the open-source Matlab tool known as Multiscale Modeling of Macromolecules}
\newacronym{ddm}{DDM}{n-Dodecyl-β-D-Maltopyranoside}
\newacronym{dppc}{DPPC}{dipalmitoylphosphatidylcholine}
\newacronym{dopa}{DOPA}{dioleoylphosphatidic acid}
\newacronym{sdr}{SDR}{software-defined radio}
\newacronym{avhe}{AVHE}{Academic Virtual Hosting Environment}
\newacronym{pcet}{PCET}{proton-coupled electron transfer}
\newacronym{rnr}{RNR}{Ribonuclease reductase}
\newacronym{iq}{IQ}{quadrature}
\newacronym{sasa}{SASA}{solvent-accessible surface area}
\newacronym{mr}{MR}{Magetic Resonance}
\newacronym{dcct}{DCCT}{domain colored coherence transfer}
\newacronym{ct}{CT}{coherence transfer}
\newacronym{lp}{LP}{Linear Prediction}
\newacronym{mw}{μw}{microwave} 
\newacronym{fid}{FID}{free induction decay}
\newacronym{rm}{RM}{reverse micelle}
\newacronym{rms}{RMs}{reverse micelles}
\newacronym{snr}{SNR}{signal to noise ratio}
\newacronym{fwhm}{FWHM}{full-width-half-maximum}
\newacronym{l2g}{L2G}{Lorentzian-to-Gaussian}
\newacronym{aot}{AOT}{Aerosol-OT}
\newacronym{svd}{SVD}{Singular Value Decomposition}
\newacronym{pca}{PCA}{Principal Component Analysis}
\newacronym{sde}{SDE}{Stokes-Debye-Einstein}
\newacronym{dls}{DLS}{Dynamic Light Scattering}
\newacronym{ctet}{\ensuremath{\text{CCl}_{4}}}{carbon tetrachloride}
\newacronym{peg}{PEG}{polyethylene glycol}
\newacronym{bsa}{BSA}{bovine serum albumin}
\newacronym{ilt}{ILT}{Inverse Laplace Transform}
\newacronym{dss}{DSS}{sodium trimethylsilylpropanesulfonate}
\newacronym{dsc}{DSC}{differential scanning
calorimetry}
\newacronym{ctab}{CTAB}{cetrimonium bromide}
\newacronym{cppvpo}{CPP-VPO}{controlled-partial-pressure vapor-pressure osmometry}
\newacronym{cpmg}{CPMG}{Car-Purcell-Meiboom-Gill}

%% file: abstract.tex
This contribution demonstrates a two dimensional
    deuterium NMR methodology for
    discriminating between
    D\textsubscript{2}O
    populations whose properties differ as a
    result of being confined inside nanoscale
    volumes.
Importantly,
    for
    reverse micelles
    (a proof-of-principal system),
    as the lengthscale
    of the confinement is changed from several
    nanometers down to less than a nanometer,
    the position of the signal peak migrates
    through the 2D spectrum,
    following a distinctive trend.
This trend most typically involves relatively
    gentle linear change in the order of magnitude
    of the NMR relaxation time for water
    confined on the scale of several
    nanometers,
    followed by a region of dramatic
    negative curvature
    (of relaxation time \textit{vs.} chemical shift)
    for water confined to lengthscales smaller
    than 1-2 nanometers.
Interestingly,
    the qualitative shape of this trend can change
    with different choices of surfactants,
    \textit{i.e.}, a different choice of the chemistry at the edges
    of the confining environment.
An important facet of this research was to
    demonstrate the relatively wide applicability
    of these techniques by showing that both:
    (1) Standard
    modern NMR instrumentation
    is capable of deploying an
    automated measurement,
    even though the choice of a deuterium nucleus
    is non standard and frequently requires
    companion proton spectra in order to reference
    the chemical shifts;
    and (2) well established (though underutilized)
    modern signal processing techniques can generate
    the resulting signal even though
    it involves the somewhat unusual combination of
    chemical shifts along one dimension and a
    distribution of relaxation times along another
    dimension.
In addition to demonstrating that this technique
    can be deployed across many samples of
    interest,
    detailed facts pertaining to the broadening
    or shifting of resulting signals upon inclusion
    of various guests molecules are also discussed.

%% file: figures.tex
\newlength{\mylinewidth}
\setlength{\mylinewidth}{3.5in}

\newcommand{\figDispersants}{\begin{figure}[tbp]
    \centering
    \subfig{fig:HexaneT1}{hexaneLong}
    \subfig{fig:IsoT1}{isoLong}
    \subfig{fig:CCl4T1}{carbontetLong}
    \begin{tabular}{cc}
        \  &
        \hspace*{0.14\mylinewidth}\includegraphics[width=0.8\mylinewidth]{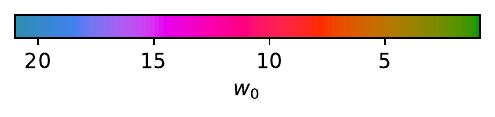}
        \vspace*{-0.15in}
        \\
        \refhexaneLong
        & \raisebox{-\height}{ 
        \includegraphics[width=0.88\mylinewidth]{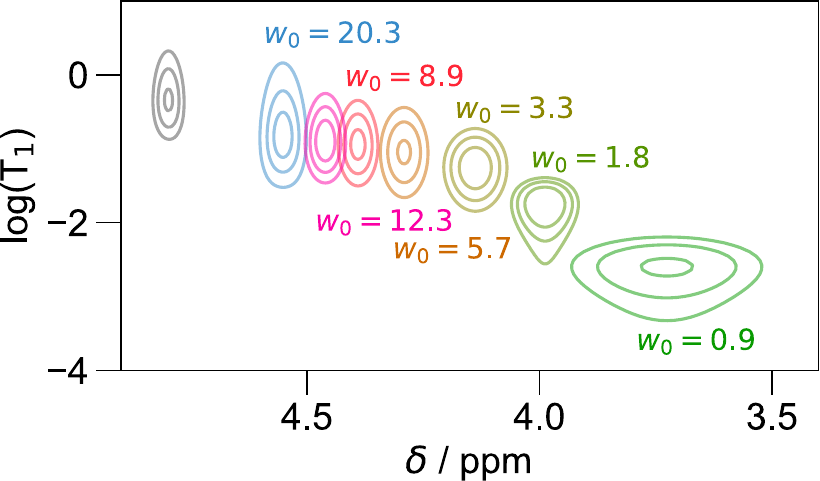}
        }
        \\
        \refisoLong
        & \raisebox{-\height}{ 
        \includegraphics[width=0.88\mylinewidth]{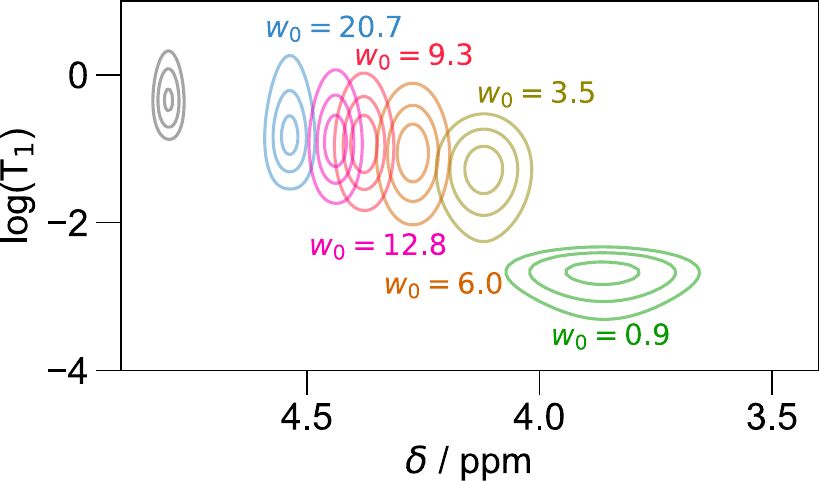}
        }
        \\
        \refcarbontetLong
        & \raisebox{-\height}{ 
        \includegraphics[width=0.88\mylinewidth]{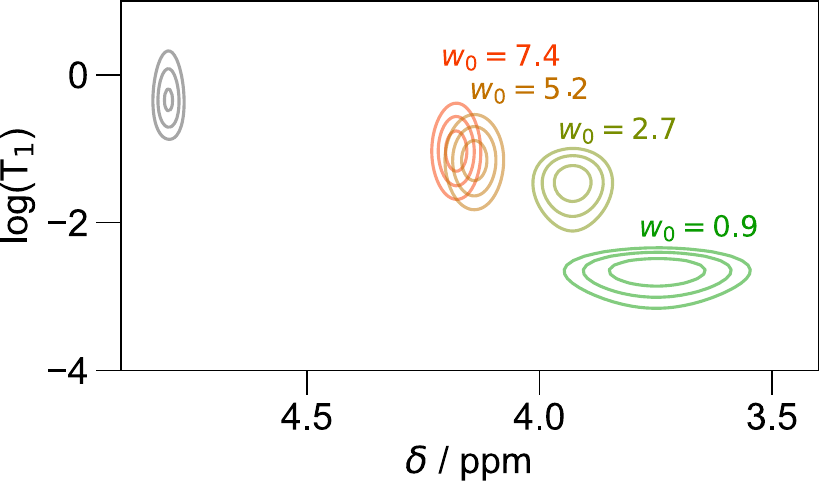}
        }
    \end{tabular}
    %
    %
    %
    \centering
    \caption{\gls{ilt} plots showing the $T_{1}$
        \textit{vs} chemical shift distribution for different \gls{aot} RMs prepared in
        \refhexaneLong\ hexane, \refisoLong\
        iso-octane for water loadings ranging
        between 0.9 to 20.3 and between 0.9 and
        7.4 for \gls{ctet} (\refcarbontetLong)}
    \label{fig:dispersants}
\end{figure}}

\newcommand{\figFlowChart}{\begin{figure}[tbp]
    \centering
    \setlength{\templen}{\dimexpr(\mylinewidth-0.25em)\relax}%
    \includegraphics[height=0.7\textheight]{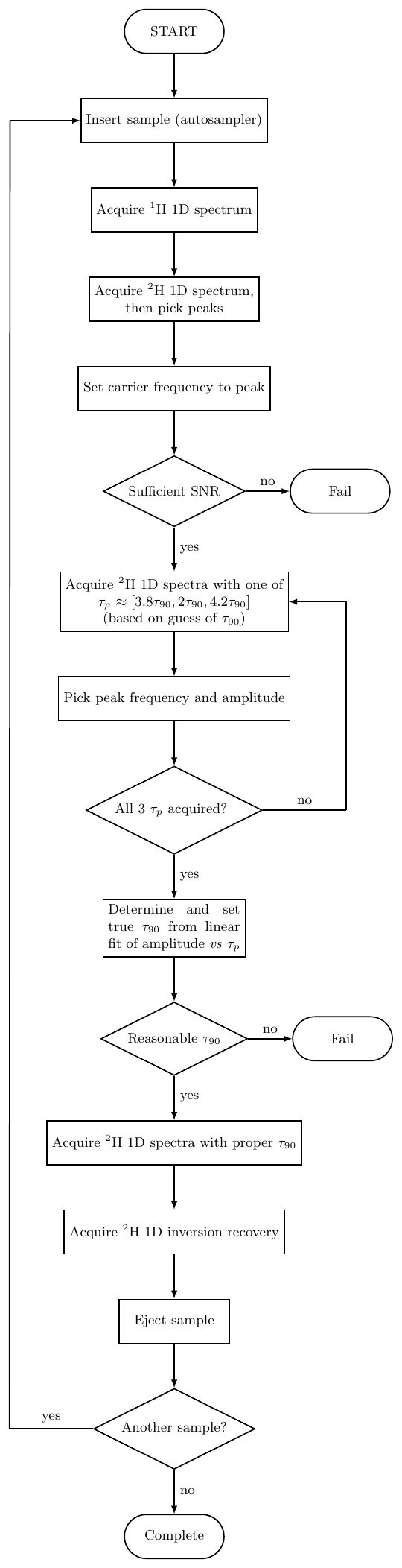}
    \caption{Flow chart describing the automated
        deuterium $T_{1}$
        characterization experiments, allowing for
        optimized pulse lengths for reliable
        $T_{1}$ measurements and chemical shift
        referencing to further the information
        attainable.
    }
    \label{fig:flowChartAU}
\end{figure}}

\newcommand{\figIgepal}{\begin{figure}[tbp]
    \centering
    \setlength{\templen}{\dimexpr(\mylinewidth-0.25em)\relax}%
    \includegraphics[width=0.9\mylinewidth]{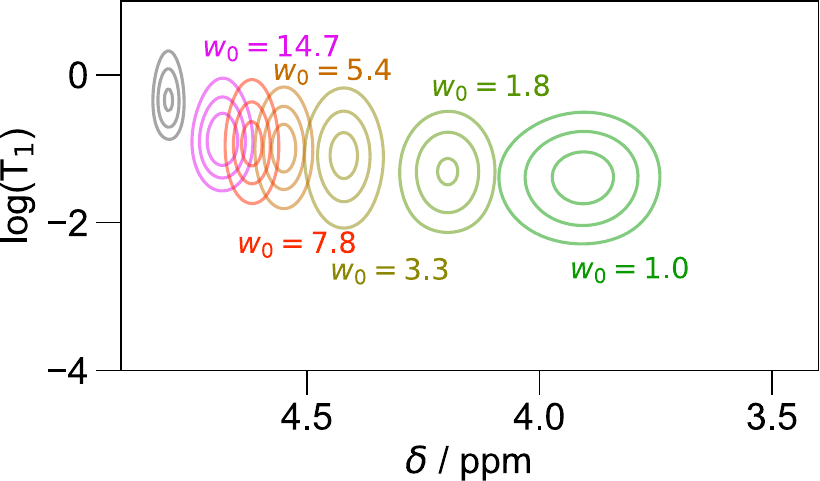}
    \caption{\gls{ilt} plots showing the $T_{1}$
        \textit{vs} chemical shift distribution of
        $\text{D}_2\text{O}$ for different Igepal RMs prepared in
        cyclohexane for water loadings ranging
        from 0.9 to 20.3.
    }
    \label{fig:IgepalT1}
\end{figure}}

\newcommand{\figIRCPMG}{\begin{figure}[tbp]
    \centering
    \includegraphics[width=0.75\mylinewidth]{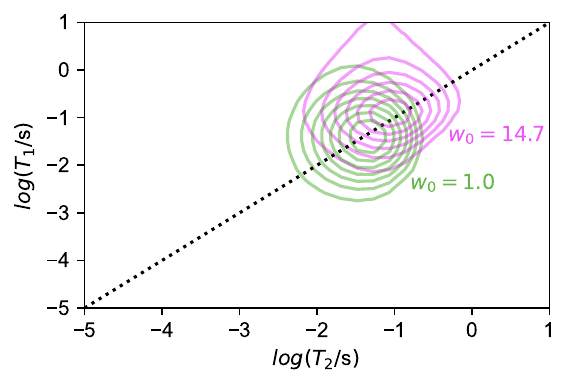}
    \centering
    \sepcomment
    \caption{Results for D$_{2}$O correlated $T_{1}$ - $T_{2}$ relaxation
        measurements, for the lowest and highest
        water loadings of Igepal/cyclohexane
        reverse micelle samples. A
        dotted line depicts values for which
        $T_2 = T_1$.
    }
    \label{fig:IRCPMG}
\end{figure}}

\newcommand{\figCTAB}{\begin{figure}[tbp]
    \centering
    \subfig{fig:waterT1CTAB}{CTABwater}
    \subfig{fig:hexanolT1CTAB}{CTABhexanol}
    \begin{tabular}{cc}
        \\
        \refCTABwater
        & \raisebox{-\height}{ 
        \includegraphics[width=0.9\mylinewidth]{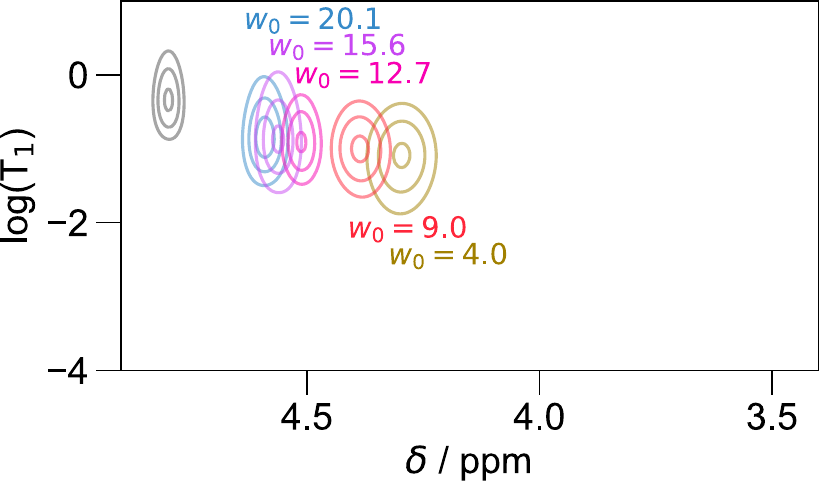}
        }
        \\
        \refCTABhexanol
        & \raisebox{-\height}{ 
        \includegraphics[width=0.9\mylinewidth]{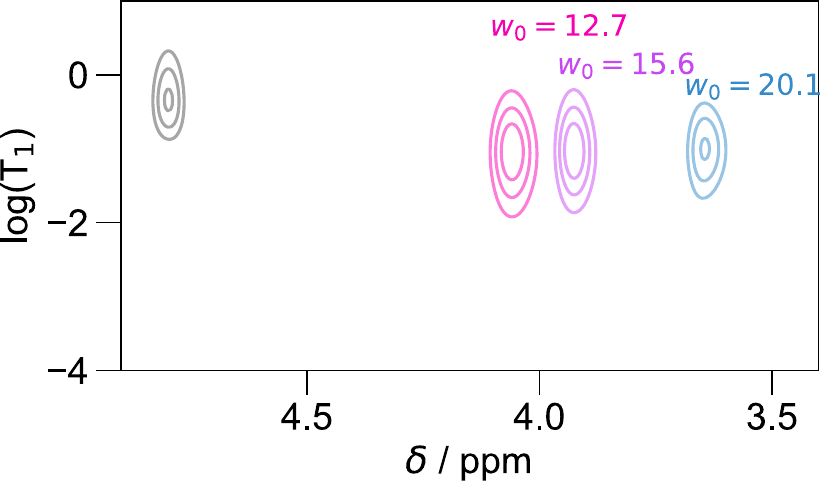}
        }
    \end{tabular}
    \centering
    \sepcomment
    \caption{Results for \gls{ctab} with hexanol
        co-surfactant in hexane for the
        \refCTABwater\ D$_{2}$O resonance and
        \refCTABhexanol\ hexanol resonance, which
        is observed in the samples above water
        loading 12. The lower plot is 10x less
        intense than the upper plot.
        Molar ratios of hexane to \gls{ctab} and hexanol
        to \gls{ctab} are 102 and 7.8, respectively, for the lowest $w_{0}$ and 152
        and 9.1, respectively, for the highest $w_{0}$.
    }
    \label{fig:CTABT1}
\end{figure}}

\newcommand{\figInclusion}{\begin{figure}[tbp]
    \centering
    \subfig{fig:InclusionGlucose}{Glucose}
    \subfig{fig:InclusionLargePEG}{LargePEG}
    \begin{tabular}{cl}
        
        \\
        \refGlucose
        & \raisebox{-\height}{ 
        \includegraphics[width=0.9\mylinewidth]{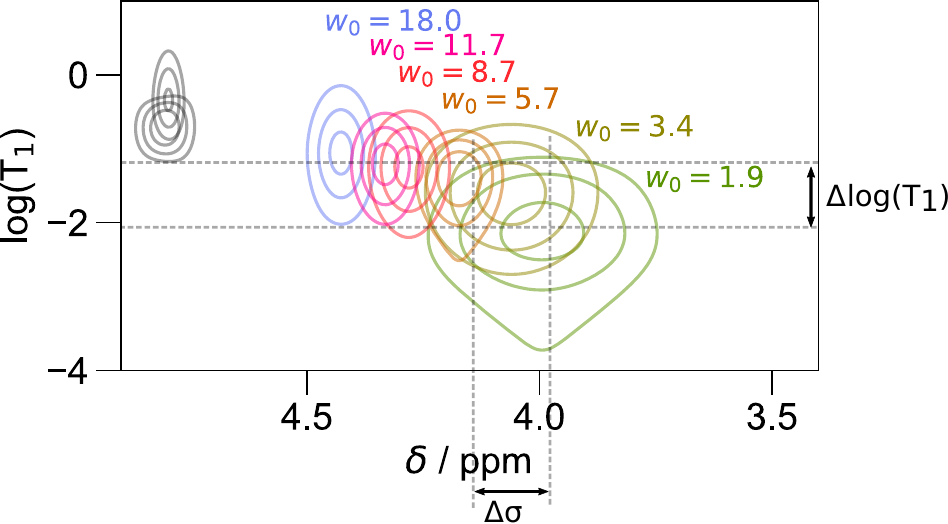}
        } 
        
        \\
        \refLargePEG
        & \raisebox{-\height}{ 
        \includegraphics[width=0.78\mylinewidth]{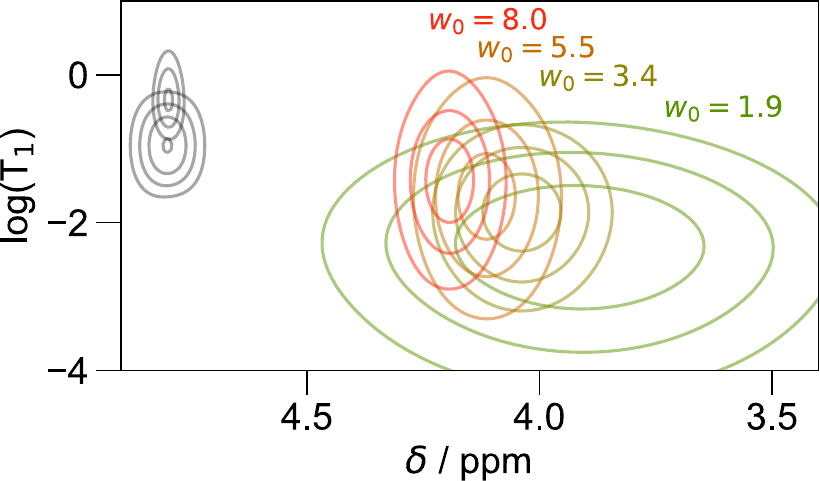}
        } 
    \end{tabular}
    \centering
    \caption{AOT RMs in isooctane prepared using
        mixtures of guest molecules in D$_{2}$O:
        \refGlucose\ 25 wt\% glucose and
        \refLargePEG\ 50 wt\% PEG-200. Reported
        water loadings use the volume of the added
        mixtures and are therefore similar
        $w_{0}$ RMs are expected to adopt a similar
        size water pool even though \eg the same
        50\% PEG solution contains about 50\%
        fewer water molecules per surfactant
        molecule.
        In both cases, the lower gray set of contours
        gives the signal for a solution of bulk water with
        osmolyte.}
    \label{fig:Inclusion}
\end{figure}}

\newcommand{\figSurf}{\begin{figure}[tbp]
    \centering
    \newlength{\templen}\setlength{\templen}{\dimexpr(\mylinewidth-0.25em)\relax}%
    \includegraphics[width=0.7\templen]{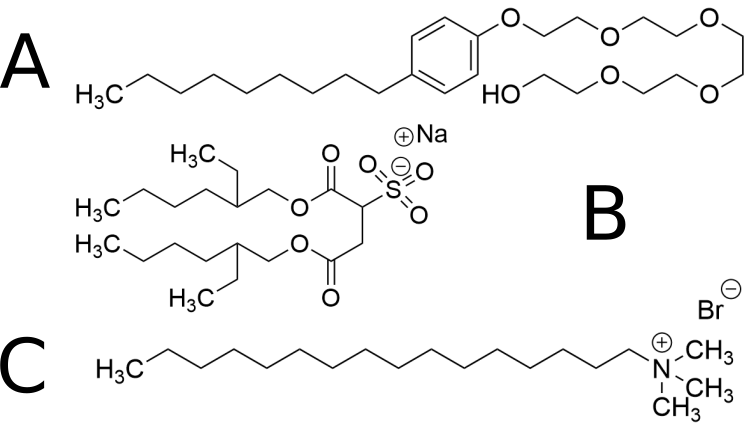}
    \caption{The three surfactants used in this
        work: (A) Igepal CO-520, (B) AOT,
    (C) \gls{ctab}.}
    \label{fig:surfactants}
\end{figure}}

\newcommand{\figBSA}{\begin{figure}[tbp]
    \centering
    \includegraphics[width=\mylinewidth]{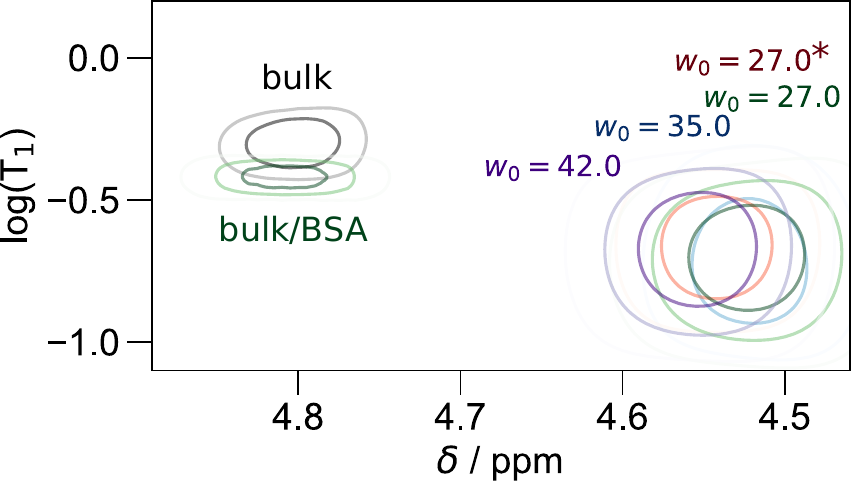}
    \caption{Bulk D$_{2}$O shown in gray with a
    bulk D$_{2}$O/BSA solution shown in green. At
    lower chemical shifts are a range of reverse
    micelles at higher water loadings ($w_{0}$ =
    27, 35, 42) containing BSA, demonstrating that
    two $T_{1}$ distributions are not observed. An
    RM without BSA of $w_{0}$ = 27 (indicated with
    asterisk) is also shown,
    demonstrating that RM-encapsulated BSA has negligible impact
    on chemical shift or $T_{1}$.
    (Because of the variety and high water loading of samples here,
    the color scheme doesn't adhere to that of the other figures in the text.)
    }
    \label{fig:BSA}
\end{figure}}

\newcommand{\figSpecDens}{\begin{figure}[tbp]
        \centering
        \includegraphics[width=\mylinewidth]{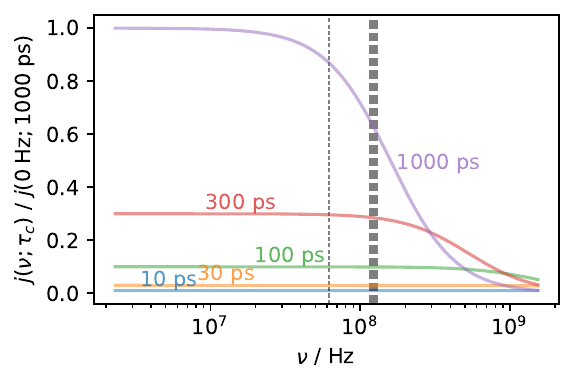}
        \caption{Displays the standard spectral density function,
            and how it varies with correlation time (different colors).
            (Curves shown here correspond to an exponential rotational correlation function.)
            The longitudinal relaxation rate
            ($R_1=1/T_1$)
            samples the spectral density at the deuterium resonance
            frequency ($\nu = \omega_D/2 \pi$, marked by the thin dashed gray line)
            and at $2\nu = \omega_D/\pi$
            (marked by the thick dashed gray line -- this value is scaled by 4),
            giving rise to \cref{eq:R1q}.
            The initial flat portion of each curve corresponds to the motional narrowing regime,
            and if the relaxation rate samples the spectral density curve in this regime,
            then \maincref{eq:R1q2} approximates the relaxation rate very well.
            As the rotational correlation time (indicated above each curve) increases,
            the spectral density of interactions arising from rotational motion spreads out in the frequency domain,
            leading to less relaxation in the motional narrowing regime.
            The values of $\nu_{rf}$ and $2\nu_{rf}$ marked here correspond to a magnetic field of 9.4~T
            ($9.4\;\text{T} \gamma_H / 2 \pi = 400 \; \text{MHz}$).
        }
        \label{fig:lorentzian}
\end{figure}}

\newcommand{\figCheckLinear}{\begin{figure}[tbp]
        \centering
        \includegraphics[width=\mylinewidth]{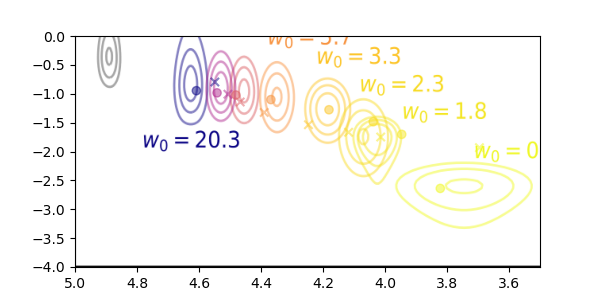}
        \caption{
            Filled circles, ``x'' markers, and contours are
            all colored according to water loading
            $w_0$
            (utilizing an earlier color scheme than the perceptually uniform
            one in the main text.)
            The ``x'' markers illustrate the best-fit
            attempt to fit a single core/shell model
            to the center of the distributions --
            where the relaxation rates
            (\textit{equiv} correlation times) average
            between the core and shell.
            The filled circles illustrate a similar
            best-fit attempt
            when the values of the relaxation time $T_1$
            (\textit{equiv.} $1/\tau_c$) are assumed to
            vary linearly with the fraction of core \vs
            shell.
            Both cases look for a linear averaging of
            chemical shielding/shift between core and
            shell.
        }
        \label{fig:checkLinear}
\end{figure}}
\newcommand{\figMetaAnal}{\begin{figure}[tbp]
        \centering
        \includegraphics[width=\mylinewidth]{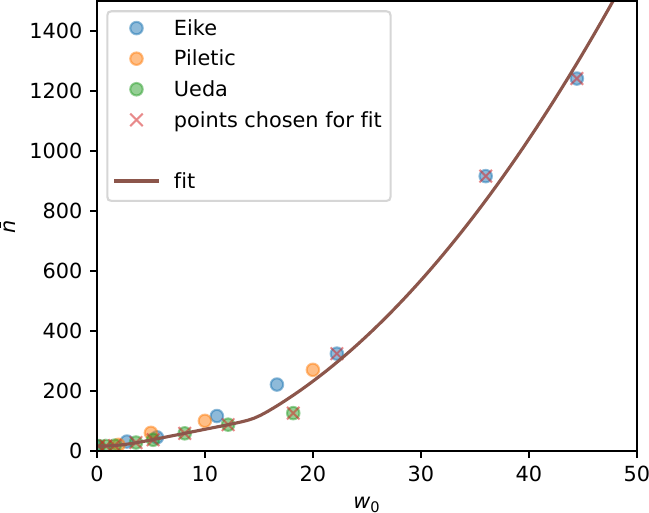}
        \caption{
            Fitting of a consensus number of water molecules as a function of reverse micelle (RM) water loading.
            Data digitized from \cite{Ueda1988MeaAggNum,Eicke1980SurNonSol,Piletic2006}
            is shown in the legend,
            as indicated by first author.
        }
        \label{fig:metaAnal}
\end{figure}}

%% file: content.tex
\section{Introduction}\label{sec:intro}
\paragraph{why observing confined water is important}
Water confined to nanometer-scale (yoctoliter and zeptoliter)
    volumes or pockets moves quite differently than bulk water.
These changes in motion tend to correlate with changes in the
    freezing of water (or lack thereof),\cite{brotoStudy1976,Boned1986ChaWatDis,Cerveny2016,quistWater1988}
    changes in excess entropy~\cite{DahanayakeFrontiers2018,Dahanayake2018,Chopra2010,Dyre2018}
    as well as changes to the type of chemistry that
    takes place in the
    water.\cite{sanfordSweet2018,Munoz-Santiburcio2021}
Therefore, routine methods that can observe alterations of water dynamics
    will soon answer questions of broad general
    interest for tracking and understanding
    thermodynamics and even the phase behavior of the
    most important types of water molecules --
    those at interfaces and inside pockets.
This is especially true if new,
    automated methods can offer insight about
    properties and behavior of water from the
    perspective of the water molecules.
Such methods
    offer the opportunity to understand
    how different solutes interact with a hydration
    layer (as modeled by a large $w_{0}$ RM) or a
    macromolecular pocket (as modeled by a small
    $w_{0}$ RM) at a very powerful, general level.
\paragraph{translational vs rotational}
While fine distinctions can be made among different modes of water diffusion,\cite{Laage2017}
    we can broadly classify them into two main categories:
First, several extant and developing methodologies investigate
    the translational diffusion of water -- \ie,
    thermal motions that displace water molecules
    on the order of a nanometer over a timescale of tens of picoseconds to a
    nanosecond,
    necessarily involving the cooperation of
    many water molecules.\cite{FranckMethEnz2018,Laage2017,Kubarych2018,Russo2004}
Second,
    more typically mature methodologies can probe the rotational diffusion of individual
    water molecules,
    and have relied on various physical
    techniques to determine the
    timescale of such motions.\cite{Okada1999,Vogel2002,Benedetto2013,debye1934part,Kaatze_rev,carlstroemWater1988a}
As the chemistry of porous and interfacial macrochemistry progresses,
    methods for interrogating rotational diffusion 
    should prove complementary with methods for interrogating
    translational diffusion.
In particular, rotational motion -- the focus of this publication --
    likely offers a better way to
    identify the presence of very slow-moving waters that could
    contribute dramatically to the free energy if they
    are released into the bulk.
In contrast, translation motion offers the opportunity to
    systematically map out relative differences between
    patches of water located at different sites on a
    macromolecule.\cite{FranckMethEnz2018}
\paragraph{²H relaxation a good probe of rotational motion}
Deuterium NMR relaxation is a
    sensitive probe for reading out
    rotational motion,
    as the quadrupolar interaction of the deuterium
    nucleus tends to provide the dominant relaxation
    mechanism.\cite{mantschDeuterium1977}
For decades,
    researchers have taken advantage
    of this capability
    to study samples
    with solid state NMR methods, most
    notably in lipid
    systems~\cite{seeligDeuterium1977,konigMolecular1994,frederickEffects2010}
    as well as to study
    variable properties of water molecules
    in other biologically relevant systems, such as
    DNA,\cite{Denisov_DNA} glucose
    solutions,\cite{moranStudy1999} and
    elastin.\cite{Sun2010}
In particular, the predominance of the quadrupolar relaxation mechanism in
    these deuterium studies generates a more accurate measure of
    intramolecular motion than the
    corresponding proton studies.
This accuracy arises because proton NMR relaxation
    reflects not only rotational motion,
    but also
    strong intermolecular dipolar interactions~\cite{carlstroemWater1988a,Hindman1973}
    as well as slow-timescale processes such as chemical exchange.\cite{Denisov_DNA}
\figSurf
\paragraph{pick out with D₂O 👈 integrate with quadrupolar}
A complication toward studying the water in these
    systems often arises from the fact that water is
    typically only a small component of the overall
    mixture; there is, in most cases, an inescapable
    amount of background signal that either arises
    from the dispersant or the surfactant.
Therefore, we propose here a routine involving
    deuterium NMR relaxometry, where D$_{2}$O is used
    in place of H$_{2}$O in sample preparation.
We demonstrate this technique
    on a series of RMs of different $w_{0}$ 
    as a means to study exclusively the water
    inside the RM.
Much attention has been paid to these systems from
    a variety of different communities.
For instance, in materials science, different
    additives, such as polymers~\cite{laiaLight2000}
    and co-surfactants~\cite{cuccoviaAnalysis2001}
    have been studied as a means to mediate certain
    properties of the system, such as droplet
    exchange and microemulsion stability.\cite{menassaHydrogen1985,nazarioNonionic1996}
In biophysical sciences, work toward
    encapsulating proteins inside these systems
    initially found relevance in protein purification
    and extraction~\cite{shiomoriEffective1995,hayesMechanism1997}
    but is now used as a unique method for studying
    the proteins themselves,\cite{wandHighresolution1998,senskeLocal2018,fuglestadChapter2019}
    presenting myriad ways to prepare these systems to
    suit the needs of the research at hand.
These RM
    systems can further serve as controlled
    environments for carrying out certain
    reactions,\cite{dasModulation2018,chaurasiyaReverse2017}
    delivering biologically active
    molecules,\cite{grooReverse2018}
    or studying confinement effects on encapsulated molecules.\cite{geethuSoft2017}
Substituting H$_{2}$O with D$_{2}$O has been shown
    to minimally affect the properties of the \gls{aot} RMs
    despite differences, such as density
    and heat capacity, between H$_{2}$O and D$_{2}$O
    \cite{dapranoVolumetric1992}.
Notably, here RMs serve as a proof of
    principal measurement for a general
    methodology applicable to other systems
    with trapped water;
    while RMs are relatively easy to prepare,
    the technique should apply equally well
    to other systems with trapped water,
    including solid mesoporous systems,
    porous and permeable polymer systems,
    and even biological systems with trapped
    or phase separated aqueous solutions.
\paragraph{identify need and that we correlate T₁ with σ}
We focus, in particular, on the fact that,
    despite its promise,
    deuterium relaxometry
    has lagged in terms of taking advantage of modern
    data-processing methodologies and modern automation
    techniques.
In particular, a few decades ago,
    Halle and others~\cite{quistWater1988}
    provided proof of principle demonstrations and
    several key applications,\cite{Denisov_DNA}
    and these rigorous techniques continue to offer
    insight into the behavior of
    water.\cite{Persson2018_range,Persson2018modifyH,Persson2018geometry,Persson2018Compressibility}
While such studies do seek a significant rigor in the understanding of
    rotational dynamics,
    they tend to do so by
    relying on specialized
    instrumentation,
    rather than capitalizing on the wide availability
    and automatability of modern commercial magnetic
    resonance instrumentation.
The work of the Boutis
    lab~\cite{Sun2011,Sun2010,Gul-E-Noor2015,Ukpebor2014,Watanabe2014}
    demonstrates the utility
    of 2D deuterium relaxometry
    on elastomeric peptides,
    and thus offers
    important clue as to how this discipline might be
    advanced.
Nonetheless, we
    still perceive a need for a
    simple and automated,
    yet powerful, measurement.
Furthermore, we are not aware of
    methodologies that explicitly correlate the
    mobility measurements attained from quadrupolar
    relaxation with structural clues about the hydrogen
    bonding matrix that can be gleaned from changes in
    diamagnetic
    shielding.\cite{wongStructure1977,Frank1973ProMagRes,maitraDetermination1984,Eicke1980SurNonSol}
Therefore, in this publication,
    we demonstrate a variant of 2D ROSY
    (relaxation-ordered) spectroscopy
    (analogous to diffusion-ordered DOSY,
    but replacing diffusive decay with relaxation
    decay).
To achieve this,
    we deploy a standard ``1.5~D \gls{ilt}''
    processing~\cite{venkataramanan2002solving}
    to display the correlation between
    the $^2$H NMR relaxation rate and the diamagnetic shielding
    (\ie chemical shift.)
We crucially combine this
    methodology with customized Bruker automation software to open
    the possibility of rapidly screening tens or even
    hundreds of samples.
\section{Theory}\label{sec:theory}
\subsection{Reverse Micelles}
While they serve as only approximate measures,
    two equations from previous literature help to
    conceptualize the size of RMs.
The radius of the water pool ($r$, in nanometers)
    can be approximated
    by~\cite{maitraDetermination1984,luisiReverse1988}:
    \begin{equation}
        r =  (0.175\;\text{nm}) w_0.
        \label{eq:waterSize}
    \end{equation}
    where $w_0$ gives the ``water loading'' --
    the molar ratio of water to surfactant.
Considering a \gls{rm} without guest molecules,
    the linear dependence of $r$ on $w_0$ is consistent with the
    basic consideration that, on the one hand, the
    volume ($V$) of the \gls{rm} interior is proportional to
    $r^3$, which is proportional to the number of
    water molecules per unit volume of solution, \ie
    $V \propto r^3 \propto [\text{H}_2\text{O}]$;
in contrast, surface area of the \gls{rm} water pool
    is proportional to the number of surfactant
    molecules per unit volume of solution, \ie
    $A \propto r^2 \propto [\text{surfactant}]$.
Various previous studies~\cite{pileticTesting2006,Eicke1976ForWatOil,Ueda1988MeaAggNum}
    document the number of AOT molecules (aggregation number $\bar{n}$) per
    \gls{rm},
    and therefore/alternately, the number of water molecules $\bar{n}w_0$.
These studies demonstrate varying levels of agreement
    and so justify some amount of meta-analysis.
In estimating a function to summarize this literature,
    we favor \gls{cppvpo} for low $w_0$ samples,
    as we expect the osmolality of the water to
    more accurately count the number of water
    molecules \textit{vs.} methods that rely in
    whole or part on the mass and/or density of
    the water
    (especially, since the AOT molecule is almost
    $25\times$ the mass of the water molecules).
The \gls{cppvpo} literature demonstrates a region
    with relatively constant $\bar{n}$,
    followed by a $\bar{n}$ that grows linearly
    with $w_0$.
For large spherical \gls{rms}
    we expect $\bar{n}\propto w^2_0$
    (and number of water molecules
        $\bar{n}w_0 \propto w^3_0$);
    as noted in,\cite{EskiciSizSizAOT2016}
    two references~\cite{Amararene2000AdiComAOT,Eicke1976ForWatOil}
    employing velocimetry \textit{vs.}
    ultracentrifugation roughly agree
    on values of $\bar{n}$ for $15<w_0<30$.
Though \cite{pileticTesting2006} offers a
    significant contribution to the field,
    we ignore the numbers presented for
    $\bar{n} w_0$ in
    \cite{pileticTesting2006},
    because they are implied to be rough estimates
    and the discrepancy with
    \cite{Eicke1976ForWatOil}
    is not explained.
%
Digitization and nonlinear fitting of these
datasets yields an equation of the form
\begin{equation}
    \begin{array}{rl}
        \displaystyle
        \exp\left( k \bar{n} \right)
        =&
        \exp\left( k a \right)
        +
        \exp\left( k (m w_0 + b) \right)\\
        \displaystyle
        &
        +
        \exp\left( k (l w^2_0 + c) \right)
    \end{array}
    \label{eq:numWaters}
\end{equation}
with $a=15.1$ controlling number of AOT molecules
in the constant-$\bar{n}$ regime,
$m=7.15$ controlling the slope of the linear region
(number of AOT molecules added per change in $w_0$),
$b=0.259$ the intercept of the linear regime,
$l=0.673$ giving the curvature of the quadratic
regime,
$c=-37.1$ the intercept of the quadratic
equation,
and $k=0.174$ controlling the curvature between
the constant, linear, and quadratic regimes.
Note that because the numbers for this fit come
from digitized figures,
the significant figures are not intended as a
claim of accuracy, but merely offer a means to
reproduce the curve to estimate $\bar{n}$.
Furthermore, while the form of the equation
roughly resembles an addition of concentration or
equilibrium terms with free energies scaling as
polynomials of $w_0$,
the rationale for the choice of functional form
arose simply in order to generate a constant
\textit{vs.} linear \textit{vs.} quadratic regime,
in order to compile the $w_0$ scaling noted in
\cite{Ueda1988MeaAggNum} \textit{vs.\@}
\cite{Eicke1976ForWatOil}.
\paragraph{guest w₀}
Next, when incorporating guest molecules,
    the definition of water loading ($w_0$)
    can become ambiguous.
Here, $w_{0}$ is defined generally as
\begin{equation}
    w_{0} = \frac{ (V_{solution})(55.2\;\text{M})}{ (\text{mol surfactant}) }
    \label{eq:w0_approx}
\end{equation}
where $V_{solution}$ includes any guest molecules,
    and 55.2~M is chosen for the molarity of pure $\text{D}_2\text{O}$
    (\textit{vs.\@} $\text{H}_2\text{O}$ which is 55.4~M).
This was referred to as the ``$w_{0}$-equivalent''
    (as opposed to water:surfactant molar
    ratio)
    preparation
    in the work from
    Wiebegna-Sandford, Levinger, and
    co-workers.\cite{sanfordSweet2018}
To the extent that addition of aqueous
    solutes inside the \gls{rm} surfactant shell
    doesn't perturb the density of aqueous solution,
    $w_{0}$-equivalent \gls{rms}
    should adopt a similar size based on the $w_0$
    value,
    regardless of the identity (or absence) of guest
    molecules.
However, Wiebegna-Sandford \etal found that
    \gls{rms} with guest molecules
    exhibit a slight size
    reduction relative to water-only RM preparations of the same $w_{0}$,
    owing to a proposed
    greater spherical nature resulting from the osmolyte
    inclusion.
\paragraph{core-shell interpolation}
Finally, we note
    (from simple geometric considerations)
    that if a core-shell model can be assumed,
    then \textit{any} property $x_{avg}$ (whether
    chemical shift, relaxation rate, \etc)
    that presents as a weighted average of the core water and
    shell water will average as:
    \begin{equation}
        x_{avg}
        = x_{shell}
        + (x_{core}-x_{shell}) \frac{\left( r_{water}-t \right)^3}{r_{water}^3}
        \label{eq:coreshell}
    \end{equation}
    where
    $t$ is the thickness of the shell,
    $x_{core}$ is the value of property for water
    molecules in the core region,
    $x_{shell}$ is the value of the property for water
    molecules in the shell region.
\subsection{$^2$H Relaxation}
\paragraph{spectral density function}
The rate of longitudinal relaxation,
    $R_1$,
    of deuterium ($I=1$) in solution is dominated
    by the quadrupolar relaxation mechanism (\cref{sec:QuadDetails}), in contrast to
    relaxation of $^{1}H$ nuclei
    ($I=\frac{1}{2}$) which is typically
    dominated by dipolar relaxation in
    solution.\cite{Hindman1973}
In the extreme narrowing limit (see \cref{fig:lorentzian}), where the
    correlation time $\tau_c$ is very fast
    relative to the resonance frequency
    $\nu_D$ (\ie $ 2 \pi \nu_D
    \tau_{c} \ll 1$),
    $J(\nu)$ evaluated at frequencies of
    $\nu_D$ and $2\nu_D$ will equal
    $2\tau_{c}$.
This condition is applicable to the overwhelming majority of samples in this work,
and simplifies \cref{eq:R1q} to
\begin{equation}
    R_1 =
    \frac{3}{8}
    \left(\frac{e^{2}Qq}{\hbar}\right)^{2}
    \tau_{c}
    \label{eq:R1q2}
\end{equation}
through which we can relate the rotational
correlation time of the deuterium nuclei $\tau_{c}$ to the corresponding
$R_1$.
Notably, under the extreme narrowing limit,
$J(0)$, $J(\nu)$, and $J(2\nu)$ all evaluate
to $2 \tau_c$ rendering the value of \cref{eq:R2q} 
equal to that of \cref{eq:R1q2}.
In other words, in the motional narrowing limit
(and only in the motional narrowing limit)
$R_1$ = $R_2$.
\paragraph{averaging of relaxation}
When considering relations to other techniques, it
    is important to note that the $\tau_c$ above can
    represent the net effect of several individual
    processes or environments.
When multiple processes contribute to the rotational
    diffusion, the rate of decay of the quadrupolar
    correlation function is additive -- \ie, the total
    correlation time takes the form
\begin{equation}
    \frac{1}{\tau_c} \approx \sum_i
    \frac{1}{\tau_{c,i}}
    \label{eq:tauCaveraging}
\end{equation}
    where the individual $\tau_{c,i}$
    give the rotational correlation times of the
    individual processes.
Note that the simultaneous presence of
    shorter timescale motions suppresses the effect of
    longer timescale motions.
For example, if the correlation function -- which
    physically represents the persistence of the water
    molecule orientation here -- decays due to both the
    tumbling of the water molecule relative to the
    entire \gls{rm},
    as well as due to the tumbling of the entire
    \gls{rm},
    then the shorter of the two correlation times will
    dominate,
    as indicated by \cref{eq:tauCaveraging}.
Similarly, in the unlikely event that the molecules
    exchange -- on a timescale faster than the correlation
    time -- between environments where they have different
    mobilities,
    then the decay of the correlation function will
    average between the two environments,
    as indicated in \cref{eq:tauCaveraging}.
This corresponds to the averaging of the inverse of the
    relaxation rates,
    or to an averaging of the relaxation times.
\paragraph{vs tau averaging}
On the other hand,
    in the more likely event that
    molecules exchange between environments on a
    timescale longer than $\tau_c$,
    but still shorter than
    the timescale of the $^2$H relaxation
    (typically order of tens of ms),
    their relaxation rate would average.
Following from \cref{eq:R1q2},
    this means an observation of $\tau_c$ \via the
    relaxation time would report an average:
    \begin{equation}
        \tau_c 
        = \sum_i a_i \tau_{c,i}
        \label{eq:R1averaging}
    \end{equation}
    where
    the individual $\tau_{c,i}$ give
    the different rotational correlation
    times that the water molecule adapts in
    each particular environment,
    and the $a_i$ give
    the fractional amounts of time for which the molecule
    resides in each environment.
\paragraph{conclusion of different types of averaging}
Importantly, $^2$H measurements that observe a single
    relaxation time do not rule out the possibility of
    fast exchange between environments with different
    correlation times.
Conversely, if a distribution is observed with
    more than one relaxation time, such an
    observation indicates the existence of two
    populations of water that remain separate on a
    timescale similar to or greater than the
    relaxation time.
\section{Experimental}
\subsection{Sample Preparation}\label{sec:Sample_Prep}
Equal volume samples of the
    lowest $w_{0}$ sample and highest $w_{0}$ samples
    ($w_{0} = 1$ and $20$, respectively)
    were prepared
    from 2.82~g AOT (6.3~mmol)
    and 141~mg of AOT (0.3~mmol),
    respectively, each in 6765 μL solvent
    (which varies in exact identity).
Vortexing yielded a completely transparent solution.
After subsequent addition of
    D$_{2}$O (115~μL, 6.3~mmol to each),
    and additional
    $3\times$ vortexing for 30 seconds,
    samples sat at room temperature until
    completely clear.
Perdeuterated hexane
    (38~μL, 0.29~mmol) was added to each solution as a
    concentration and chemical shift reference.
Equal volumes of high and
    low $w_{0}$ samples were mixed
    to generate a third sample
    (yielding averaging of the reciprocal of the water loadings).
Subsequent samples were prepared by
    mixing two previously prepared sample mixtures,
    always using equal volumes.
(Thus, while mixing adusts
    $w_0 = \left( w_{0,1}^{-1}+w_{0,2}^{-1}
    \right)^{-1}$,
    the concentration of water per total solution
    volume remains fixed.)
After addition of TMS
    ($\sim 1\;\text{μL}$),
    each sample was vortexed $3\times$ 30~s.
In general, thermodynamic stability of nanoemulsions can be
    identified by the turbidity
    of the solution~\cite{deSolution1995} and only
    clear and transparent solutions were analyzed here.
NMR samples were then prepared by adding 600~μL of 
    each mixture to separate 5~mm tubes and flame sealing.
\paragraph{guest molecule prep}
For guest molecule studies, solutions of the guest
    molecule in D$_{2}$O were prepared at the stated
    concentrations, and the appropriate volume of
    solution was substituted for pure D$_{2}$O.
For the \gls{bsa} samples, the D$_{2}$O solution also contained 50~mM Tris (buffer)
    and the pH was adjusted to 7.4, following
    established preparation procedure.\cite{Marzola1991}
\subsection{NMR Experiments}\label{sec:NMRExp}
Data was collected using a Bruker Avance III HD
    400 MHz spectrometer with a room temperature smart
    probe.
Deuterium experiments
    were run on the lock channel
    at 300 K after auto-shimming with a CDCl$_{3}$
    standard.
\paragraph{automation}
When acquiring inversion recovery experiments,
    it is important that the pulse tip angles
    and exact resonance frequency are
    properly calibrated;
    here, reference 1D spectra,
    both $^2$H and $^1$H
    (from TMS, surfactant, and solvent protons, and residual
    HOD protons),
    were also desired.
To automate data acquisition, an AU program was
    developed to acquire a single $^{1}$H scan,
    followed by a series of $^{2}$H scans to determine
    the optimal transmitter frequency and pulse width,
    in order to run an inversion recovery experiment
    with optimized parameters
    (\cref{fig:flowChartAU}).
\paragraph{figure}
\figFlowChart 
\paragraph{acquisition parameters}
Given the short (450~ms) $T_1$ of $^{2}$H for D$_{2}$O
    at room temperature (compared to 3.6 s for
        the $^1\text{H}$ of
        H$_{2}$O)~\cite{mantschDeuterium1977}
    compounded with the shortened
    relaxation times for water inside a RM, a maximum repetition delay of 1 s
    satisfies the requirement of 3-5x $T_1$ for these RM samples.
Typical inversion recovery experiments
    (customized pulse sequence~\cite{beatonModernized2022}~\cref{sec:AUandPPG})
    involved 16 indirect steps with an 8
    step phase cycle, acquiring 4096 complex
    points at 614~Hz spectral width
    along the direct dimension
    with only one scan per phase cycle step.
\paragraph{field referencing and data proc}
TMS provides a measure
    of the field strength ($B_0$) in the $^{1}$H spectra taken
    immediately before the inversion recovery
    experiments.
By calibrating the TMS peak in the $^{1}$H
    spectrum to 0 ppm, and using the relative magnitude
    of the gyromagnetic ratios for the $^{1}$H and
    $^{2}$H experiments, one is able to effectively
    reference the position of the $^{2}$H peaks
    relative to the field readout for the TMS peak
    (see \cref{sec:susceptibility} for more details).
The resulting 2D NMR data were processed using code developed
    in house,\cite{beatonModernized2022,pySpecData}
    with the \gls{ilt} performed via adaptation
    of the Butler Reeds Dawson (BRD)
    algorithm.\cite{butlerEstimating1981,venkataramanan2002solving}
Whether the data is 1.5D or 2D,
    we utilize the basis set compression of~\cite{venkataramanan2002solving}
    for greater efficiency.
\paragraph{colorscheme}
In order to present the contours for different
    water loadings ($w_0$),
    the perceptually uniform ``Lab''
    color space~\cite{skimageColor}
    assisted in designing a custom colormap
    (above \cref{fig:HexaneT1}).
The vector distance of $L,$ $a,$ and $b$ components
    in the Lab space corresponds
    to accepted perceptual difference in the
    corresponding colors.
This allows a choice of color map where
    the variation in water loading values
    corresponds to the perceptual variation in the
    colors.
Specifically, we choose the lightness ($L$) component
    to be the same for
    all water loading values,
    while the magnitude of gradient
    ($\sqrt{ \frac{\partial a}{\partial w_0} +
    \frac{\partial b}{\partial w_0}}$)
    in the color components
    $a$ and $b$ was set equal across the chosen
    colormap.
Thus,
    to match the modernization of the data
    acquisition and processing,
    the details of the data presentation also
    capitalize on the widely available
    color-theory tools.
\section{Results and Discussions}\label{sec:results}
\paragraph{what we offer that is new}
The new automated technique presented here enables a systematic study
	of the water pools in RMs as a function of $w_{0}$,
	for RMs prepared with different surfactants,
    different dispersants, and different additives,
    monitoring each system
    for evidence of correlated changes in rotational mobility and average
    hydrogen bonding strength.
In particular,
    it explores
    a series of different \gls{rm} samples
    in order to test the hypothesis that
    distinct and noticeable changes in the pattern of
    correlated rotational mobility and hydrogen bonding
    will occur as the size of macromolecular structures
    (here the size of the RM water pool)
    crosses length scales comparable to the
    correlation length of
    water.\cite{Chandler2007,Chandler2005,Cerveny2016}
\paragraph{a summary of RMs}
\Gls{rms} are thermodynamically
    stable mixtures comprising droplets of water
    encapsulated by surfactant
	(occasionally with added co-surfactant)
	and dispersed in an apolar
    medium.
As a demonstration system for this technique,
    \gls{rms} offer a unique opportunity to construct water
    pools with a continuous range of different sizes,
    while keeping the chemistry at the outer edge of
    the pool the same for all sizes.
One can rapidly generate a wide range of
    \gls{rms}
    without advanced synthetic capabilities,
    and they serve as useful nanocontainers for
    chemical reactions,
    as well as adjustable and controlled model systems
    for the study of chemical confinement.
The water loading also correlates to the size of
    the water pool, with sizes ranging from less than
    1 nm to slightly more than 10 nm at low
    ($w_0 = 1$) and high ($w_0 = 60$) water loadings,
    respectively.\cite{maitraDetermination1984,luisiReverse1988}
The water pools inside these \gls{rms} have been
    studied for several decades by a range of
    techniques, including NMR,\cite{wongStructure1977,demarco1HNMR1986,carlstroemWater1988a}
    neutron scattering,\cite{fredaElastic2001,harphamWater2004}
    light scattering,\cite{zulaufInverted1979}
    dielectric relaxation,\cite{dangeloHighfrequency1995,dangeloDynamics1996}
    infrared spectroscopy,\cite{pileticDynamics2005,pileticTesting2006,fennWater2011,henselMolecular2017}
    and ESR.\cite{haeringCharacterization1988,hauserInteraction1989,baglioniElectron1991,wassermanSpin1994}
These various methods all offer support for a general
    increase in the bulk-like nature of the water
    within this water pool as the water loading
	(\ie, $w_0$, water to surfactant molar ratio),
    increases.
However, frequently, studies may be confined to one
	type of \gls{rm} system at a time,
	or else focus on \gls{rm} prepared only with a
	subset of possible $w_0$.
\subsection{Trend of Correlated Measurement \textit{vs.} $w_0$}
\figDispersants
\paragraph{describe shape of trend}
An increase in $w_{0}$
    induces
    a correlated increase in chemical shift and
    $T_1$ in all cases
    (hexane, iso-octane, and \gls{ctet}),
    as shown in
    \cref{fig:HexaneT1,fig:IsoT1,fig:CCl4T1}.
In all cases, we observe a rather sharp increase
    in the order of magnitude of $T_1$ from the lowest water loadings up to
    $w_{0}\sim 3-5$.
At $w_{0}\sim 3-5$,
    the signals transition
    into a regime of gentle increase of $T_1$ that
    extends up to
    the highest water loadings.
Both the diamagnetic shielding
    and the $T_1$ approach, but never quite reach,
    the bulk values.
(\gls{aot} preparations were also stable to higher
    $w_0$ than shown here, and acquired,
    but these merely extend the trend seen here,
    and cannot be compared to other surfactants
    where such high $w_0$ cannot be achieved.)
Furthermore, a qualitative extension of the
    observed trend appears to closely,
    but never quite exactly,
    converge on bulk water.
This serves as a reminder that even the
    largest \gls{rm} water pools are small on
    a bulk scale,
    and may be subject to effects,
    including but not limited to:
    pH modulation by the
    surfactant head groups,
    different rates of proton exchange,
    local variations in the magnetic
    susceptibility,
    or non-spherical shape distributions,
    all of which might contribute slight
    variations to the diamagnetic shielding
    or longitudinal relaxation.
\paragraph{compare to previous literature}
An extremely popular and widely studied
    surfactant, \gls{aot} readily forms
    \gls{rms} in a three component mixture of
    surfactant, dispersant, and water.\cite{Eicke1980SurNonSol,deSolution1995}
In previous literature, it has not been atypical
    to develop results and conclusions from
    data that focus entirely on
    \gls{rms} in the higher water loading regime
    that here corresponds to the regime of gentle
    increase in $T_1$.
Previous results support the trend in D$_{2}$O
    $T_1$ measurements in this regime, both in
    \gls{aot}~\cite{carlstroemWater1988a} and in other
    surfactant
    systems.\cite{hansenHighresolution1974}
However, the data of \cref{fig:HexaneT1,fig:IsoT1,fig:CCl4T1}
    highlight the value of expanding this viewpoint
    both to observe the correlation of
    chemical shift and $T_1$,
    as well as to span the entire range of
    thermodynamically stable preparations of varying
    $w_{0}$,
    including low $w_0$ values.
\paragraph{inflection point real and crossover to more cooperative}
We propose that
    these measurements lends credence to the idea that the
    chemical/physical environment inside the \gls{rms}
    truly undergoes a crossover between two limiting
    cases.
At the extreme of low $w_{0}$,
    identified here by a signal trend with very negative
    curvature,
    individual water molecules interact tightly with
    the surfactant via hydrogen bonding or
    charge-dipole interactions.
The correlated change in chemical shift and $T_1$
    tells a consistent story indicating that
    confinement leads to a dramatically non-bulk-like
    environment where hydrogen bonds are possibly broken and
    where rotational dynamics are slowed.
At the extreme of
    high $w_{0}$,
    identified here as the regime of more
    gently increasing $T_1$ and chemical
    shift,
    waters are expected to interact overwhelmingly with
    other water molecules
    \textit{via} hydrogen bonding,
    leading to more ``bulk-like'' behavior.
The interactions with other water molecules are expected
    to be cooperative in nature,
    while the interactions with the surfactant are
    potentially far more localized.
It is worth noting that the crossover
    between these two regimes
    corresponds roughly to important
    \gls{rm} size thresholds in other studies,
    such as water loadings
    where confined
    water does not freeze ($w_0\approx 6.5$),
    and where other dramatic
    effects of confinement are
    evident.\cite{biswasNonmonotonic2012,Boned1986ChaWatDis,Lee2014InfPumPro}
It is also worth noting how this points to
    the value of a strategy not of attempting
    extreme rigor in the interpretation of
    individual measurements,
    but rather in employing modern technology
    to automate many measurements and to
    observe the correlated changes of more
    than one measurement.
\subsection{Effect of Different Dispersants}\label{sec:disp}
\paragraph{dispersants affect water 👈 results}
The identity of the dispersant is also an important
    factor in determining the stability of RM
    system.\cite{deSolution1995}
Careful combination of dispersants provide a
    nuanced control of phase
    stability,\cite{myakonkayaControl2009}
    emphasizing the complex relationship
    dependence of \gls{rm} stability on dispersant. 
Dispersant choice also has been shown to impact the shape of
    \gls{rm}s,\cite{balakrishnanSmallAngle2008} and such shape fluctuations have been shown
    to complicate data interpretation and analysis.\cite{carlstroemShape1989}
\paragraph{hexane and isooctane similar -- differences in phase due to surfactant, not water}
Despite the ability of isooctane to accommodate
    a larger $w_{0}$ than hexane before turning
    into (thermodynamically unstable, \textit{i.e.}, non-\gls{rm})
    emulsions,\cite{lamesaPhase1992,deSolution1995}
    \cref{fig:IsoT1} shows that
    up to a $w_{0}$ of
    approximately 20,
    the correlated measurement reports
    no dramatic difference in the trend in mobility
    \vs hydrogen bonding environment of the water pool when
    suspended in isooctane \vs hexane.
Thus,
    these data would support the result that
    free energy contributions from the water
    itself likely cannot rationalize the differing
    phase stability
    and that interactions between the solvent and surfactant
    likely likely drive the differences in stability of the two
    systems.
(Likely, differences arise from the ability of each
    to support different surfactant curvatures.)%
\begin{table}
\begin{center}
    \caption{Values for $\tau_{c}$ determined from the $T_1$ measurements for
    various RM samples.\\
    \textbf{Legend}: $^{\dagger}$ includes hexanol as co-surfactant;
    $^{\ast}$ includes 25\%-weight glucose as guest molecule;
    $^{\wedge}$ includes 50\%-weight PEG-200 as guest molecule.
    }
    \label{tab:table1}
\begin{tabular}{|c|c|c|c|c|}											
\hline\hline
\textbf{Surfactant} & \textbf{Solute/Guest} & \textbf{Dispersant} & \textbf{$w_0$} & \textbf{$\tau_c$} (ps) \\
\hline
AOT              &                    & Isooctane   & 0.9  & 1276.9 \\
AOT              &                    & Isooctane   & 3.5  & 80.6   \\
AOT              &                    & Isooctane   & 6    & 16.1   \\
AOT              &                    & Isooctane   & 9.3  & 12.8   \\
AOT              &                    & Isooctane   & 12.8 & 9.0    \\
AOT              &                    & Isooctane   & 20.7 & 8.1    \\
AOT              &                    & Hexane      & 0.9  & 805.7  \\
AOT              &                    & Hexane      & 1.8  & 80.6   \\
AOT              &                    & Hexane      & 2.3  & 57.0   \\
AOT              &                    & Hexane      & 3.3  & 22.7   \\
AOT              &                    & Hexane      & 5.7  & 16.1   \\
AOT              &                    & Hexane      & 8.9  & 10.1   \\
AOT              &                    & Hexane      & 12.3 & 9.0    \\
AOT              &                    & Hexane      & 20.3 & 6.4    \\
AOT              &                    & \gls{ctet}   & 0.9  & 805.7  \\
AOT              &                    & \gls{ctet}   & 1.6  & 160.8  \\
AOT              &                    & \gls{ctet}   & 2.7  & 71.8   \\
AOT              &                    & \gls{ctet}   & 5.2  & 20.2   \\
AOT              &                    & \gls{ctet}   & 7.4  & 12.8   \\
Igepal           &                    & Cyclohexane & 1    & 25.5   \\
Igepal           &                    & Cyclohexane & 1.8  & 22.7   \\
Igepal           &                    & Cyclohexane & 3.3  & 10.1   \\
Igepal           &                    & Cyclohexane & 5.4  & 9.0    \\
Igepal           &                    & Cyclohexane & 7.8  & 8.1    \\
Igepal           &                    & Cyclohexane & 14.7 & 5.7    \\
CTAB$^{\dagger}$ &                    & Hexane      & 4    & 25.5   \\
CTAB$^{\dagger}$ &                    & Hexane      & 9    & 22.7   \\
CTAB$^{\dagger}$ &                    & Hexane      & 12.7 & 10.1   \\
CTAB$^{\dagger}$ &                    & Hexane      & 15.6 & 9.0    \\
CTAB$^{\dagger}$ &                    & Hexane      & 20.1 & 8.1    \\
AOT              & glucose$^{\ast}$   & Isooctane   & 1.9  & 160.8  \\
AOT              & glucose$^{\ast}$   & Isooctane   & 3.4  & 80.6   \\
AOT              & glucose$^{\ast}$   & Isooctane   & 5.7  & 32.1   \\
AOT              & glucose$^{\ast}$   & Isooctane   & 8.7  & 20.2   \\
AOT              & glucose$^{\ast}$   & Isooctane   & 11.7 & 12.8   \\
AOT              & glucose$^{\ast}$   & Isooctane   & 18   & 10.1   \\
AOT              & PEG-200$^{\wedge}$ & Isooctane   & 1.9  & 202.4  \\
AOT              & PEG-200$^{\wedge}$ & Isooctane   & 3.4  & 101.4  \\
AOT              & PEG-200$^{\wedge}$ & Isooctane   & 5.5  & 80.6   \\
AOT              & PEG-200$^{\wedge}$ & Isooctane   & 8    & 25.5   \\
\hline
\end{tabular}
\end{center}
\end{table}
\paragraph{chemical shift differences of all 3 dispersants}
This is not to say that there is no observable difference
    upon change of dispersant:
\Gls{rms} prepared with equivalent $w_{0}$
    in the different dispersants
    do display some differences in the
    chemical shift positions.
In particular,
    while maintaining a qualitatively similar 2D correlation pattern,
    the range of chemical shifts of $\text{D}_2\text{O}$
    inside \gls{rms} dispersed in hexane expands relative to those
    from \gls{rms} dispersed in isooctane.
The chemical shifts
    of $w_0=1$ and $w_0=1.9$ correspond to
    slightly greater magnetic shielding in hexane \vs isooctane,
    which could suggest slightly weaker average hydrogen bonding,
    while at $w_0\approx 20$,
    the opposite is true.
Meanwhile,
    \gls{ctet} shows peaks with
    more shielding (relative to both isooctane and hexane)
    for all $w_0$,
    again suggesting the water pool has a more frustrated
    hydrogen-bonding matrix.
While it is possible the suggested changes in
    hydrogen-bonding could provide insight into
    the change in the stability of the RMs,
    it is also possible that they arise from
    differences in ellipticity of the different
    \gls{rm} systems.
It is also possible
    that stronger electric fields at
    the \gls{ctet} RM interface due to interactions
    between the surfactant and \gls{ctet}, or less
    intercalation of \gls{ctet} into the surfactant
    layer lead to less shape fluctuation and
    longer-lived water/AOT interactions
    that in turn reduce the
    ability of water to hydrogen bond with other water
    molecules.
Finally, it is important to acknowledge that
    changes in average hydrogen bonding
    strength or duration might not be the only reasons
    for changes in chemical shift,
    especially in smaller \gls{rms};
    differences in polarizability may arise
    from more detailed changes to the motion
    of electrons within the sulfate groups,
    \textit{etc.}
Therefore, we will interpret
    changes in chemical shift broadly as a ``frustration''
    of the hydrogen-bonding matrix,
    simply reflecting that
    effects leading to changes in the
    magnetic shielding
    at the deuteron
    (or even the presence of ring currents, \etc)
    correspond to an
    alteration away from the
    transiently hydrogen-bonded structures
    typically formed by bulk-like water.
\paragraph{CCl₄ is mostly confined, all the time}
It is known that \gls{ctet} cannot support $w_{0}$
    larger than approximately
    10,\cite{fennWater2011} owed to its high
    polarizability, leading to strong interactions
    between it and the surfactant headgroup
    dipole.\cite{Eicke1980SurNonSol}
In spite of this, \gls{ctet} continues to be widely
    used for RM studies due to its spectroscopic
    silence in important measurements, such as IR
    spectroscopy,\cite{henselMolecular2017,tanDynamics2005,fredaHydration2002}
    neutron
    scattering,\cite{fredaHydration2002,fredaElastic2001}
    and dielectric
    relaxation~.\cite{fredaHydration2002,dangeloHighfrequency1995}
It is worth mentioning that,
    comparing to
    \cref{fig:IsoT1,fig:HexaneT1}
    all stable \gls{ctet} \gls{rms}
    fall in the regime where the trend of
    signals \vs $w_0$ exhibits a significant
    negative curvature.
That is, even the largest \gls{ctet}
    \gls{rms} that can be made have not yet
    entered the regime where the order of magnitude 
    of rotational motion changes gently as a function of the
    hydrogen bonding strength.
One notable interpretation of this result is that
    \gls{ctet} \gls{rms} never enter the regime 
    where they can be thought of even
    approximately as possessing a separate
    hydration layer ``shell''
    and more freely moving water ``core'';
    rather all values of $w_0$ possible with
    \gls{ctet} dispersant likely correspond to
    all water molecules
    experiencing some level of confinement.
As various seminal infrared and dielectric
    studies~\cite{dangeloHighfrequency1995,dangeloDynamics1996}
    utilized the \gls{ctet} system as a model system,
    the interpretation of those studies would
    be likewise colored by the realization that
    all the water under study was subject to what
    appears here as identifiable
    confinement.
\subsection{Effect of Different Surfactants}\label{sec:surf}
\figIgepal
\figCTAB
\paragraph{surfactant affects water 👈 results}
Choice of surfactant has been shown to
    impact the behavior of the RM water pools, with
    differences noted in the behavior of water in
    AOT, an anionic surfactant,
    compared to Igepal, a neutral surfactant.\cite{moilanenConfinement2007}
To further complicate comparative measurements,
    some surfactants such as \gls{ctab}, (cationic)
    or Triton X-100 (neutral)
    typically are used in the presence of
    co-surfactants,\cite{ciszewskiTakes2019,kaushikOptimizing2007}
    or particular dispersant
    preparations~\cite{lipgensPercolation1998,moilanenConfinement2007}
    -- where these choices in sample preparation
    usually center around the desire to achieve higher
    water loadings ($w_{0}$)s while still maintaining
    thermodynamically stable \gls{rm} microemulsions.
Small differences in the surfactant structure,
    such as the number of ethylene oxide units in
    Triton X-100, can additionally alter the stability
    of RM preparations.\cite{zhangReverse2020}
\paragraph{introduce study of various surfactant systems}
%
%
This technique can also interrogate how the water matrix responds to
    substitution of various surfactants:
    anionic AOT surfactant
    \vs neutral Igepal CO-520
    \vs cationic \gls{ctab} surfactant.
The literature guides the experiments
    in the choice of a typical dispersant for each surfactant:
    isooctane for AOT,\cite{tamamushiFormation1980,demarco1HNMR1986,amicoInfrared1995}
    cyclohexane for Igepal,\cite{vanderloopStructure2012,sedgwickWhat2009}
    and hexane for \gls{ctab}.\cite{giustiniMicrostructure1996a,millsNMR2014,fuglestadCharacterization2016}
Additionally, for the \gls{ctab} surfactant,
    an added single chain alcohol co-surfactant (here, hexanol,
    though other choices are
    available~\cite{fuglestadChapter2019,marquesMeasurement2014,cuccoviaAnalysis2001,ciszewskiTakes2019,liStudies1997,millsNMR2014}),
    acts to stabilize the \gls{rm} phase.
\paragraph{comparing CTAB to AOT 👈 this was tacked to hexanol comments -- move this up where we are more generally commenting on different surfactants}
As compared to water in the AOT systems, the \gls{ctab}
    water exhibits similar chemical shifts and $T_1$ correlations for the same
    $w_{0}$ values,
    in spite of the opposite charge of the headgroup
    and the presence of the uncharged alcohol co-surfactant.
\paragraph{introduce hexanol resonance}
Notably, \gls{ctab} \gls{rms} yield two deuterium resonances
    for $w_0$ above 12,
    which converge to a single resonance below
    $w_0\approx 12$.
The lower intensity, more shielded (lower ppm)
    resonance likely arises from a deuterated alcohol group
    generated by exchange
    between the D$_2$O and the hexanol co-surfactant.
Because the two peaks have $\sim 10-$fold different intensities,
    \cref{fig:hexanolT1CTAB} shows this hexanol resonance
    (with the higher ppm range signal set to zero),
    whereas
    \cref{fig:waterT1CTAB} shows the D$_{2}$O peak,
    (with the lower ppm range signal set to zero).
\paragraph{discuss hexanol resonance}
The hexanol peak has a shorter relaxation time
    than D$_{2}$O for the same $w_{0}$,
    which is consistent with the expectation of very
    slow exchange and very slow rotational correlation times for
    the nuclei at this position.
As the $w_{0}$ increases, the $T_1$ of the
    hexanol peak also increases,
    though only slightly.
Given that the hexanol resides at the interface of
    the surfactant and RM, these observations suggest there is more
    mobility at this interface for larger $w_{0}$,
    which may be an important factor
    toward deciding the phase stability of the system.
Interestingly, the hexanol
    resonance exhibits increased shielding (lower
    chemical shifts) as the $w_{0}$
    increases -- anti-correlated with the effect
    observed for the D$_2$O resonance.
Thus, as the water pool grows in size,
    the increase in D$_2$O chemical shift suggests an
    increase in the hydrogen bonding strength present
    in the internal water pool while,
    at the same time, the decrease in the
    chemical shift of the hexanol resonance with
    increasing $w_{0}$ suggests a decrease in
    hydrogen bonding strength of the hexanol group.
Such an effect likely represents a ``decoupling'' of
    the hydrogen bonding matrix of the water from the
    alcohol groups at the surface.
As the $w_{0}$ increases,
    the opportunity for water molecules to hydrogen
    bond to other water molecules
    improves,
    while at low $w_{0}$,
    the hexanol appears to be more strongly
    hydrogen bonding to the internal water pool.
As the $w_{0}$ increases and the water pool
    grows in size, the role of hexanol in hydrogen
    bonding to the water becomes less
    pronounced,
    providing interesting insight into the role of
    this co-surfactant.
\paragraph{ability to resolve two resonances}
In general, these hexanol results also point to the
    capability of this technique to resolve multiple
    types of deuterons,
    when they are present in distinct populations that
    do not exchange on the timescale of $T_1$.
In that sense, these results also emphasize that
    the deuterons in all the other samples studied here
    are, in fact, generally exchanging on the timescale
    of $T_1$ (here, tens to hundreds of ms),
    or else present in populations with very similar
    properties.
In other words, the measurements of all other systems
    actively agree with the
    typical picture of nanoemulsions as
    thermodynamically stable systems with uniform
    populations of aggregates (\ie, \gls{rms}) that all
    present similar sized pools of water with similar
    properties.
\subsection{Choice of Surfactant can Induce Qualitative Differences in Water Properties}
\paragraph{compare Igepal vs. others 👉 this is a bit redundant}
Overall,
    the surfactant appears to exhibit a greater effect
    on the water dynamics than the dispersant.
The importance of surfactant composition is underscored by
    the significantly different correlation between the
    chemical shift and $T_1$ values observed for
    Igepal (\cref{fig:IgepalT1}) \vs the other surfactants
    (\textit{esp.} \cref{fig:dispersants}) studied here.
As shown in~\cref{fig:IgepalT1},
    the signals moves overall up and to the
    left,
    relative to \cref{fig:dispersants}.
Given these increases in both $T_1$ and
    chemical shift relative to AOT in hexane or
    isooctane,
the water in
    Igepal RMs appears to be engaged in greater
    hydrogen-bonding and greater rotational
    motion.
The switch from sulfate to polyethoxylate head groups
    stabilizes more ``bulk like''
    behavior of water at room temperature.
\paragraph{}
Perhaps most noticeably of all,
    the results show a qualitative difference
    in
    the shape of the trend traced
    by the signal as a function of $w_0$:
While the trend in \gls{aot} signals (\cref{fig:dispersants})
    has a negative curvature,
    the trend in the
    Igepal signals (\cref{fig:IgepalT1})
    has a positive curvature.
In the case of Igepal,
    the presence of the bulky
    polyethoxylate groups would be expected to alter
    the dynamics and transient structures
    of the water at the lowest water loadings,
    with the polyethoxylate
    sidechains contributing additional hydrogen bond
    acceptor sites.
Notably, there is an increase
    in the $T_1$ by an approximate order of
    magnitude and the
    linewidth is correspondingly halved, from 0.9 ppm in AOT/isooctane to
    0.4 ppm in Igepal.
The increased mobility may arise from the
    ability to more quickly jump between the
    larger variety of H-bond acceptor sites
    available (as compared to the
    comparatively limited
    sites offered by \gls{aot}'s sulfate).
Meanwhile, the chemical shift still indicates that,
    like for \gls{aot}, the
    hydrogen bonding matrix remains
    frustrated in the confined environment,
    likely since the ethoxylate groups do not
    also provide additional H-bond donor
    sites.
The different curvature observed here,
    likely arising from the  combination of the increased mobility
    and unresolved frustration,
    importantly emphasize that the water
    pool inside very small Igepal \gls{rms} presents very
    different properties \vs the pool inside
    small \gls{aot} \gls{rms}.
%
\paragraph{low w₀ Ctab 👈 needs to come after introduction of Igepal}
The straight chain of \gls{ctab} does not support the high
    curvature of smaller reverse micelles,
    and only forms stable RM phase
    for $w_0$ greater than about $w_0\ge 5$.
This obviously hampers the ability
    to compare \gls{ctab} to other surfactants across lower $w_{0}$ values,
    where the obvious, qualitative differences are evident
    when comparing the signal from AOT \vs Igepal \gls{rms}.
Because of this, the data neither supports nor
    invalidates the hypothesis that the unexpected
    mobility of the water in low $w_{0}$ Igepal
    RMs
    arises from the presence of the \gls{peg}-like
    groups, versus the hypothesis that it simply
    arises from the lack of charge of the
    headgroups.
On the other hand,
    at higher $w_0$,
    noting that the correspondence of data in \cref{fig:CTABT1} with those
    in~\cref{fig:IsoT1} and~\cref{fig:IgepalT1} can be ascertained from the color
    scheme, which is identical for identical $w_0$,
    comparisons can be made.
While the results do indicate some differences in
    average hydrogen bonding strength,
    the mobility and overall position in the 2D
    correlation plot are similar for \gls{ctab} and other
    charged surfactant systems.
\paragraph{contrast Igepal vs. AOT 👈 move together with other Igepal results}
These results highlight the extremely close-range
    interactions between water and the \gls{rm} headgroups,
    and may arise from two independent causes.
In the first case,
    they may indicate that interactions between water and
    charged AOT headgroups restrict the water
    motion and that relieving this interaction improves
    the mobility.
AOT is a sulfonate with an extremely low pKa.
The sulfonate restricts the
    water,\cite{handeExploration2016}
    and is present in the \gls{rm} core in very high local
    concentrations.
In fact, for context,
    at even the highest $w_{0}$ values
    employed here,
    a similar ratio of sodium sulfate to water would constitute a
    super-saturated solution.
Conversely, it is possible that
    the ethoxylate units in Igepal
    contribute to an expansion of the water hydrogen
    bonding matrix,
    thereby effectively reducing the level of
    confinement.
The polyethoxylate portion of Igepal resembles
    a \gls{peg} chain,
    which engages in particularly fruitful hydrogen
    bonding with the water matrix.\cite{Ensing2019}
The molecular weight of this chain is 237~Da.
At lower water loadings,
    the mass of polyethoxylate inside the core of the \gls{rm} likely exceeds
    that of the water.
Even though \gls{peg} has been hypothesized to encourage an
    ``iceberg-like'' cage structure of water
    molecules,\cite{Daley2017}
    it could still be acting here to effectively expand
    the size of the water pool.
\subsection{Dynamics do not Follow Core-Shell Model}
The water inside RMs is known to exhibit different
    properties depending on the water pool size,
    ranging from very restricted motion at small sizes
    to a more typical,
    rapid, bulk-like motion
    at larger RM sizes.\cite{biswasAnomalous2018}
These water pools are frequently assumed to follow a
    core/shell structure, with
    water in the core exhibiting more bulk-like
    properties and water in the shell exhibiting
    properties of interfacial water with restricted
    motion.\cite{pileticTesting2006,levingerUltrafast2009}
Notably, however, such conclusions arise from
    measurements above a certain $w_{0}$.
Further complicating the picture,
    different techniques point to different
    \gls{rm} sizes (values of $w_0$) at
    which the bulk-like properties arise or become dominant.
For example, dielectric relaxation dispersion
    studies have indicated this bulk-like cross over
    occurs above $w_{0} = 6$~\cite{dangeloHighfrequency1995}
    whereas ESR indicates bulk-like dominance
    above $w_{0} = 10$,\cite{hauserInteraction1989}
    and established dominance above $20$.\cite{wassermanSpin1994}
IR data indicates
    bulk-like dominance above $w_{0} = 16.5$,\cite{fayerAnalysis2010}
    yet IR supplemented
    with MD points to $w_{0} = 7.5$.\cite{biswasNonmonotonic2012}
Meanwhile, even \gls{dls} offers input here,
    demonstrating that the water inside
    \gls{rms} of $w_0\ge 7$ demonstrates a
    freezing transition when cooled into what
    is now known as ``no man's land,''
    while water inside smaller \gls{rms}
    demonstrates no such transition.\cite{Boned1986ChaWatDis}
\paragraph{}
Overall,
    the correlated trends observed here argue
    against the validity of a ``core-shell'' model
    that extends to low ($w_0<6$) water loadings.
While higher water loadings demonstrate relaxation times and chemical
    shifts that converge close to the value of bulk water,
one of the most distinctive features uncovered by these
    measurements is that
the trend of the signals as a function of
    $w_0$
    in \cref{fig:dispersants}
    display dramatic change in slope near $w_0\approx 3-4$.
Recalling that the $y$-axis here represents roughly the logarithm of the
    rotational correlation time,
    this change in slope represents the point at which the water mobility
    effectively approaches zero.
\paragraph{fit τ to a core-shell}
We tested the origin of this dramatic change
    in slope against
    models based on a core/shell hypothesis.
In the first model,
    two separate environments,
    \eg, a core of ``bulk-like'' water and a shell of
    surfactant hydrating water,
    water molecules exchanging between two environments
    on a timescale longer than the rotational
    correlation time, $\tau_c$ would display an averaged rate of
    relaxation;
because the rate of relaxation is proportional to
    $\tau_c$ (\cref{eq:R1q2}),
    this corresponds to an averaged $\tau_c$.
An attempt to fit the correlated chemical shift and
    relaxation rates to such a core shell model
    failed to find a reasonable interpolation between
    different correlation times (and relaxation rates)
    \via \cref{eq:coreshell,eq:waterSize} --
    as the general form of the trend doesn't match the
    data,
    regardless of the choice of parameters.
In fact, as shown in \cref{fig:checkLinear},
    the curvature of the correlation expected by such
    a model is opposite the curvature demonstrated in
    the correlated trend for all surfactants except
    Igepal.
\paragraph{fit to a different model}
To underscore this point,
    we also attempted a model where the $T_1$ varies
    linearly with the fraction of core \vs shell water
    molecules.
As noted previously,
    a core-shell argument for averaging
    of $T_1\propto 1/\tau_c$ likely lacks
    physical motivation;
    requiring as it does exchange of water molecules
    between environments on a timescale faster than
    $\tau_c$.
Therefore, while the resulting fit cannot support a linear
    interpolation of both diamagnetic shielding and $T_1$
    based on an averaging of core and shell waters,
    the general form of the resulting curve matches the
    data far better than a model that assumes averaging
    of $R_1$ (and $\tau_c$).
Therefore, we rather propose that the trend arises
    from changes to the collective network of water
    that roughly interpolate the rate of decay of the
    rotational correlation function in a fashion that
    scales with \gls{rm} size.
This potentially includes an overall slow-down of all
    rotational motion of the water molecules,
    and
    (consistent with suggestions in earlier literature)
    leaving only a correlation mechanism arising
    from the rotation of entire the \gls{rm} itself.
Overall, the presence,
    in the $\log(T_1)$ vs shielding correlation,
    of a region where the slope changes
    dramatically,
    transitioning from a more
    linear trend
    into a negative curvature trend,
    at the very least
    argues strongly against the
    presence of a distinct ``shell'' layer of hydration
    water and core layer of more bulk-like water --
    especially for water loadings in the
    region of significant negative curvature.
\subsection{Effect of Different Guest Molecules}\label{sec:guest}
\paragraph{👈(move to results) why load stuff into RMs?}
The novel applications of RMs as nanocarriers of
    biologically active
    molecules~\cite{grooReverse2018} or as
    nanoreactors for catalytic
    reactions~\cite{tangActivation2015} necessitates
    investigation of the effect of inclusion molecules
    on the water dynamics of the internal water pools.
Furthermore, encapsulation of proteins inside RMs
    can afford interesting perspectives on their
    dynamics in confined
    environments.\cite{wandHighresolution1998,vanhornReverse2009,marquesMeasurement2014}
RMs can also be
    used as vehicles to extract proteins from complex
    mixtures.\cite{shiomoriEffective1995,nicotProteins1996}
\paragraph{why study viscogens}
In particular, such studies provide an opportunity to
    understand
    the extent to which guest molecules frustrate
    the hydrogen-bonding matrix and dynamics of water,
    and to what extent such a frustration varies as a
    function of confinement.
For example,
    if inclusion of viscogens reduces the rotational
    mobility of water in the bulk,
    one can ask whether or not this effect compounds the
    slow-down of water in the already motionally-restricted 
    environment inside RMs with low
    $w_0$,
    to what extent such a slow-down correlates with a
    frustration in attempts at hydrogen bonding
    (as manifest by the chemical shielding),
    and to what extent the inclusion of viscogens in
    bulk water mimics or differs from the effect of
    confining water inside soft nanoenclosures.
\figInclusion

\subsubsection{Glucose}\label{sec:glucose}
\paragraph{effect of glucose on viscosity}
Comparing AOT RMs in isooctane prepared with a
    solution of 25 wt\% glucose (\cref{fig:InclusionGlucose}) to
    simple solutions of AOT RMs in isooctane
    (\cref{fig:IsoT1}), there is a noted
    decrease in $T_1$,
    as well as a broadening in
    linewidth, especially at low $w_{0}$, consistent with a reduction in
    the rotational mobility of D$_{2}$O.
Bulk water
    measurements containing 20-30\% glucose display a
    viscosity $1.68-2.48\times$ greater than bulk
    water.\cite{telisViscosity2007}
Meanwhile,
    for \gls{rms} above $w_0\approx 5$,
    \cref{eq:R1q2} predicts correlation times
    for the water inside the \gls{rms} with a correlation
    time roughly $2\times$ slower upon incorporation of the
    glucose.
Therefore, the slowdown in rotational mobility from
    addition of viscogens appears roughly additive with
    the slowdown due to confinement
    for \gls{rms} above $w_0\approx 5$.
However, below this threshold,
    glucose makes a subtle to non-existent perturbation,
    implying that interactions with the surfactant dominate.
At low water loading,
    glucose serves mainly to broaden lines
    along both dimensions
    (increasing both $\Delta \log (T_1)$ and $\Delta \sigma$ in
    \cref{fig:InclusionGlucose}),
    an effect that may indicate a greater diversity in the
    structures of individual \gls{rms},
    as will be discussed in greater detail with the PEG results.
\paragraph{glucose disrupts H-bonding as we see from σ}
Comparing similar-sized RMs with glucose
    (\cref{fig:InclusionGlucose}) to 
    AOT \gls{rms} without glucose
    (\cref{fig:IsoT1}),
    the signal coming from high $w_0$ \gls{rms}
    with included glucose shift to more strongly shielded
    resonances,
    indicative of a frustration of the hydrogen bonding
    network that is, again, additive with the effects
    of confinement by the surfactant.
Characterization of osmolyte-loaded RMs, such as those containing glucose, has
    recently been investigated with high-field
    NMR.\cite{sanfordSweet2018}
The impact of the confined water environment on these molecules was also
    explored.\cite{millerNanoconfinement2021}
In relation to those studies,
    the experimental tools presented here may
    help to explain
    the perturbation to the
    typical ratio of the β:α anomers of glucose
    recently observed within a
    \gls{rm}.\cite{sanfordSweet2018}
\subsubsection{Polyethylene glycol (PEG)}\label{sec:peg}
\paragraph{PEG inclusion 👈 results}
Provided its molecular weight is
    sufficiently small ($<1\;\text{kDa}$), \gls{peg} does not perturb
    the RM size or
    geometry.\cite{mehtaTemperatureinduced2006}
Typically the minimal $w_{0}$ that can
    accommodate a given polymer should provide a water
    pool radius about equal to the
    hydrodynamic radius of the polymer;
    failure
    to meet this condition results in the formation
    of insoluble precipitates in the solution,
    otherwise known as
    ``necklacing''.\cite{meierKerr1997,laiaLight2000}
Similarly, here,
    an inability to form stable \gls{rms} above $w_0= 8.0$
    may be indicative of increasing quantities
    of \gls{peg} destabilizing the \gls{rm}. 
Following this caution,
    \Gls{peg} as a guest molecule has been
    shown to lead to interesting results in RM systems.
For example, water-soluble polymers such as \gls{peg}
    have been shown to facilitate mixing of
    the individual water
    pools.\cite{mehtaTemperatureinduced2006,fanEffect2020a}
\paragraph{stuff is slower moving in 50\% PEG, and we might see polydispersity, which is interesting}
Here (\cref{fig:InclusionLargePEG}),
    we sought a large weight percentage of
    \gls{peg} in order to ascertain the effects of a
    high concentration of polymer and to specifically compare
    to the water inside Igepal \gls{rms};
    therefore,
    \gls{peg}-200 was incorporated at 50~wt\%.
The solutions encapsulated in \gls{rms} of various sizes in
    \cref{fig:InclusionLargePEG} display a decrease in $T_1$
    relative \gls{rms} without guest molecules (\cref{fig:dispersants}).
Note that the 50 wt\% PEG-200 control (bulk solution)
    shows a shorter $T_1$ along with a wider
    distribution along the $T_1$ and chemical shift dimensions
    \vs bulk D$_2$O.
\paragraph{compare based on number of waters}
Remembering that \cref{fig:Inclusion}
    (as elsewhere here) 
    labels the samples using the
    definition of $w_0$ 
    based on aqueous solution volume
    (\ie, ``$w_0$-equivalent''),
    note that for these 50\% \gls{peg} data,
    the water:surfactant molecular ratio
    is approximately $w_0/2$,
    while in \cref{fig:dispersants},
    the water:surfactant molecular ratio is $w_0$.
Indeed,
    the \gls{rms} containing 50\% \gls{peg}
    display $T_1$
    similar to,
    though still slightly shorter than,
    ``empty'' AOT reverse micelles
    (\ie, without any guest molecules,
    \cref{fig:IsoT1})
    of half the $w_0$.
Thus, they demonstrate a rotational mobility
    similar to,
    though still slightly slower than,
    empty \gls{rms} with half the $w_0$.
The chemical shift of the \gls{peg}-loaded \gls{rms}
    similarly resembles empty \gls{rms} with half the
    $w_0$.
\paragraph{start to talk about actual polydispersity}
Notably,
    the high concentration
    of \gls{peg} also induces a
    broadening
    of the peaks along both dimensions.
This broadening may indicate a greater diversity
    of water micro-environments accompanying differences
    in the number or conformations of the
    \gls{peg} macromolecules loaded into each \gls{rm}.
Specifically, note that
    because $T_2>T_1$ and the linewidth must be at
    least $1/\pi T_2$,
    a $T_1$ of $10^{-1.5}\;\text{s}$ would demand a
    linewidth of at least the $\Delta \sigma$ displayed
    in \cref{fig:InclusionGlucose};
    this explains nearly all of the line broadening
    of the glucose-containing \gls{rms} in
    \cref{fig:InclusionGlucose},
    but not that of the \gls{peg} in
    \cref{fig:InclusionLargePEG},
    which likely indicates a true heterogeneity along
    both chemical shielding and $T_1$ dimensions,
    tied to heterogeneity in the aggregate structures.
At these concentrations,
    there is one \gls{peg} molecule for every 0.4~kDa of
    solution, so that at very low water loadings,
    we expect only one (or a handful) of \gls{peg}
    molecules to be included.
As a reference, compare 0.4~kDa to the fact
    that \cref{eq:numWaters}
    (combined with 18~g/molecule)
    predicts that
    a \gls{rm} of $w_0=1.0$ will contain
    only 0.3~kDa of solution,
    while a \gls{rm} of $w_0=2.8$
    will contain 0.8~kDa.
This implies a significant variability
    when constructing \gls{rms},
    as some will contain a \gls{peg} molecule,
    and some will not.
The broadening of the 2D peaks
    observed, especially
    at lower water loadings,
    likely arises from the resulting
    heterogeneity of
    \gls{rm} structure,
    where some \gls{rms} contain \gls{peg} molecules,
    while others do not.
This heterogeneity may also indicate that the
    considerable size of the \gls{peg}
    molecule relative to the \gls{rm} might be
    interfering with the typical structure,
    so that even \gls{rms} of the same constitution
    could exhibit different conformations.
\subsubsection{Bovine Serum Album (BSA)}
\figBSA
\paragraph{BSA inclusion 👈 results}
\gls{bsa} is a widely available and commonly studied protein.
Often used
    to mimic the effects of molecular crowding in the
    cell,
    \gls{bsa} offered here a unique opportunity to study
    the hydration layer of a protein
    and how it responds to encapsulation within the
    \gls{rm},
    and to see
    if multiple RM distributions would be detected by either $T_1$ or chemical
    shift resolution.
Previous studies that tracked 
    changes to 2D NMR (HSQC) spectra of ubiquitin as a
    measure of the level and character of macromolecular crowding
    reported that \gls{bsa} interfaces
    (in crowded solution)
    and \gls{rm} interfaces
    (\ie, encapsulating the ubiquitin)
    both yielded a similar crowding effect.\cite{vanhornReverse2009}
This leads to the base expectation that in
    \gls{rms} with \gls{bsa} guest molecules,
    the surfactant and \gls{bsa} surfaces will yield an
    additive effect when confining the water.
A related serum albumin protein, Human Serum
    Albumin (HSA), was incorporated into AOT RMs and
    the environment studied by ESR spectroscopy as a
    function of $w_{0}$ and was found to
    restrict the rotational diffusion of the protein
    at smaller $w_{0}$~\cite{Marzola1991};
    this observation also points to an additive
    retardation of diffusion when comparing HSA in
    aqueous solution \vs inside the \gls{rm}. 
\paragraph{setup of BSA experiment + hypothesis}
Here, the impact of 600~μM (4 wt/v\%) \gls{bsa} on the properties of the
    internal water pool was investigated.
Note that at this concentration,
    there is approximately 1 \gls{bsa} molecule per every
    $\sim 1700$~kDa of solution;
    while the sizes of the water pools for $w_0=27,$
    $35,$ and $42$ are
    $\sim 220$~kDa, $\sim 500$~kDa, and $\sim 870$~kDa (\cref{eq:numWaters}).
Therefore, we isolate this study to higher $w_{0}$ values
    than the other studies in this paper,
    since moving to lower $w_0$ would result
    in solutions where the number of RM aggregates far
    exceeded the number of \gls{bsa} molecules;
    already,
    from the estimated sizes above,
    $w_0=27,$ $35,$ and $42$
    correspond to approximately
    7.6, 3.4, and 1.9 \gls{rm} per \gls{bsa} molecule.
\paragraph{}
As 4\% seems to be a small concentration,
    for a sense of scale,
    it is worth considering two spheres,
    equal in density,
    that differ in weight by 100\% to 4\%
    -- \ie the most simplistic model for \gls{bsa} molecules
    loaded one per \gls{rm}.
Under these conditions, the radius
    of the smaller sphere would be $\sim 34\%$ of the
    larger sphere,
    while the surface area would be $\sim 12\%$.
To form a slightly more accurate estimate of the
    relative surface areas of \gls{rm} \vs \gls{bsa} in
    these samples,
note that at $w_0=42,$ \cref{eq:waterSize} indicates the surface
    area of the surfactant inside the reverse micelle will
    be $678\;\text{nm}^2$.
At $w_0=27$,
    it predicts a surface are of $280\;\text{nm}^2$.
Meanwhile,
    \gls{bsa} presents a \gls{sasa}
    of about $300\;\text{nm}^2$
    (notably larger than the spherical prediction
        above,
        due to surface roughness).
Thus,
    differing methods of estimation
    predict that
    loading \gls{bsa} into a \gls{rm} increases the surface
    area of the water by anywhere from 12\% to 93\% of the
    surface area in the absence of \gls{bsa}.
Based on a strictly core-shell argument,
    and following on the previous observations that 
    frustration of the water matrix due to confinement
    \vs inclusion of guest molecules seemed to be
    additive,
    one would expect water inside \gls{bsa}-loaded reverse
    micelles to behave somewhat differently as
    compared to unloaded micelles.
\paragraph{3 percent BSA very small impact}
As seen in~\cref{fig:BSA}, the presence of \gls{bsa}
    only subtly impacts the properties of
    the internal water pool observed by this
    measurement,
    with a very slight shift to slower rotational
    mobility (shorter $T_1$) and more frustrated hydrogen bonding
    (lower chemical shift) 
    \vs
    the ``empty'' D$_{2}$O \gls{rms}.
Specifically focusing on $w_0=27,$
    where one might expect a \gls{bsa} molecule
    loaded into roughly 50\% of the \gls{rms},
    significant
    additional separation or broadening of the peak
    along the $T_1$ and/or chemical shift dimensions does
    not occur,
    even though both broadening and 2
    separate 2D peaks have been observed in
    other measurements reported here.
This may be owed to negligibly small differences
    in the $T_1$
    of RMs containing \gls{bsa} vs those that do not,
    though the reason for the lack of two populations
    of water with different chemical shielding remains
    unclear.
Overall, if \gls{bsa} remains roughly centered on
    the \gls{rm} here and traps the water between the
    surface of the \gls{bsa} and the surface of the
    \gls{rm},
    the resulting geometry does not seem to further
    frustrate the hydrogen bonding matrix of the water.
Notably, even at this low concentration,
    a core/shell model would lead one to believe this
    geometry leads to a significant increase in more
    ``hydration''-like water (\ie an ``inner shell'')
    that would result in noticeable changes to chemical
    shielding and $T_1$
    -- however, if such an increase is present, it is
    extremely subtle.
\subsection{Details of Interpretation}
\paragraph{convert to τc 👈 group with 2D in "interpretation"?}
From the $T_1$ values measured in these experiments, approximate values of
    $\tau_{c}$ were determined using~\cref{eq:R1q} and tabulated
    in~\cref{tab:table1}.
The quadrupolar coupling constant
    230~kHz\cite{mantschDeuterium1977} was used in all cases.
The exact quadrupolar coupling constants in these \gls{rm} experiments
    may show some amount of variation given that the
    quadrupolar coupling constant decreases with an increase in O-H bond
    length.\cite{cumminsEffect1985a}
Despite the fact that the quadrupolar coupling constants have been
    determined from MD simulations,\cite{eggenbergerInitio1992a}
    questions arise as to accuracy of the water model in
    question~\cite{roppRotational2001} in such determinations.
%
%
Furthermore, illustrated by \cref{fig:lorentzian},
    correlation times over a few hundred nanoseconds
    will actually exhibit a relaxation time somewhat shorter than the
    relaxation indicated by the motional narrowing limit approximation
    (\cref{eq:R1q2}).
Therefore, \cref{tab:table1} simply approximates for all cases
    as the quadrupolar coupling constant for bulk D$_{2}$O.
Accurate determination of the quadrupolar coupling constants,
    and an
    accounting for deviations from \cref{eq:R1q2} at longer
    values of $\tau_c$,
    would yield greater accuracy in the determination of the values
    of $\tau_c$ (\cref{tab:table1}) in these
    confined water environments.
Such potential studies represent important future directions.
The current investigation
    focuses less on the detailed refinement of the models used to
    interpret the
    data,
    and chooses to accept a simplified interpretation.
It can then proceed to interrogate stark/qualitative (and still valid)
    trends in how the signal
    migrates through the 2D correlation plot
    as the size of the confinement changes.
\paragraph{2D: verify motional narrowing 👈 gather in interpretation -- also call extension?}
\figIRCPMG
A full 2D relaxation-relaxation correlation
    (\ie 2D relaxometry)
    experiment can test whether or not
    a particular population of $\text{D}_2\text{O}$
    exhibits relaxation in the motional narrowing
    regime.
Specifically, as noted in the text after \cref{eq:R1q2},
    $R_2=R_1$, \ie, $T_1=T_2$, if and only if
    the correlation time falls in the motional
    narrowing regime
    (\ie, much faster than the resonance frequency).
Note that such a 2D relaxometry experiment
    (generated from relaxation decays in both
    dimensions)
    differs from the 1.5D relaxometry experiments
    throughout the rest of this paper,
    where relaxation decay supplies one dimension
    (the $T_1$),
    while a Fourier transform of resonant oscillations
    (\ie, standard chemical shift) supplies the other
    dimension of the correlation.
In this 2D relaxometry experiment,
    a stroposcopic experiment replaces the
    standard direct detection dimension ($t_2$).
As an example,
    we test whether or not
    the water inside Igepal remains in the motional
    narrowing regime for all values of $w_0$.
In \cref{fig:IRCPMG}, correlated $T_1 - T_{2}$
    measurements were carried out for Igepal 
    reverse micelles for small and large $w_{0}$:
    $w_0=1$ and $w_0=14.7$, respectively.
In both cases, the $T_1-T_{2}$ distribution is
    unimodal and approximately centered about the
    $T_1 = T_{2}$ line, indicating that the D$_{2}$O
    is in the extreme narrowing limit.
Small off-centering of the signal or small peaks in
    other part of the correlation map
    may be considered artefactual.\cite{songT12002}
\section{Conclusions}\label{sec:conc}
\paragraph{statement of utility and impact}
The correlated change in water chemical shift and $T_1$,
    in general,
    tells a consistent story indicating that
    confinement leads to a dramatically non-bulk-like
    environment when water is confined inside small \gls{rms},
    where the hydrogen bonding matrix of the water
    pool is frustrated and
    where the water rotational dynamics are slowed.
Here, automated measurements
    of water confined under a range of conditions
    probe
    the 2D correlation between relaxation and chemical shift.
These measurements observe
    changes spanning a few orders of magnitude
    of relaxation times and chemical shift variations
    of $\sim25\%$.
Water, of course, comprises a dynamic structure of
    fluctuating hydrogen bonds,
    constantly swapped and rearranged as the water
    undergoes not only rotational,
    but also translational diffusion.
Interestingly, these results here observe an inflection
    point in the shape swept out by the correlated measurements near $w_0=3$ to
    $5$ ($w_0\equiv
    [\text{H}_2\text{O}]/[\text{surfactant}]$)
    with a fair degree of consistency.
Note that at this point,
    \cref{eq:numWaters}
    expects
    \gls{rms} to
    contain somewhere between 70 and almost 200 water molecules,
    so that an apparent breakdown in the dynamic
    structure of
    water at this $w_0$ proves quite interesting.
This evidence also supports the rationale behind \cref{eq:numWaters},
    which expects this lower number of water moleucules,
    in comparison to, \eg, \cite{pileticTesting2006},
    which expects hundreds of water molecules.
The notable exception to this result was for the Igepal
    surfactant,
    which likely forms a structure with polyethoxylate
    groups oriented inwards, surrounded by, and hydrogen bonded to,
    the water.
For water confined inside Igepal \gls{rms},
    while changes in diamagnetic shielding indicating a
    change in average hydrogen bonding strength remain
    prominent,
    it is not correlated with as prominent a
    retardation of the rotational diffusion,
    giving rise to a correlation of a fundamentally
    different character.
\paragraph{combine with other techniques}
Here, the technique was demonstrated in
    \gls{rms}, as they provide systems where the levels
    of confinement can be easily adjusted,
    and where response of the water matrix to the
    incorporation of guest molecules can be easily
    interrogated.
However, this technique is ready both for wide-scale
    application to various confined water systems,
    and for integration with standard measurement
    suites.
Future applications of this deuterium NMR
    screening technique might span,
    for instance,
    samples ranging from structured porous silicates
    through to membraneless organelles or other
    liquid-liquid phase separation systems of
    biological relevance.
\paragraph{next steps}
This publication focused on
    the deployment of an automated technique,
    thus providing a method to screen or
    survey differences in confined water
    pockets and through this focus was able
    to demonstrate striking qualitative
    differences between different
    \gls{rm} systems.
Note that
    it also lays the groundwork for the clear
    subsequent steps.
In with regard to further investigation of
    the nuances of how to best interpret
    changes in $T_1$ and chemical shift under
    such extreme circumstances,
    it presents the possibility of further
    deploying these measurements,
    alongside the $T_1$ \vs $T_2$ test demonstrated
    here,
    to more exactly quantify the correlation time \vs
    quadrupolar coupling constants for the smaller
    pockets of confined water that were studied here.
Standard models (\cref{fig:lorentzian}) indicate that
    changes to relaxation rates,
    and evidence of departure from the motional
    narrowing regime,
    should be observable for the smallest $w_0$ AOT
    reverse micelles that were observed here.
This departure should enable subsequent studies
    to experimentally tweeze apart
    the correlation time from
    any variations in the quadrupolar coupling
    constant, and would offer insight into the exact
    timescales of diffusion inside these smallest
    reverse micelles, as well as the impact of the
    electric field gradients imposed by ionic
    sidechains
    under various extreme confinements.
Similarly,
    when considering protein guest molecules,
    a clear next step would involve the study of water
    exposed to protein surface in at a
    higher ratio
    of \gls{sasa} to \gls{rm} water molecules
    (without inducing transitions in the protein
    system).
In particular,
    future studies employing the method introduced here
    can determine the
    utility of freezing the encapsulated \gls{bsa} in
    order to ``shed'' the water~\cite{vanhornReverse2009}
    or of the ``evaporation-injection'' method~\cite{fuglestadChapter2019},
    both of which offer promise of encapsulating large macromolecules
    with this method inside smaller (lower $w_0$) \gls{rms}.

%% file: supp_content.tex
\section{Meta-Analysis and Fitting of $\bar{n}$}
\figMetaAnal
As indicated in the text,
    we performed a meta-analysis
    where we digitized data from
    a series of
    previous publications pertaining to
    $\bar{n}$ (the aggregation number,
    or number of surfactant molecules
    per individual reverse micelle).
\Cref{fig:metaAnal}
    shows the data digitized from previous
    publications and fit to the form indicated by
    \maincref{eq:numWaters}.
The points are chosen for the least squares fit
    following the reasoning that vapor-pressure
    osmometry data serves as a better guide at
    lower water loading,
    while data that relies on centrifugation and
    other techniques assuming a constant density
    serve as better guides at higher water
    loading.
As indicated in \cite{Ueda1988MeaAggNum},
    the osmometry data is expected to have less
    value for higher water loading,
    explaining the one significant outlier from
    the fit here.
See the main text for more details.
\section{Background on Quadrupolar Relaxation}\label{sec:QuadDetails}
Quadrupolar relaxation,
    for deuterium ($I = 1$), is given
    by~\cite{abragamPrinciples1961quadRelax}
\begin{equation}
    R_1 =
    \frac{3}{80}
    \left(\frac{e^{2}Qq}{\hbar}\right)^{2}
    \left[J(\nu_D) +
    4J(2\nu_D)\right]
    \label{eq:R1q}
\end{equation}
where $e$ is the elementary charge, $Q$ is the
    quadrupolar moment of the nucleus, $q$
    represents the electric field gradient at the
    nucleus,
    $\nu_D$ is the deuterium resonance frequency,
    and $J(\nu)$ is the reduced spectral density
    function
    (the Fourier transform of a normalized
    auto-correlation function that describes the
    rotational motion).\cite{slichterPrinciples1989}
Similarly, the rate of transverse relaxation,
$R_{2}$, is given by,
\begin{equation}
    R_{2} =
    \frac{1}{160}
    \left(\frac{e^{2}Qq}{\hbar}\right)^{2}
    \left[9J(0) +
    15J(\nu_D) +
    6J(2\nu_D)\right]
    \label{eq:R2q}
\end{equation}
\figSpecDens
\section{Test of Core-Shell Model}
\figCheckLinear
The model of \cref{fig:checkLinear} assumes weighted core and shell
    contributions to the $T_1$ time and to the
    chemical shift.
The values shown here interpolate between a
    chemical shift of 3.82 for shell water
    and 4.72 for core water,
    and between $\log(T_1/\text{s})$ of -2.64 for shell
    water and -0.89 for core water.
The thickness of the shell water that
    yields this fit is 0.15~nm,
    notably less that a layer of water,
    and likely also arguing against the validity of a
    core-shell model.
Also, as noted in the main text,
    while averaging of relaxation rates ($T_1^{-1}\propto \tau_c$)
    of core \vs shell water molecules
    represents a physically reasonable situation,
    averaging of the relaxation times ($T_1\propto \tau_c^{-1}$) does not.
\section{Susceptibility}\label{sec:susceptibility}
The susceptibility of different RM solutions varies.
On the same day,
    the relative susceptibility is given by the
    relative frequencies of the TMS resonance.
The absolute susceptibility can be determined
    by looking at the experiments where a capillary
    of water with \gls{dss} was added into the sample.
The susceptibility of water is known,
    allowing us to calculate the exact field,
    and then the TMS frequency of the RM solution
    gives the susceptibility of the RM solution.
\section{A note on proton exchange in glucose}
\paragraph{comment on proton exchange with glucose 👈 figure out where to put this -- maybe just SI}
Because it is a significant recent finding,
    it is worth discussing that
    glucose, like other solutes,
    when encapsulated in \gls{rms}
    experiences a
    decrease in the rate of proton exchange
    (relative to the rates in bulk aqueous solution)
    that exhibits a dependence on $w_0$.\cite{sanfordNanoconfinement2016}
Initiation of proton exchange is a rare event, but
    once initiated, it can propagate via standard
    proton exchange mechanism (\eg, Grotthuss).
Assuming that such mechanisms cannot propagate through
    the aprotic solvent,
    it's expected that exchange should be dramatically
    reduced in reverse micelles for all cases.
Regardless, the rates of proton exchange in all cases
    are expected to be slower than the relaxation rates
    observed here and, therefore, not to affect the
    present measurements.
\section{Relation to Other Types of Measurements}
Note that studies of \gls{rms} by \gls{dls}
    have been performed in the past,
    though have not been repeated here,
    because we expect the polydispersity measurements
    by \gls{dls} to be similar or weaker reliability
    compared to the $T_1$ distributions presented here,
    since both rely on analysis of multivariate
    exponential decays.
Furthermore, \gls{dls} is generally considered
    optimal for larger sized particles and has shown
    significant discrepancy with the measurements of
    particle diffusion garnered from \gls{nmr}
    diffusion measurements,
    which are expected to be
    quite accurate and more optimal for smaller
    particle sizes.
It is worth noting that significant groundwork has
    also been laid to relate the present types of
    measurements to molecular dynamics
    simulations.\cite{Honegger2019NQR}
\section{Low Weight Percent PEG-200}\label{sec:pegLow}
Comparing \gls{aot} \gls{rms} in isooctane prepared with a
    solution of 1 w\% PEG-200 to
    simple solutions of \gls{aot} \gls{rms} in isooctane, \gls{rms} containing PEG-200
    exhibit slightly shorter $T_{1}$ as those without.
This indicates a slight reduction in the
    rotational mobility in the presence of the PEG
    solution which is expected given the small
    quantity of PEG added to the solution.
Additionally the chemical shifts in the presence
    of PEG are shifted to slightly lower frequencies.
This is indicative of the D$_{2}$O engaging in
    less hydrogen bonding to the water matrix, which
    is again expected as it would be engaging with the
    PEG molecule.
The linewidths with PEG inclusion are slightly
    larger than those without, which is consistent
    with the observed slight decrease in $T_{1}$. 
\section{Automation and Pulse Program Scripts}\label{sec:AUandPPG}
\subsection{Automation Program}
The automation program (``AU'' program) is as follows:
{\footnotesize%
\input{auProgramFormatted}}
\subsection{Pulse Program}
The pulse program is a standard inversion recovery, but
employs (1) utilization of the lock channel for
acquisition and (2) separately stored phase cycles
following the recipe given
previously.\cite{beatonModernized2022}
{\footnotesize%
\input{brukerPPG}}
\section{Inverse Laplace Transform}\label{sec:ipynb}
A PDF export of an explanatory jupyter notebook that converts
raw data from these programs to a correlated
distribution of $T_1$ \textit{vs.} chemical shift is
attached at the end of this supporting information document.
(The data in the resulting HDF file is read into
subsequent scripts that compile and
display contours from multiple experiments.)

%% file: auProgramFormatted.tex
\begin{Verbatim}[commandchars=\\\{\}]
\PY{k+kt}{FILE}\PY{+w}{ }\PY{o}{*}\PY{n}{logfp}\PY{p}{;}
\PY{k+kt}{char}\PY{+w}{ }\PY{n}{logpath}\PY{p}{[}\PY{l+m+mi}{1000}\PY{p}{]}\PY{p}{;}
\PY{k+kt}{char}\PY{+w}{ }\PY{n}{titlestr}\PY{p}{[}\PY{l+m+mi}{1000}\PY{p}{]}\PY{p}{;}
\PY{k+kt}{double}\PY{+w}{ }\PY{n}{peakFreqHz}\PY{p}{,}\PY{+w}{ }\PY{n}{peakFreqPPM}\PY{p}{,}\PY{+w}{ }\PY{n}{peakIntensity}\PY{p}{,}\PY{+w}{ }\PY{n}{maxpsh}\PY{p}{,}
\PY{+w}{    }\PY{n}{maxpsp}\PY{p}{,}\PY{+w}{ }\PY{n}{maxips}\PY{p}{,}\PY{+w}{ }\PY{n}{first\PYZus{}maxpsp}\PY{p}{;}
\PY{k+kt}{double}\PY{+w}{ }\PY{n}{sf}\PY{p}{,}\PY{+w}{ }\PY{n}{sfo1}\PY{p}{,}\PY{+w}{ }\PY{n}{o1}\PY{p}{;}
\PY{k+kt}{float}\PY{+w}{ }\PY{n}{p1}\PY{p}{;}
\PY{c+c1}{// p1\PYZus{}list should be length\PYZhy{}3 array for pulse}
\PY{c+c1}{// lengths around 180 time}
\PY{c+c1}{// nutation\PYZus{}peaks should be length\PYZhy{}3 array with}
\PY{c+c1}{// peak intensities (sign matters)}
\PY{c+c1}{// bit for linear regression}
\PY{c+c1}{// pull y\PYZhy{}intercept/2 for p1}
\PY{k+kt}{float}\PY{+w}{ }\PY{n}{p1\PYZus{}list}\PY{p}{[}\PY{l+m+mi}{3}\PY{p}{]}\PY{+w}{ }\PY{o}{=}\PY{+w}{ }\PY{p}{\PYZob{}}\PY{l+m+mf}{660.0}\PY{p}{,}\PY{+w}{ }\PY{l+m+mf}{720.0}\PY{p}{,}\PY{+w}{ }\PY{l+m+mf}{780.0}\PY{p}{\PYZcb{}}\PY{p}{;}
\PY{k+kt}{double}\PY{+w}{ }\PY{n}{nutation\PYZus{}peaks}\PY{p}{[}\PY{l+m+mi}{3}\PY{p}{]}\PY{p}{;}
\PY{k+kt}{int}\PY{+w}{ }\PY{n}{noofscans}\PY{p}{,}\PY{+w}{ }\PY{n}{pscal\PYZus{}save}\PY{p}{,}\PY{+w}{ }\PY{n}{i}\PY{p}{,}\PY{+w}{ }\PY{n}{j}\PY{p}{,}\PY{+w}{ }\PY{n}{numPeaks}\PY{p}{;}
\PY{k+kt}{int}\PY{+w}{ }\PY{n}{ns}\PY{p}{;}
\PY{k+kt}{float}\PY{+w}{ }\PY{n}{repDelay}\PY{p}{;}
\PY{k+kt}{float}\PY{+w}{ }\PY{n}{my\PYZus{}d1}\PY{p}{;}
\PY{k+kt}{int}\PY{+w}{ }\PY{n}{stillgoing}\PY{p}{;}
\PY{c+cp}{\PYZsh{}}\PY{c+cp}{define REAL double}
\PY{c+cp}{\PYZsh{}}\PY{c+cp}{define MIN\PYZus{}DOUBLE 1e\PYZhy{}9}

\PY{n}{REAL}\PY{+w}{ }\PY{n}{p1\PYZus{}list\PYZus{}forreg}\PY{p}{[}\PY{l+m+mi}{3}\PY{p}{]}\PY{p}{;}
\PY{c+c1}{// the following are just function definitions,}
\PY{c+c1}{// with the code given at the bottom, after the quit}
\PY{c+c1}{// statement}
\PY{k+kt}{int}\PY{+w}{ }\PY{n+nf}{linreg}\PY{p}{(}\PY{k+kt}{int}\PY{p}{,}\PY{+w}{ }\PY{k}{const}\PY{+w}{ }\PY{n}{REAL}\PY{+w}{ }\PY{o}{*}\PY{p}{,}\PY{+w}{ }\PY{k}{const}\PY{+w}{ }\PY{n}{REAL}\PY{+w}{ }\PY{o}{*}\PY{p}{,}\PY{+w}{ }\PY{n}{REAL}\PY{+w}{ }\PY{o}{*}\PY{p}{,}
\PY{+w}{           }\PY{n}{REAL}\PY{+w}{ }\PY{o}{*}\PY{p}{,}\PY{+w}{ }\PY{n}{REAL}\PY{+w}{ }\PY{o}{*}\PY{p}{)}\PY{p}{;}
\PY{k+kt}{void}\PY{+w}{ }\PY{n+nf}{settitle}\PY{p}{(}\PY{k+kt}{char}\PY{+w}{ }\PY{o}{*}\PY{p}{)}\PY{p}{;}
\PY{n}{stillgoing}\PY{+w}{ }\PY{o}{=}\PY{+w}{ }\PY{l+m+mi}{1}\PY{p}{;}

\PY{c+c1}{//\PYZob{}\PYZob{}\PYZob{} open log file}
\PY{n}{GETCURDATA}\PY{p}{;}\PY{+w}{ }\PY{c+c1}{// pulls the info about the foreground}
\PY{+w}{            }\PY{c+c1}{// dataset}
\PY{p}{(}\PY{k+kt}{void}\PY{p}{)}\PY{n}{sprintf}\PY{p}{(}\PY{n}{logpath}\PY{p}{,}\PY{+w}{ }\PY{l+s}{\PYZdq{}}\PY{l+s}{\PYZpc{}s/\PYZpc{}s/\PYZpc{}d/logfile}\PY{l+s}{\PYZdq{}}\PY{p}{,}\PY{+w}{ }\PY{n}{disk}\PY{p}{,}\PY{+w}{ }\PY{n}{name}\PY{p}{,}
\PY{+w}{              }\PY{n}{expno}\PY{p}{)}\PY{p}{;}
\PY{k}{if}\PY{+w}{ }\PY{p}{(}\PY{p}{(}\PY{n}{logfp}\PY{+w}{ }\PY{o}{=}\PY{+w}{ }\PY{n}{fopen}\PY{p}{(}\PY{n}{logpath}\PY{p}{,}\PY{+w}{ }\PY{l+s}{\PYZdq{}}\PY{l+s}{wt}\PY{l+s}{\PYZdq{}}\PY{p}{)}\PY{p}{)}\PY{+w}{ }\PY{o}{=}\PY{o}{=}\PY{+w}{ }\PY{n+nb}{NULL}\PY{p}{)}\PY{+w}{ }\PY{p}{\PYZob{}}
\PY{+w}{  }\PY{n}{Proc\PYZus{}err}\PY{p}{(}\PY{n}{DEF\PYZus{}ERR\PYZus{}OPT}\PY{p}{,}\PY{+w}{ }\PY{l+s}{\PYZdq{}}\PY{l+s}{Can\PYZsq{}t open \PYZpc{}s}\PY{l+s+se}{\PYZbs{}n}\PY{l+s}{\PYZpc{}s}\PY{l+s}{\PYZdq{}}\PY{p}{,}\PY{+w}{ }\PY{n}{logpath}\PY{p}{,}
\PY{+w}{           }\PY{n}{strerror}\PY{p}{(}\PY{n}{errno}\PY{p}{)}\PY{p}{)}\PY{p}{;}
\PY{+w}{  }\PY{k}{return}\PY{+w}{ }\PY{l+m+mi}{0}\PY{p}{;}
\PY{p}{\PYZcb{}}
\PY{n}{fprintf}\PY{p}{(}\PY{n}{logfp}\PY{p}{,}
\PY{+w}{        }\PY{l+s}{\PYZdq{}}\PY{l+s}{hello! I am a log file, and I live in expno \PYZpc{}d }\PY{l+s}{\PYZdq{}}
\PY{+w}{        }\PY{l+s}{\PYZdq{}}\PY{l+s}{\PYZhy{}\PYZhy{} new version}\PY{l+s+se}{\PYZbs{}n}\PY{l+s}{\PYZdq{}}\PY{p}{,}
\PY{+w}{        }\PY{n}{expno}\PY{p}{)}\PY{p}{;}
\PY{c+c1}{//\PYZcb{}\PYZcb{}\PYZcb{}}

\PY{c+c1}{//\PYZob{}\PYZob{}\PYZob{} 1H experiment}
\PY{n}{RPAR}\PY{p}{(}\PY{l+s}{\PYZdq{}}\PY{l+s}{ab\PYZus{}1H\PYZus{}zg}\PY{l+s}{\PYZdq{}}\PY{p}{,}\PY{+w}{ }\PY{l+s}{\PYZdq{}}\PY{l+s}{all}\PY{l+s}{\PYZdq{}}\PY{p}{)}\PY{p}{;}
\PY{n}{fprintf}\PY{p}{(}\PY{n}{logfp}\PY{p}{,}
\PY{+w}{        }\PY{l+s}{\PYZdq{}}\PY{l+s}{Beginning 1H NMR 1D experiment in \PYZpc{}d...}\PY{l+s+se}{\PYZbs{}n}\PY{l+s}{\PYZdq{}}\PY{p}{,}
\PY{+w}{        }\PY{n}{expno}\PY{p}{)}\PY{p}{;}
\PY{n}{RPROCNO}\PY{p}{(}\PY{l+m+mi}{1}\PY{p}{)}\PY{p}{;}\PY{+w}{ }\PY{c+c1}{// sets procno}
\PY{n}{SETCURDATA}\PY{+w}{  }\PY{c+c1}{// pull the information for a particular}
\PY{+w}{           }\PY{c+c1}{// dataset \PYZhy{}\PYZhy{} in contrast,}
\PY{+w}{           }\PY{c+c1}{//            getcurdata pulls the foreground}
\PY{+w}{           }\PY{c+c1}{//            dataset}
\PY{+w}{               }\PY{n}{settitle}\PY{p}{(}\PY{l+s}{\PYZdq{}}\PY{l+s}{NMR experiment w/ standard }\PY{l+s}{\PYZdq{}}
\PY{+w}{                        }\PY{l+s}{\PYZdq{}}\PY{l+s}{parameters}\PY{l+s}{\PYZdq{}}\PY{p}{)}\PY{p}{;}
\PY{n}{RGA}\PY{+w}{ }\PY{n}{ZG\PYZus{}OVERWRITE}
\PY{+w}{    }\PY{c+c1}{// ERRORABORT returns from AU or AU subroutine}
\PY{+w}{    }\PY{c+c1}{//            with value of AUERR (if it is less than}
\PY{+w}{    }\PY{c+c1}{//            0) there are a couple diff options}
\PY{+w}{    }\PY{c+c1}{//            (table 2.22 in AU programming manual)}
\PY{+w}{        }\PY{n}{ERRORABORT}\PY{p}{;}
\PY{n}{EF}\PY{p}{;}\PY{+w}{ }\PY{c+c1}{// FFT w/ exponential apodization}
\PY{n}{ERRORABORT}\PY{p}{;}
\PY{n}{APK}\PY{p}{;}\PY{+w}{ }\PY{c+c1}{// auto phase (0 and 1)}
\PY{n}{fprintf}\PY{p}{(}\PY{n}{logfp}\PY{p}{,}\PY{+w}{ }\PY{l+s}{\PYZdq{}}\PY{l+s}{Finished 1H NMR 1D experiment.}\PY{l+s+se}{\PYZbs{}n}\PY{l+s}{\PYZdq{}}\PY{p}{)}\PY{p}{;}
\PY{c+c1}{//\PYZcb{}\PYZcb{}\PYZcb{}}

\PY{c+c1}{//\PYZob{}\PYZob{}\PYZob{} 2H experiment for peak picking}
\PY{n}{IEXPNO}\PY{p}{;}
\PY{n}{SETCURDATA}\PY{p}{;}\PY{+w}{ }\PY{c+c1}{// see comment below about IEXPNO}
\PY{n}{sprintf}\PY{p}{(}\PY{n}{titlestr}\PY{p}{,}\PY{+w}{ }\PY{l+s}{\PYZdq{}}\PY{l+s}{2H to find resonance}\PY{l+s}{\PYZdq{}}\PY{p}{)}\PY{p}{;}
\PY{n}{settitle}\PY{p}{(}\PY{n}{titlestr}\PY{p}{)}\PY{p}{;}
\PY{n}{RPAR}\PY{p}{(}\PY{l+s}{\PYZdq{}}\PY{l+s}{ab\PYZus{}2H\PYZus{}zg}\PY{l+s}{\PYZdq{}}\PY{p}{,}\PY{+w}{ }\PY{l+s}{\PYZdq{}}\PY{l+s}{all}\PY{l+s}{\PYZdq{}}\PY{p}{)}\PY{p}{;}
\PY{n}{fprintf}\PY{p}{(}\PY{n}{logfp}\PY{p}{,}
\PY{+w}{        }\PY{l+s}{\PYZdq{}}\PY{l+s}{Beginning 2H NMR 1D experiment for peak }\PY{l+s}{\PYZdq{}}
\PY{+w}{        }\PY{l+s}{\PYZdq{}}\PY{l+s}{picking in \PYZpc{}d...}\PY{l+s+se}{\PYZbs{}n}\PY{l+s}{\PYZdq{}}\PY{p}{,}
\PY{+w}{        }\PY{n}{expno}\PY{p}{)}\PY{p}{;}
\PY{n}{sf}\PY{+w}{ }\PY{o}{=}\PY{+w}{ }\PY{l+m+mf}{0.0}\PY{p}{;}
\PY{n}{sfo1}\PY{+w}{ }\PY{o}{=}\PY{+w}{ }\PY{l+m+mf}{0.0}\PY{p}{;}
\PY{n}{o1}\PY{+w}{ }\PY{o}{=}\PY{+w}{ }\PY{l+m+mf}{0.0}\PY{p}{;}
\PY{n}{p1}\PY{+w}{ }\PY{o}{=}\PY{+w}{ }\PY{l+m+mf}{0.0}\PY{p}{;}
\PY{c+c1}{//\PYZob{}\PYZob{}\PYZob{} pull all info}
\PY{n}{REXPNO}\PY{p}{(}\PY{n}{expno}\PY{p}{)}\PY{p}{;}
\PY{n}{RPROCNO}\PY{p}{(}\PY{n}{procno}\PY{p}{)}\PY{p}{;}
\PY{n}{SETCURDATA}
\PY{c+c1}{//\PYZcb{}\PYZcb{}\PYZcb{}}
\PY{n}{FETCHPAR}\PY{p}{(}\PY{l+s}{\PYZdq{}}\PY{l+s}{SFO1}\PY{l+s}{\PYZdq{}}\PY{p}{,}\PY{+w}{ }\PY{o}{\PYZam{}}\PY{n}{sfo1}\PY{p}{)}\PY{p}{;}
\PY{n}{FETCHPAR}\PY{p}{(}\PY{l+s}{\PYZdq{}}\PY{l+s}{O1}\PY{l+s}{\PYZdq{}}\PY{p}{,}\PY{+w}{ }\PY{o}{\PYZam{}}\PY{n}{o1}\PY{p}{)}\PY{p}{;}
\PY{n}{fprintf}\PY{p}{(}\PY{n}{logfp}\PY{p}{,}\PY{+w}{ }\PY{l+s}{\PYZdq{}}\PY{l+s}{here is sfo1: \PYZpc{}f}\PY{l+s+se}{\PYZbs{}n}\PY{l+s}{\PYZdq{}}\PY{p}{,}\PY{+w}{ }\PY{n}{sfo1}\PY{p}{)}\PY{p}{;}
\PY{n}{fprintf}\PY{p}{(}\PY{n}{logfp}\PY{p}{,}\PY{+w}{ }\PY{l+s}{\PYZdq{}}\PY{l+s}{here is o1: \PYZpc{}f}\PY{l+s+se}{\PYZbs{}n}\PY{l+s}{\PYZdq{}}\PY{p}{,}\PY{+w}{ }\PY{n}{o1}\PY{p}{)}\PY{p}{;}
\PY{c+c1}{// pull + set number of scans}
\PY{n}{FETCHPAR}\PY{p}{(}\PY{l+s}{\PYZdq{}}\PY{l+s}{NS}\PY{l+s}{\PYZdq{}}\PY{p}{,}\PY{+w}{ }\PY{o}{\PYZam{}}\PY{n}{noofscans}\PY{p}{)}\PY{p}{;}
\PY{n}{fprintf}\PY{p}{(}\PY{n}{logfp}\PY{p}{,}\PY{+w}{ }\PY{l+s}{\PYZdq{}}\PY{l+s}{I got \PYZpc{}d scans}\PY{l+s+se}{\PYZbs{}n}\PY{l+s}{\PYZdq{}}\PY{p}{,}\PY{+w}{ }\PY{n}{noofscans}\PY{p}{)}\PY{p}{;}
\PY{n}{ns}\PY{+w}{ }\PY{o}{=}\PY{+w}{ }\PY{l+m+mi}{1}\PY{p}{;}
\PY{n}{STOREPAR}\PY{p}{(}\PY{l+s}{\PYZdq{}}\PY{l+s}{NS}\PY{l+s}{\PYZdq{}}\PY{p}{,}\PY{+w}{ }\PY{n}{ns}\PY{p}{)}\PY{p}{;}
\PY{n}{FETCHPAR}\PY{p}{(}\PY{l+s}{\PYZdq{}}\PY{l+s}{NS}\PY{l+s}{\PYZdq{}}\PY{p}{,}\PY{+w}{ }\PY{o}{\PYZam{}}\PY{n}{noofscans}\PY{p}{)}\PY{p}{;}
\PY{n}{fprintf}\PY{p}{(}\PY{n}{logfp}\PY{p}{,}\PY{+w}{ }\PY{l+s}{\PYZdq{}}\PY{l+s}{after change, I got \PYZpc{}d scans}\PY{l+s+se}{\PYZbs{}n}\PY{l+s}{\PYZdq{}}\PY{p}{,}
\PY{+w}{        }\PY{n}{noofscans}\PY{p}{)}\PY{p}{;}
\PY{c+c1}{// pull + set r.d.}
\PY{n}{FETCHPAR}\PY{p}{(}\PY{l+s}{\PYZdq{}}\PY{l+s}{D1}\PY{l+s}{\PYZdq{}}\PY{p}{,}\PY{+w}{ }\PY{o}{\PYZam{}}\PY{n}{repDelay}\PY{p}{)}\PY{p}{;}
\PY{n}{fprintf}\PY{p}{(}\PY{n}{logfp}\PY{p}{,}\PY{+w}{ }\PY{l+s}{\PYZdq{}}\PY{l+s}{I got d1: \PYZpc{}f}\PY{l+s+se}{\PYZbs{}n}\PY{l+s}{\PYZdq{}}\PY{p}{,}\PY{+w}{ }\PY{n}{repDelay}\PY{p}{)}\PY{p}{;}
\PY{n}{my\PYZus{}d1}\PY{+w}{ }\PY{o}{=}\PY{+w}{ }\PY{l+m+mf}{2.5}\PY{p}{;}
\PY{n}{STOREPAR}\PY{p}{(}\PY{l+s}{\PYZdq{}}\PY{l+s}{D1}\PY{l+s}{\PYZdq{}}\PY{p}{,}\PY{+w}{ }\PY{n}{my\PYZus{}d1}\PY{p}{)}\PY{p}{;}
\PY{n}{FETCHPAR}\PY{p}{(}\PY{l+s}{\PYZdq{}}\PY{l+s}{D1}\PY{l+s}{\PYZdq{}}\PY{p}{,}\PY{+w}{ }\PY{o}{\PYZam{}}\PY{n}{repDelay}\PY{p}{)}\PY{p}{;}
\PY{n}{fprintf}\PY{p}{(}\PY{n}{logfp}\PY{p}{,}\PY{+w}{ }\PY{l+s}{\PYZdq{}}\PY{l+s}{after change, I got d1: \PYZpc{}f}\PY{l+s+se}{\PYZbs{}n}\PY{l+s}{\PYZdq{}}\PY{p}{,}\PY{+w}{ }\PY{n}{repDelay}\PY{p}{)}\PY{p}{;}
\PY{n}{RGA}\PY{p}{;}
\PY{n}{ZG\PYZus{}OVERWRITE}\PY{p}{;}
\PY{n}{ERRORABORT}\PY{p}{;}
\PY{n}{EF}\PY{p}{;}
\PY{n}{ERRORABORT}\PY{p}{;}
\PY{n}{APK}\PY{p}{;}
\PY{c+c1}{// PSCAL appears to be for vertical scaling. We want to}
\PY{c+c1}{// set}
\PY{c+c1}{//       it to \PYZdq{}global.\PYZdq{}}
\PY{n}{FETCHPAR}\PY{p}{(}\PY{l+s}{\PYZdq{}}\PY{l+s}{PSCAL}\PY{l+s}{\PYZdq{}}\PY{p}{,}\PY{+w}{ }\PY{o}{\PYZam{}}\PY{n}{pscal\PYZus{}save}\PY{p}{)}\PY{p}{;}
\PY{n}{STOREPAR}\PY{p}{(}\PY{l+s}{\PYZdq{}}\PY{l+s}{PSCAL}\PY{l+s}{\PYZdq{}}\PY{p}{,}\PY{+w}{ }\PY{l+m+mi}{6}\PY{p}{)}\PY{p}{;}
\PY{n}{fprintf}\PY{p}{(}\PY{n}{logfp}\PY{p}{,}\PY{+w}{ }\PY{l+s}{\PYZdq{}}\PY{l+s}{pscal\PYZus{}save is \PYZpc{}d}\PY{l+s+se}{\PYZbs{}n}\PY{l+s}{\PYZdq{}}\PY{p}{,}\PY{+w}{ }\PY{n}{pscal\PYZus{}save}\PY{p}{)}\PY{p}{;}
\PY{n}{PP}\PY{p}{;}\PY{+w}{ }\PY{c+c1}{// this tells it to run peak picking}
\PY{n}{ERRORABORT}\PY{p}{;}
\PY{n}{numPeaks}\PY{+w}{ }\PY{o}{=}\PY{+w}{ }\PY{n}{readPeakList}\PY{p}{(}\PY{n}{PROCPATH}\PY{p}{(}\PY{l+m+mi}{0}\PY{p}{)}\PY{p}{)}\PY{p}{;}
\PY{n}{fprintf}\PY{p}{(}\PY{n}{logfp}\PY{p}{,}\PY{+w}{ }\PY{l+s}{\PYZdq{}}\PY{l+s}{I find \PYZpc{}d peaks}\PY{l+s+se}{\PYZbs{}n}\PY{l+s}{\PYZdq{}}\PY{p}{,}\PY{+w}{ }\PY{n}{numPeaks}\PY{p}{)}\PY{p}{;}
\PY{c+c1}{// \PYZob{}\PYZob{}\PYZob{} store the max peak intensity, frequency, and ppm}
\PY{c+c1}{// in maxip[s,h,p]}
\PY{n}{maxips}\PY{+w}{ }\PY{o}{=}\PY{+w}{ }\PY{l+m+mf}{0.0}\PY{p}{;}
\PY{n}{maxpsh}\PY{+w}{ }\PY{o}{=}\PY{+w}{ }\PY{l+m+mf}{0.0}\PY{p}{;}
\PY{k}{for}\PY{+w}{ }\PY{p}{(}\PY{n}{i}\PY{+w}{ }\PY{o}{=}\PY{+w}{ }\PY{l+m+mi}{0}\PY{p}{;}\PY{+w}{ }\PY{n}{i}\PY{+w}{ }\PY{o}{\PYZlt{}}\PY{+w}{ }\PY{n}{numPeaks}\PY{p}{;}\PY{+w}{ }\PY{n}{i}\PY{o}{+}\PY{o}{+}\PY{p}{)}\PY{+w}{ }\PY{p}{\PYZob{}}
\PY{+w}{  }\PY{n}{peakIntensity}\PY{+w}{ }\PY{o}{=}\PY{+w}{ }\PY{n}{getPeakIntensity}\PY{p}{(}\PY{n}{i}\PY{p}{)}\PY{p}{;}
\PY{+w}{  }\PY{n}{peakFreqHz}\PY{+w}{ }\PY{o}{=}\PY{+w}{ }\PY{n}{getPeakFreqHz}\PY{p}{(}\PY{n}{i}\PY{p}{)}\PY{p}{;}
\PY{+w}{  }\PY{n}{peakFreqPPM}\PY{+w}{ }\PY{o}{=}\PY{+w}{ }\PY{n}{getPeakFreqPPM}\PY{p}{(}\PY{n}{i}\PY{p}{)}\PY{p}{;}
\PY{+w}{  }\PY{k}{if}\PY{+w}{ }\PY{p}{(}\PY{n}{peakIntensity}\PY{+w}{ }\PY{o}{\PYZgt{}}\PY{+w}{ }\PY{n}{maxips}\PY{p}{)}\PY{+w}{ }\PY{p}{\PYZob{}}
\PY{+w}{    }\PY{n}{maxips}\PY{+w}{ }\PY{o}{=}\PY{+w}{ }\PY{n}{peakIntensity}\PY{p}{;}
\PY{+w}{    }\PY{n}{maxpsh}\PY{+w}{ }\PY{o}{=}\PY{+w}{ }\PY{n}{peakFreqHz}\PY{p}{;}
\PY{+w}{    }\PY{n}{maxpsp}\PY{+w}{ }\PY{o}{=}\PY{+w}{ }\PY{n}{peakFreqPPM}\PY{p}{;}
\PY{+w}{    }\PY{n}{maxpsp}\PY{+w}{ }\PY{o}{=}\PY{+w}{ }\PY{n}{peakFreqPPM}\PY{p}{;}
\PY{+w}{  }\PY{p}{\PYZcb{}}
\PY{p}{\PYZcb{}}
\PY{n}{freePeakList}\PY{p}{(}\PY{p}{)}\PY{p}{;}
\PY{c+c1}{// \PYZcb{}\PYZcb{}\PYZcb{}}
\PY{n}{FETCHPAR}\PY{p}{(}\PY{l+s}{\PYZdq{}}\PY{l+s}{SF}\PY{l+s}{\PYZdq{}}\PY{p}{,}\PY{+w}{ }\PY{o}{\PYZam{}}\PY{n}{sf}\PY{p}{)}\PY{p}{;}
\PY{n}{fprintf}\PY{p}{(}\PY{n}{logfp}\PY{p}{,}\PY{+w}{ }\PY{l+s}{\PYZdq{}}\PY{l+s}{I got \PYZpc{}f for SF}\PY{l+s+se}{\PYZbs{}n}\PY{l+s}{\PYZdq{}}\PY{p}{,}\PY{+w}{ }\PY{n}{sf}\PY{p}{)}\PY{p}{;}
\PY{n}{sfo1}\PY{+w}{ }\PY{o}{=}\PY{+w}{ }\PY{n}{sf}\PY{+w}{ }\PY{o}{+}\PY{+w}{ }\PY{n}{maxpsh}\PY{+w}{ }\PY{o}{*}\PY{+w}{ }\PY{l+m+mf}{1.0e\PYZhy{}6}\PY{p}{;}
\PY{n}{fprintf}\PY{p}{(}\PY{n}{logfp}\PY{p}{,}\PY{+w}{ }\PY{l+s}{\PYZdq{}}\PY{l+s}{I got \PYZpc{}f for SFO1 to set}\PY{l+s+se}{\PYZbs{}n}\PY{l+s}{\PYZdq{}}\PY{p}{,}\PY{+w}{ }\PY{n}{sfo1}\PY{p}{)}\PY{p}{;}
\PY{n}{STOREPAR}\PY{p}{(}\PY{l+s}{\PYZdq{}}\PY{l+s}{SFO1}\PY{l+s}{\PYZdq{}}\PY{p}{,}\PY{+w}{ }\PY{n}{sfo1}\PY{p}{)}\PY{p}{;}
\PY{n}{FETCHPAR}\PY{p}{(}\PY{l+s}{\PYZdq{}}\PY{l+s}{SFO1}\PY{l+s}{\PYZdq{}}\PY{p}{,}\PY{+w}{ }\PY{o}{\PYZam{}}\PY{n}{sfo1}\PY{p}{)}\PY{p}{;}
\PY{n}{FETCHPAR}\PY{p}{(}\PY{l+s}{\PYZdq{}}\PY{l+s}{O1}\PY{l+s}{\PYZdq{}}\PY{p}{,}\PY{+w}{ }\PY{o}{\PYZam{}}\PY{n}{o1}\PY{p}{)}\PY{p}{;}
\PY{n}{FETCHPAR}\PY{p}{(}\PY{l+s}{\PYZdq{}}\PY{l+s}{P1}\PY{l+s}{\PYZdq{}}\PY{p}{,}\PY{+w}{ }\PY{o}{\PYZam{}}\PY{n}{p1}\PY{p}{)}\PY{p}{;}
\PY{n}{fprintf}\PY{p}{(}\PY{n}{logfp}\PY{p}{,}\PY{+w}{ }\PY{l+s}{\PYZdq{}}\PY{l+s}{I set SFO1 to \PYZpc{}f}\PY{l+s+se}{\PYZbs{}n}\PY{l+s}{\PYZdq{}}\PY{p}{,}\PY{+w}{ }\PY{n}{sfo1}\PY{p}{)}\PY{p}{;}
\PY{n}{fprintf}\PY{p}{(}\PY{n}{logfp}\PY{p}{,}\PY{+w}{ }\PY{l+s}{\PYZdq{}}\PY{l+s}{I set O1 to \PYZpc{}f}\PY{l+s+se}{\PYZbs{}n}\PY{l+s}{\PYZdq{}}\PY{p}{,}\PY{+w}{ }\PY{n}{o1}\PY{p}{)}\PY{p}{;}
\PY{n}{fprintf}\PY{p}{(}\PY{n}{logfp}\PY{p}{,}\PY{+w}{ }\PY{l+s}{\PYZdq{}}\PY{l+s}{I set SF to \PYZpc{}f}\PY{l+s+se}{\PYZbs{}n}\PY{l+s}{\PYZdq{}}\PY{p}{,}\PY{+w}{ }\PY{n}{sf}\PY{p}{)}\PY{p}{;}
\PY{n}{fprintf}\PY{p}{(}
\PY{+w}{    }\PY{n}{logfp}\PY{p}{,}
\PY{+w}{    }\PY{l+s}{\PYZdq{}}\PY{l+s}{I get P1 of \PYZpc{}f (which comes from the par file)}\PY{l+s+se}{\PYZbs{}n}\PY{l+s}{\PYZdq{}}\PY{p}{,}
\PY{+w}{    }\PY{n}{p1}\PY{p}{)}\PY{p}{;}
\PY{c+c1}{// pull + set number of scans}
\PY{n}{FETCHPAR}\PY{p}{(}\PY{l+s}{\PYZdq{}}\PY{l+s}{NS}\PY{l+s}{\PYZdq{}}\PY{p}{,}\PY{+w}{ }\PY{o}{\PYZam{}}\PY{n}{noofscans}\PY{p}{)}\PY{p}{;}
\PY{n}{fprintf}\PY{p}{(}\PY{n}{logfp}\PY{p}{,}\PY{+w}{ }\PY{l+s}{\PYZdq{}}\PY{l+s}{I got \PYZpc{}d scans}\PY{l+s+se}{\PYZbs{}n}\PY{l+s}{\PYZdq{}}\PY{p}{,}\PY{+w}{ }\PY{n}{noofscans}\PY{p}{)}\PY{p}{;}
\PY{n}{ns}\PY{+w}{ }\PY{o}{=}\PY{+w}{ }\PY{l+m+mi}{1}\PY{p}{;}
\PY{n}{STOREPAR}\PY{p}{(}\PY{l+s}{\PYZdq{}}\PY{l+s}{NS}\PY{l+s}{\PYZdq{}}\PY{p}{,}\PY{+w}{ }\PY{n}{ns}\PY{p}{)}\PY{p}{;}
\PY{n}{FETCHPAR}\PY{p}{(}\PY{l+s}{\PYZdq{}}\PY{l+s}{NS}\PY{l+s}{\PYZdq{}}\PY{p}{,}\PY{+w}{ }\PY{o}{\PYZam{}}\PY{n}{noofscans}\PY{p}{)}\PY{p}{;}
\PY{n}{fprintf}\PY{p}{(}\PY{n}{logfp}\PY{p}{,}\PY{+w}{ }\PY{l+s}{\PYZdq{}}\PY{l+s}{after change, I got \PYZpc{}d scans}\PY{l+s+se}{\PYZbs{}n}\PY{l+s}{\PYZdq{}}\PY{p}{,}
\PY{+w}{        }\PY{n}{noofscans}\PY{p}{)}\PY{p}{;}
\PY{c+c1}{// pull + set r.d.}
\PY{n}{FETCHPAR}\PY{p}{(}\PY{l+s}{\PYZdq{}}\PY{l+s}{D1}\PY{l+s}{\PYZdq{}}\PY{p}{,}\PY{+w}{ }\PY{o}{\PYZam{}}\PY{n}{repDelay}\PY{p}{)}\PY{p}{;}
\PY{n}{fprintf}\PY{p}{(}\PY{n}{logfp}\PY{p}{,}\PY{+w}{ }\PY{l+s}{\PYZdq{}}\PY{l+s}{I got d1: \PYZpc{}f}\PY{l+s+se}{\PYZbs{}n}\PY{l+s}{\PYZdq{}}\PY{p}{,}\PY{+w}{ }\PY{n}{repDelay}\PY{p}{)}\PY{p}{;}
\PY{n}{STOREPAR}\PY{p}{(}\PY{l+s}{\PYZdq{}}\PY{l+s}{D1}\PY{l+s}{\PYZdq{}}\PY{p}{,}\PY{+w}{ }\PY{n}{my\PYZus{}d1}\PY{p}{)}\PY{p}{;}
\PY{n}{FETCHPAR}\PY{p}{(}\PY{l+s}{\PYZdq{}}\PY{l+s}{D1}\PY{l+s}{\PYZdq{}}\PY{p}{,}\PY{+w}{ }\PY{o}{\PYZam{}}\PY{n}{repDelay}\PY{p}{)}\PY{p}{;}
\PY{n}{fprintf}\PY{p}{(}\PY{n}{logfp}\PY{p}{,}\PY{+w}{ }\PY{l+s}{\PYZdq{}}\PY{l+s}{after change, I got d1: \PYZpc{}f}\PY{l+s+se}{\PYZbs{}n}\PY{l+s}{\PYZdq{}}\PY{p}{,}\PY{+w}{ }\PY{n}{repDelay}\PY{p}{)}\PY{p}{;}
\PY{n}{ZG\PYZus{}OVERWRITE}\PY{p}{;}
\PY{n}{ERRORABORT}\PY{p}{;}
\PY{n}{EFP}\PY{p}{;}
\PY{n}{APK}\PY{p}{;}
\PY{c+c1}{//\PYZcb{}\PYZcb{}\PYZcb{}}

\PY{c+c1}{// \PYZob{}\PYZob{}\PYZob{} loop for nutation curve}
\PY{k}{for}\PY{+w}{ }\PY{p}{(}\PY{n}{i}\PY{+w}{ }\PY{o}{=}\PY{+w}{ }\PY{l+m+mi}{0}\PY{p}{;}\PY{+w}{ }\PY{n}{i}\PY{+w}{ }\PY{o}{\PYZlt{}}\PY{+w}{ }\PY{l+m+mi}{3}\PY{p}{;}\PY{+w}{ }\PY{n}{i}\PY{o}{+}\PY{o}{+}\PY{p}{)}\PY{+w}{ }\PY{p}{\PYZob{}}
\PY{+w}{  }\PY{k}{if}\PY{+w}{ }\PY{p}{(}\PY{n}{stillgoing}\PY{+w}{ }\PY{o}{\PYZgt{}}\PY{+w}{ }\PY{l+m+mi}{0}\PY{p}{)}\PY{+w}{ }\PY{p}{\PYZob{}}
\PY{+w}{    }\PY{n}{fprintf}\PY{p}{(}\PY{n}{logfp}\PY{p}{,}
\PY{+w}{            }\PY{l+s}{\PYZdq{}}\PY{l+s}{**********************************}\PY{l+s+se}{\PYZbs{}n}\PY{l+s}{\PYZdq{}}\PY{p}{)}\PY{p}{;}
\PY{+w}{    }\PY{n}{fprintf}\PY{p}{(}\PY{n}{logfp}\PY{p}{,}
\PY{+w}{            }\PY{l+s}{\PYZdq{}}\PY{l+s}{Beginning point \PYZpc{}i of nutation curve, the }\PY{l+s}{\PYZdq{}}
\PY{+w}{            }\PY{l+s}{\PYZdq{}}\PY{l+s}{experiment }\PY{l+s}{\PYZdq{}}
\PY{+w}{            }\PY{l+s}{\PYZdq{}}\PY{l+s}{will be in expno=\PYZpc{}d...}\PY{l+s+se}{\PYZbs{}n}\PY{l+s}{\PYZdq{}}\PY{p}{,}
\PY{+w}{            }\PY{n}{i}\PY{p}{,}\PY{+w}{ }\PY{n}{expno}\PY{+w}{ }\PY{o}{+}\PY{+w}{ }\PY{l+m+mi}{1}\PY{p}{)}\PY{p}{;}
\PY{+w}{    }\PY{n}{FETCHPAR}\PY{p}{(}\PY{l+s}{\PYZdq{}}\PY{l+s}{P1}\PY{l+s}{\PYZdq{}}\PY{p}{,}\PY{+w}{ }\PY{o}{\PYZam{}}\PY{n}{p1}\PY{p}{)}\PY{p}{;}
\PY{+w}{    }\PY{n}{fprintf}\PY{p}{(}\PY{n}{logfp}\PY{p}{,}\PY{+w}{ }\PY{l+s}{\PYZdq{}}\PY{l+s}{I had p1 of \PYZpc{}f}\PY{l+s+se}{\PYZbs{}n}\PY{l+s}{\PYZdq{}}\PY{p}{,}\PY{+w}{ }\PY{n}{p1}\PY{p}{)}\PY{p}{;}
\PY{+w}{    }\PY{n}{p1}\PY{+w}{ }\PY{o}{=}\PY{+w}{ }\PY{n}{p1\PYZus{}list}\PY{p}{[}\PY{n}{i}\PY{p}{]}\PY{p}{;}
\PY{+w}{    }\PY{n}{fprintf}\PY{p}{(}\PY{n}{logfp}\PY{p}{,}\PY{+w}{ }\PY{l+s}{\PYZdq{}}\PY{l+s}{I want p1 to be \PYZpc{}f}\PY{l+s+se}{\PYZbs{}n}\PY{l+s}{\PYZdq{}}\PY{p}{,}\PY{+w}{ }\PY{n}{p1}\PY{p}{)}\PY{p}{;}
\PY{+w}{    }\PY{n}{STOREPAR}\PY{p}{(}\PY{l+s}{\PYZdq{}}\PY{l+s}{P1}\PY{l+s}{\PYZdq{}}\PY{p}{,}\PY{+w}{ }\PY{n}{p1}\PY{p}{)}\PY{p}{;}
\PY{+w}{    }\PY{n}{FETCHPAR}\PY{p}{(}\PY{l+s}{\PYZdq{}}\PY{l+s}{P1}\PY{l+s}{\PYZdq{}}\PY{p}{,}\PY{+w}{ }\PY{o}{\PYZam{}}\PY{n}{p1}\PY{p}{)}\PY{p}{;}
\PY{+w}{    }\PY{n}{fprintf}\PY{p}{(}\PY{n}{logfp}\PY{p}{,}\PY{+w}{ }\PY{l+s}{\PYZdq{}}\PY{l+s}{I set p1 to \PYZpc{}f}\PY{l+s+se}{\PYZbs{}n}\PY{l+s}{\PYZdq{}}\PY{p}{,}\PY{+w}{ }\PY{n}{p1}\PY{p}{)}\PY{p}{;}
\PY{+w}{    }\PY{c+c1}{// pull + set number of scans}
\PY{+w}{    }\PY{n}{FETCHPAR}\PY{p}{(}\PY{l+s}{\PYZdq{}}\PY{l+s}{NS}\PY{l+s}{\PYZdq{}}\PY{p}{,}\PY{+w}{ }\PY{o}{\PYZam{}}\PY{n}{noofscans}\PY{p}{)}\PY{p}{;}
\PY{+w}{    }\PY{n}{fprintf}\PY{p}{(}\PY{n}{logfp}\PY{p}{,}\PY{+w}{ }\PY{l+s}{\PYZdq{}}\PY{l+s}{I got \PYZpc{}d scans}\PY{l+s+se}{\PYZbs{}n}\PY{l+s}{\PYZdq{}}\PY{p}{,}\PY{+w}{ }\PY{n}{noofscans}\PY{p}{)}\PY{p}{;}
\PY{+w}{    }\PY{n}{ns}\PY{+w}{ }\PY{o}{=}\PY{+w}{ }\PY{l+m+mi}{1}\PY{p}{;}
\PY{+w}{    }\PY{n}{STOREPAR}\PY{p}{(}\PY{l+s}{\PYZdq{}}\PY{l+s}{NS}\PY{l+s}{\PYZdq{}}\PY{p}{,}\PY{+w}{ }\PY{n}{ns}\PY{p}{)}\PY{p}{;}
\PY{+w}{    }\PY{n}{FETCHPAR}\PY{p}{(}\PY{l+s}{\PYZdq{}}\PY{l+s}{NS}\PY{l+s}{\PYZdq{}}\PY{p}{,}\PY{+w}{ }\PY{o}{\PYZam{}}\PY{n}{noofscans}\PY{p}{)}\PY{p}{;}
\PY{+w}{    }\PY{n}{fprintf}\PY{p}{(}\PY{n}{logfp}\PY{p}{,}\PY{+w}{ }\PY{l+s}{\PYZdq{}}\PY{l+s}{after change, I got \PYZpc{}d scans}\PY{l+s+se}{\PYZbs{}n}\PY{l+s}{\PYZdq{}}\PY{p}{,}
\PY{+w}{            }\PY{n}{noofscans}\PY{p}{)}\PY{p}{;}
\PY{+w}{    }\PY{c+c1}{// pull + set r.d.}
\PY{+w}{    }\PY{n}{FETCHPAR}\PY{p}{(}\PY{l+s}{\PYZdq{}}\PY{l+s}{D1}\PY{l+s}{\PYZdq{}}\PY{p}{,}\PY{+w}{ }\PY{o}{\PYZam{}}\PY{n}{repDelay}\PY{p}{)}\PY{p}{;}
\PY{+w}{    }\PY{n}{fprintf}\PY{p}{(}\PY{n}{logfp}\PY{p}{,}\PY{+w}{ }\PY{l+s}{\PYZdq{}}\PY{l+s}{I got d1: \PYZpc{}f}\PY{l+s+se}{\PYZbs{}n}\PY{l+s}{\PYZdq{}}\PY{p}{,}\PY{+w}{ }\PY{n}{repDelay}\PY{p}{)}\PY{p}{;}
\PY{+w}{    }\PY{n}{STOREPAR}\PY{p}{(}\PY{l+s}{\PYZdq{}}\PY{l+s}{D1}\PY{l+s}{\PYZdq{}}\PY{p}{,}\PY{+w}{ }\PY{n}{my\PYZus{}d1}\PY{p}{)}\PY{p}{;}
\PY{+w}{    }\PY{n}{FETCHPAR}\PY{p}{(}\PY{l+s}{\PYZdq{}}\PY{l+s}{D1}\PY{l+s}{\PYZdq{}}\PY{p}{,}\PY{+w}{ }\PY{o}{\PYZam{}}\PY{n}{repDelay}\PY{p}{)}\PY{p}{;}
\PY{+w}{    }\PY{n}{fprintf}\PY{p}{(}\PY{n}{logfp}\PY{p}{,}\PY{+w}{ }\PY{l+s}{\PYZdq{}}\PY{l+s}{after change, I got d1: \PYZpc{}f}\PY{l+s+se}{\PYZbs{}n}\PY{l+s}{\PYZdq{}}\PY{p}{,}
\PY{+w}{            }\PY{n}{repDelay}\PY{p}{)}\PY{p}{;}
\PY{+w}{    }\PY{n}{IEXPNO}\PY{p}{;}
\PY{+w}{    }\PY{n}{SETCURDATA}\PY{p}{;}\PY{+w}{ }\PY{c+c1}{// pg 15 \PYZhy{}\PYZhy{} IEXPNO changes current}
\PY{+w}{                }\PY{c+c1}{// dataset, but doesn\PYZsq{}t}
\PY{+w}{    }\PY{c+c1}{//             \PYZdq{}make it available\PYZdq{} \PYZhy{}\PYZhy{} must be}
\PY{+w}{    }\PY{c+c1}{//             followed by SETCURDATA for other}
\PY{+w}{    }\PY{c+c1}{//             commands that access the current}
\PY{+w}{    }\PY{c+c1}{//             dataset (I believe that here that}
\PY{+w}{    }\PY{c+c1}{//             includes our settitle function)}
\PY{+w}{    }\PY{c+c1}{// \PYZob{}\PYZob{}\PYZob{} set processing params and title unique to}
\PY{+w}{    }\PY{c+c1}{// nutation}
\PY{+w}{    }\PY{n}{sprintf}\PY{p}{(}\PY{n}{titlestr}\PY{p}{,}\PY{+w}{ }\PY{l+s}{\PYZdq{}}\PY{l+s}{nutation step \PYZpc{}d}\PY{l+s}{\PYZdq{}}\PY{p}{,}\PY{+w}{ }\PY{n}{i}\PY{p}{)}\PY{p}{;}
\PY{+w}{    }\PY{n}{settitle}\PY{p}{(}\PY{n}{titlestr}\PY{p}{)}\PY{p}{;}
\PY{+w}{    }\PY{n}{STOREPAR}\PY{p}{(}\PY{l+s}{\PYZdq{}}\PY{l+s}{LB}\PY{l+s}{\PYZdq{}}\PY{p}{,}\PY{+w}{ }\PY{l+m+mf}{5.0}\PY{p}{)}\PY{p}{;}
\PY{+w}{    }\PY{n}{STOREPAR}\PY{p}{(}\PY{l+s}{\PYZdq{}}\PY{l+s}{PSIGN}\PY{l+s}{\PYZdq{}}\PY{p}{,}\PY{+w}{ }\PY{l+m+mi}{2}\PY{p}{)}\PY{p}{;}\PY{+w}{ }\PY{c+c1}{// \PYZdq{}both\PYZdq{}}
\PY{+w}{    }\PY{c+c1}{// \PYZcb{}\PYZcb{}\PYZcb{}}
\PY{+w}{    }\PY{n}{ZG\PYZus{}OVERWRITE}\PY{p}{;}
\PY{+w}{    }\PY{n}{EFP}\PY{+w}{ }\PY{c+c1}{// exp FT and apply PHC0 and PHC1 \PYZhy{}\PYZhy{} don\PYZsq{}t think}
\PY{+w}{        }\PY{c+c1}{// it alters PHC0 and PHC1}
\PY{+w}{        }\PY{n+nf}{if}\PY{+w}{ }\PY{p}{(}\PY{n}{i}\PY{+w}{ }\PY{o}{=}\PY{o}{=}\PY{+w}{ }\PY{l+m+mi}{0}\PY{p}{)}\PY{p}{\PYZob{}}
\PY{+w}{            }\PY{n}{APK}\PY{+w}{ }\PY{c+c1}{// this alters PHC0 and PHC1}
\PY{+w}{        }\PY{p}{\PYZcb{}}\PY{+w}{ }\PY{n}{FETCHPAR}\PY{p}{(}\PY{l+s}{\PYZdq{}}\PY{l+s}{PSCAL}\PY{l+s}{\PYZdq{}}\PY{p}{,}\PY{+w}{ }\PY{o}{\PYZam{}}\PY{n}{pscal\PYZus{}save}\PY{p}{)}\PY{p}{;}
\PY{+w}{    }\PY{n}{STOREPAR}\PY{p}{(}\PY{l+s}{\PYZdq{}}\PY{l+s}{PSCAL}\PY{l+s}{\PYZdq{}}\PY{p}{,}\PY{+w}{ }\PY{l+m+mi}{6}\PY{p}{)}\PY{p}{;}
\PY{+w}{    }\PY{n}{fprintf}\PY{p}{(}\PY{n}{logfp}\PY{p}{,}\PY{+w}{ }\PY{l+s}{\PYZdq{}}\PY{l+s}{pscal\PYZus{}save is \PYZpc{}d}\PY{l+s+se}{\PYZbs{}n}\PY{l+s}{\PYZdq{}}\PY{p}{,}\PY{+w}{ }\PY{n}{pscal\PYZus{}save}\PY{p}{)}\PY{p}{;}
\PY{+w}{    }\PY{n}{PP}\PY{p}{;}\PY{+w}{ }\PY{c+c1}{// peak pick}
\PY{+w}{    }\PY{n}{ERRORABORT}\PY{p}{;}
\PY{+w}{    }\PY{n}{numPeaks}\PY{+w}{ }\PY{o}{=}\PY{+w}{ }\PY{n}{readPeakList}\PY{p}{(}\PY{n}{PROCPATH}\PY{p}{(}\PY{l+m+mi}{0}\PY{p}{)}\PY{p}{)}\PY{p}{;}
\PY{+w}{    }\PY{n}{fprintf}\PY{p}{(}\PY{n}{logfp}\PY{p}{,}\PY{+w}{ }\PY{l+s}{\PYZdq{}}\PY{l+s}{I find \PYZpc{}d peaks}\PY{l+s+se}{\PYZbs{}n}\PY{l+s}{\PYZdq{}}\PY{p}{,}\PY{+w}{ }\PY{n}{numPeaks}\PY{p}{)}\PY{p}{;}
\PY{+w}{    }\PY{c+c1}{// store the max peak intensity, frequency, and ppm}
\PY{+w}{    }\PY{c+c1}{// in maxip[s,h,p]}
\PY{+w}{    }\PY{n}{maxips}\PY{+w}{ }\PY{o}{=}\PY{+w}{ }\PY{l+m+mf}{0.0}\PY{p}{;}
\PY{+w}{    }\PY{n}{maxpsh}\PY{+w}{ }\PY{o}{=}\PY{+w}{ }\PY{l+m+mf}{0.0}\PY{p}{;}
\PY{+w}{    }\PY{k}{for}\PY{+w}{ }\PY{p}{(}\PY{n}{j}\PY{+w}{ }\PY{o}{=}\PY{+w}{ }\PY{l+m+mi}{0}\PY{p}{;}\PY{+w}{ }\PY{n}{j}\PY{+w}{ }\PY{o}{\PYZlt{}}\PY{+w}{ }\PY{n}{numPeaks}\PY{p}{;}\PY{+w}{ }\PY{n}{j}\PY{o}{+}\PY{o}{+}\PY{p}{)}\PY{+w}{ }\PY{p}{\PYZob{}}
\PY{+w}{      }\PY{n}{peakIntensity}\PY{+w}{ }\PY{o}{=}\PY{+w}{ }\PY{n}{getPeakIntensity}\PY{p}{(}\PY{n}{j}\PY{p}{)}\PY{p}{;}
\PY{+w}{      }\PY{n}{peakFreqHz}\PY{+w}{ }\PY{o}{=}\PY{+w}{ }\PY{n}{getPeakFreqHz}\PY{p}{(}\PY{n}{j}\PY{p}{)}\PY{p}{;}
\PY{+w}{      }\PY{n}{peakFreqPPM}\PY{+w}{ }\PY{o}{=}\PY{+w}{ }\PY{n}{getPeakFreqPPM}\PY{p}{(}\PY{n}{j}\PY{p}{)}\PY{p}{;}
\PY{+w}{      }\PY{n}{fprintf}\PY{p}{(}\PY{n}{logfp}\PY{p}{,}\PY{+w}{ }\PY{l+s}{\PYZdq{}}\PY{l+s}{check peak intensity \PYZpc{}f at \PYZpc{}f}\PY{l+s+se}{\PYZbs{}n}\PY{l+s}{\PYZdq{}}\PY{p}{,}
\PY{+w}{              }\PY{n}{peakIntensity}\PY{p}{,}\PY{+w}{ }\PY{n}{peakFreqPPM}\PY{p}{)}\PY{p}{;}
\PY{+w}{      }\PY{n}{fprintf}\PY{p}{(}\PY{n}{logfp}\PY{p}{,}\PY{+w}{ }\PY{l+s}{\PYZdq{}}\PY{l+s}{comparing \PYZpc{}f to \PYZpc{}f}\PY{l+s+se}{\PYZbs{}n}\PY{l+s}{\PYZdq{}}\PY{p}{,}
\PY{+w}{              }\PY{n}{abs}\PY{p}{(}\PY{n}{peakIntensity}\PY{p}{)}\PY{p}{,}\PY{+w}{ }\PY{n}{abs}\PY{p}{(}\PY{n}{maxips}\PY{p}{)}\PY{p}{)}\PY{p}{;}
\PY{+w}{      }\PY{k}{if}\PY{+w}{ }\PY{p}{(}\PY{n}{abs}\PY{p}{(}\PY{n}{peakIntensity}\PY{p}{)}\PY{+w}{ }\PY{o}{\PYZgt{}}\PY{+w}{ }\PY{n}{abs}\PY{p}{(}\PY{n}{maxips}\PY{p}{)}\PY{p}{)}\PY{+w}{ }\PY{p}{\PYZob{}}
\PY{+w}{        }\PY{n}{maxips}\PY{+w}{ }\PY{o}{=}\PY{+w}{ }\PY{n}{peakIntensity}\PY{p}{;}
\PY{+w}{        }\PY{n}{maxpsh}\PY{+w}{ }\PY{o}{=}\PY{+w}{ }\PY{n}{peakFreqHz}\PY{p}{;}
\PY{+w}{        }\PY{n}{maxpsp}\PY{+w}{ }\PY{o}{=}\PY{+w}{ }\PY{n}{peakFreqPPM}\PY{p}{;}
\PY{+w}{      }\PY{p}{\PYZcb{}}
\PY{+w}{    }\PY{p}{\PYZcb{}}
\PY{+w}{    }\PY{k}{if}\PY{+w}{ }\PY{p}{(}\PY{n}{i}\PY{+w}{ }\PY{o}{=}\PY{o}{=}\PY{+w}{ }\PY{l+m+mi}{0}\PY{p}{)}\PY{+w}{ }\PY{p}{\PYZob{}}
\PY{+w}{      }\PY{n}{first\PYZus{}maxpsp}\PY{+w}{ }\PY{o}{=}\PY{+w}{ }\PY{n}{maxpsp}\PY{p}{;}
\PY{+w}{    }\PY{p}{\PYZcb{}}\PY{+w}{ }\PY{k}{else}\PY{+w}{ }\PY{p}{\PYZob{}}
\PY{+w}{      }\PY{k}{if}\PY{+w}{ }\PY{p}{(}\PY{n}{abs}\PY{p}{(}\PY{n}{first\PYZus{}maxpsp}\PY{+w}{ }\PY{o}{\PYZhy{}}\PY{+w}{ }\PY{n}{maxpsp}\PY{p}{)}\PY{+w}{ }\PY{o}{\PYZgt{}}\PY{+w}{ }\PY{l+m+mf}{1.0}\PY{p}{)}\PY{+w}{ }\PY{p}{\PYZob{}}
\PY{+w}{        }\PY{n}{fprintf}\PY{p}{(}\PY{n}{logfp}\PY{p}{,}\PY{+w}{ }\PY{l+s}{\PYZdq{}}\PY{l+s}{I\PYZsq{}m exiting because the }\PY{l+s}{\PYZdq{}}
\PY{+w}{                       }\PY{l+s}{\PYZdq{}}\PY{l+s}{frequency deviation }\PY{l+s}{\PYZdq{}}
\PY{+w}{                       }\PY{l+s}{\PYZdq{}}\PY{l+s}{is too large}\PY{l+s+se}{\PYZbs{}n}\PY{l+s}{\PYZdq{}}\PY{p}{)}\PY{p}{;}
\PY{+w}{        }\PY{n}{stillgoing}\PY{+w}{ }\PY{o}{=}\PY{+w}{ }\PY{l+m+mi}{0}\PY{p}{;}
\PY{+w}{      }\PY{p}{\PYZcb{}}
\PY{+w}{    }\PY{p}{\PYZcb{}}
\PY{+w}{  }\PY{p}{\PYZcb{}}
\PY{+w}{  }\PY{k}{if}\PY{+w}{ }\PY{p}{(}\PY{n}{stillgoing}\PY{+w}{ }\PY{o}{\PYZgt{}}\PY{+w}{ }\PY{l+m+mi}{0}\PY{p}{)}\PY{+w}{ }\PY{p}{\PYZob{}}
\PY{+w}{    }\PY{n}{fprintf}\PY{p}{(}\PY{n}{logfp}\PY{p}{,}
\PY{+w}{            }\PY{l+s}{\PYZdq{}}\PY{l+s}{for nutation step \PYZpc{}d, the peak parameters }\PY{l+s}{\PYZdq{}}
\PY{+w}{            }\PY{l+s}{\PYZdq{}}\PY{l+s}{are ips=\PYZpc{}f }\PY{l+s}{\PYZdq{}}
\PY{+w}{            }\PY{l+s}{\PYZdq{}}\PY{l+s}{psh=\PYZpc{}f psp=\PYZpc{}f}\PY{l+s+se}{\PYZbs{}n}\PY{l+s}{\PYZdq{}}\PY{p}{,}
\PY{+w}{            }\PY{n}{i}\PY{p}{,}\PY{+w}{ }\PY{n}{maxips}\PY{p}{,}\PY{+w}{ }\PY{n}{maxpsh}\PY{p}{,}\PY{+w}{ }\PY{n}{maxpsp}\PY{p}{)}\PY{p}{;}
\PY{+w}{    }\PY{n}{freePeakList}\PY{p}{(}\PY{p}{)}\PY{p}{;}
\PY{+w}{    }\PY{n}{fprintf}\PY{p}{(}\PY{n}{logfp}\PY{p}{,}
\PY{+w}{            }\PY{l+s}{\PYZdq{}}\PY{l+s}{For my first nutation, I set p1 to \PYZpc{}f}\PY{l+s+se}{\PYZbs{}n}\PY{l+s}{\PYZdq{}}\PY{p}{,}
\PY{+w}{            }\PY{n}{p1}\PY{p}{)}\PY{p}{;}
\PY{+w}{    }\PY{n}{fprintf}\PY{p}{(}
\PY{+w}{        }\PY{n}{logfp}\PY{p}{,}
\PY{+w}{        }\PY{l+s}{\PYZdq{}}\PY{l+s}{For my first nutation, I get peak max of \PYZpc{}f}\PY{l+s+se}{\PYZbs{}n}\PY{l+s}{\PYZdq{}}\PY{p}{,}
\PY{+w}{        }\PY{n}{maxips}\PY{p}{)}\PY{p}{;}
\PY{+w}{    }\PY{n}{nutation\PYZus{}peaks}\PY{p}{[}\PY{n}{i}\PY{p}{]}\PY{+w}{ }\PY{o}{=}\PY{+w}{ }\PY{n}{maxips}\PY{p}{;}
\PY{+w}{  }\PY{p}{\PYZcb{}}
\PY{p}{\PYZcb{}}
\PY{c+c1}{// \PYZcb{}\PYZcb{}\PYZcb{}}

\PY{c+c1}{// \PYZob{}\PYZob{}\PYZob{} find the 90 time from the list of 3 peaks}
\PY{k}{if}\PY{+w}{ }\PY{p}{(}\PY{n}{stillgoing}\PY{+w}{ }\PY{o}{\PYZgt{}}\PY{+w}{ }\PY{l+m+mi}{0}\PY{p}{)}\PY{+w}{ }\PY{p}{\PYZob{}}
\PY{+w}{  }\PY{n}{fprintf}\PY{p}{(}\PY{n}{logfp}\PY{p}{,}
\PY{+w}{          }\PY{l+s}{\PYZdq{}}\PY{l+s}{*** *** *** *** *** *** *** *** ***}\PY{l+s+se}{\PYZbs{}n}\PY{l+s}{\PYZdq{}}\PY{p}{)}\PY{p}{;}
\PY{+w}{  }\PY{n}{fprintf}\PY{p}{(}\PY{n}{logfp}\PY{p}{,}\PY{+w}{ }\PY{l+s}{\PYZdq{}}\PY{l+s}{result of nutation curve:}\PY{l+s+se}{\PYZbs{}n}\PY{l+s}{\PYZdq{}}\PY{p}{)}\PY{p}{;}
\PY{+w}{  }\PY{k}{for}\PY{+w}{ }\PY{p}{(}\PY{n}{i}\PY{+w}{ }\PY{o}{=}\PY{+w}{ }\PY{l+m+mi}{0}\PY{p}{;}\PY{+w}{ }\PY{n}{i}\PY{+w}{ }\PY{o}{\PYZlt{}}\PY{+w}{ }\PY{l+m+mi}{3}\PY{p}{;}\PY{+w}{ }\PY{n}{i}\PY{o}{+}\PY{o}{+}\PY{p}{)}\PY{+w}{ }\PY{p}{\PYZob{}}
\PY{+w}{    }\PY{n}{fprintf}\PY{p}{(}\PY{n}{logfp}\PY{p}{,}\PY{+w}{ }\PY{l+s}{\PYZdq{}}\PY{l+s}{p1=\PYZpc{}f}\PY{l+s+se}{\PYZbs{}t}\PY{l+s}{height=\PYZpc{}f}\PY{l+s+se}{\PYZbs{}n}\PY{l+s}{\PYZdq{}}\PY{p}{,}\PY{+w}{ }\PY{n}{p1\PYZus{}list}\PY{p}{[}\PY{n}{i}\PY{p}{]}\PY{p}{,}
\PY{+w}{            }\PY{n}{nutation\PYZus{}peaks}\PY{p}{[}\PY{n}{i}\PY{p}{]}\PY{p}{)}\PY{p}{;}
\PY{+w}{  }\PY{p}{\PYZcb{}}
\PY{+w}{  }\PY{k}{if}\PY{+w}{ }\PY{p}{(}\PY{n}{nutation\PYZus{}peaks}\PY{p}{[}\PY{l+m+mi}{2}\PY{p}{]}\PY{+w}{ }\PY{o}{\PYZgt{}}\PY{+w}{ }\PY{l+m+mi}{0}\PY{p}{)}\PY{+w}{ }\PY{p}{\PYZob{}}
\PY{+w}{    }\PY{n}{fprintf}\PY{p}{(}\PY{n}{logfp}\PY{p}{,}\PY{+w}{ }\PY{l+s}{\PYZdq{}}\PY{l+s}{I\PYZsq{}m going to quit b/c the 3rd }\PY{l+s}{\PYZdq{}}
\PY{+w}{                   }\PY{l+s}{\PYZdq{}}\PY{l+s}{nutation peak is positive!}\PY{l+s+se}{\PYZbs{}n}\PY{l+s}{\PYZdq{}}\PY{p}{)}\PY{p}{;}
\PY{+w}{    }\PY{n}{stillgoing}\PY{+w}{ }\PY{o}{=}\PY{+w}{ }\PY{l+m+mi}{0}\PY{p}{;}
\PY{+w}{  }\PY{p}{\PYZcb{}}
\PY{p}{\PYZcb{}}
\PY{k}{if}\PY{+w}{ }\PY{p}{(}\PY{n}{stillgoing}\PY{+w}{ }\PY{o}{\PYZgt{}}\PY{+w}{ }\PY{l+m+mi}{0}\PY{p}{)}\PY{+w}{ }\PY{p}{\PYZob{}}
\PY{+w}{  }\PY{n}{REAL}\PY{+w}{ }\PY{n}{m}\PY{p}{,}\PY{+w}{ }\PY{n}{b}\PY{p}{,}\PY{+w}{ }\PY{n}{r2}\PY{p}{;}
\PY{+w}{  }\PY{k}{for}\PY{+w}{ }\PY{p}{(}\PY{n}{i}\PY{+w}{ }\PY{o}{=}\PY{+w}{ }\PY{l+m+mi}{0}\PY{p}{;}\PY{+w}{ }\PY{n}{i}\PY{+w}{ }\PY{o}{\PYZlt{}}\PY{+w}{ }\PY{l+m+mi}{3}\PY{p}{;}\PY{+w}{ }\PY{n}{i}\PY{o}{+}\PY{o}{+}\PY{p}{)}\PY{+w}{ }\PY{p}{\PYZob{}}
\PY{+w}{    }\PY{n}{p1\PYZus{}list\PYZus{}forreg}\PY{p}{[}\PY{n}{i}\PY{p}{]}\PY{+w}{ }\PY{o}{=}\PY{+w}{ }\PY{p}{(}\PY{n}{REAL}\PY{p}{)}\PY{n}{p1\PYZus{}list}\PY{p}{[}\PY{n}{i}\PY{p}{]}\PY{p}{;}
\PY{+w}{  }\PY{p}{\PYZcb{}}
\PY{+w}{  }\PY{n}{linreg}\PY{p}{(}\PY{l+m+mi}{3}\PY{p}{,}\PY{+w}{ }\PY{n}{p1\PYZus{}list\PYZus{}forreg}\PY{p}{,}\PY{+w}{ }\PY{n}{nutation\PYZus{}peaks}\PY{p}{,}\PY{+w}{ }\PY{o}{\PYZam{}}\PY{n}{m}\PY{p}{,}\PY{+w}{ }\PY{o}{\PYZam{}}\PY{n}{b}\PY{p}{,}\PY{+w}{ }\PY{o}{\PYZam{}}\PY{n}{r2}\PY{p}{)}\PY{p}{;}
\PY{+w}{  }\PY{n}{p1}\PY{+w}{ }\PY{o}{=}\PY{+w}{ }\PY{o}{\PYZhy{}}\PY{n}{b}\PY{+w}{ }\PY{o}{/}\PY{+w}{ }\PY{n}{m}\PY{+w}{ }\PY{o}{/}\PY{+w}{ }\PY{l+m+mf}{2.}\PY{p}{;}
\PY{+w}{  }\PY{n}{IEXPNO}\PY{p}{;}
\PY{+w}{  }\PY{n}{fprintf}\PY{p}{(}\PY{n}{logfp}\PY{p}{,}\PY{+w}{ }\PY{l+s}{\PYZdq{}}\PY{l+s}{m=\PYZpc{}f b=\PYZpc{}f r=\PYZpc{}g}\PY{l+s+se}{\PYZbs{}n}\PY{l+s}{\PYZdq{}}\PY{p}{,}\PY{+w}{ }\PY{n}{m}\PY{p}{,}\PY{+w}{ }\PY{n}{b}\PY{p}{,}\PY{+w}{ }\PY{n}{r2}\PY{p}{)}\PY{p}{;}
\PY{+w}{  }\PY{n}{fprintf}\PY{p}{(}\PY{n}{logfp}\PY{p}{,}\PY{+w}{ }\PY{l+s}{\PYZdq{}}\PY{l+s}{going to set p1 to \PYZpc{}f in \PYZpc{}d}\PY{l+s+se}{\PYZbs{}n}\PY{l+s}{\PYZdq{}}\PY{p}{,}\PY{+w}{ }\PY{n}{p1}\PY{p}{,}
\PY{+w}{          }\PY{n}{expno}\PY{p}{)}\PY{p}{;}
\PY{+w}{  }\PY{n}{STOREPAR}\PY{p}{(}\PY{l+s}{\PYZdq{}}\PY{l+s}{P1}\PY{l+s}{\PYZdq{}}\PY{p}{,}\PY{+w}{ }\PY{n}{p1}\PY{p}{)}\PY{p}{;}
\PY{+w}{  }\PY{n}{ZG\PYZus{}OVERWRITE}\PY{p}{;}
\PY{+w}{  }\PY{n}{ERRORABORT}\PY{p}{;}
\PY{+w}{  }\PY{n}{EFP}\PY{p}{;}
\PY{+w}{  }\PY{n}{APK}\PY{p}{;}
\PY{+w}{  }\PY{k}{if}\PY{+w}{ }\PY{p}{(}\PY{n}{p1}\PY{+w}{ }\PY{o}{\PYZgt{}}\PY{+w}{ }\PY{l+m+mi}{800}\PY{+w}{ }\PY{o}{|}\PY{o}{|}\PY{+w}{ }\PY{n}{p1}\PY{+w}{ }\PY{o}{\PYZlt{}}\PY{+w}{ }\PY{l+m+mi}{0}\PY{p}{)}\PY{+w}{ }\PY{p}{\PYZob{}}
\PY{+w}{    }\PY{n}{fprintf}\PY{p}{(}\PY{n}{logfp}\PY{p}{,}\PY{+w}{ }\PY{l+s}{\PYZdq{}}\PY{l+s}{that\PYZsq{}s an invalid value for }\PY{l+s}{\PYZdq{}}
\PY{+w}{                   }\PY{l+s}{\PYZdq{}}\PY{l+s}{p1!!!}\PY{l+s+se}{\PYZbs{}n}\PY{l+s}{I\PYZsq{}m going to quit!!!}\PY{l+s+se}{\PYZbs{}n}\PY{l+s}{\PYZdq{}}\PY{p}{)}\PY{p}{;}
\PY{+w}{    }\PY{n}{STOREPAR}\PY{p}{(}\PY{l+s}{\PYZdq{}}\PY{l+s}{P1}\PY{l+s}{\PYZdq{}}\PY{p}{,}\PY{+w}{ }\PY{l+m+mf}{360.0}\PY{p}{)}\PY{p}{;}
\PY{+w}{    }\PY{n}{stillgoing}\PY{+w}{ }\PY{o}{=}\PY{+w}{ }\PY{l+m+mi}{0}\PY{p}{;}
\PY{+w}{  }\PY{p}{\PYZcb{}}
\PY{p}{\PYZcb{}}
\PY{c+c1}{// \PYZcb{}\PYZcb{}\PYZcb{}}

\PY{k}{if}\PY{+w}{ }\PY{p}{(}\PY{n}{stillgoing}\PY{p}{)}\PY{+w}{ }\PY{p}{\PYZob{}}
\PY{+w}{  }\PY{c+c1}{//\PYZob{}\PYZob{}\PYZob{} 2H IR}
\PY{+w}{  }\PY{n}{IEXPNO}
\PY{+w}{  }\PY{n}{SETCURDATA}
\PY{+w}{  }\PY{n}{fprintf}\PY{p}{(}\PY{n}{logfp}\PY{p}{,}\PY{+w}{ }\PY{l+s}{\PYZdq{}}\PY{l+s}{**********************************}\PY{l+s+se}{\PYZbs{}n}\PY{l+s}{\PYZdq{}}\PY{p}{)}\PY{p}{;}
\PY{+w}{  }\PY{n}{fprintf}\PY{p}{(}\PY{n}{logfp}\PY{p}{,}
\PY{+w}{          }\PY{l+s}{\PYZdq{}}\PY{l+s}{Incrementing experiment number to \PYZpc{}d...}\PY{l+s+se}{\PYZbs{}n}\PY{l+s}{\PYZdq{}}\PY{p}{,}
\PY{+w}{          }\PY{n}{expno}\PY{p}{)}\PY{p}{;}
\PY{+w}{  }\PY{n}{RPAR}\PY{p}{(}\PY{l+s}{\PYZdq{}}\PY{l+s}{ab\PYZus{}2H\PYZus{}T1\PYZus{}logSpace}\PY{l+s}{\PYZdq{}}\PY{p}{,}\PY{+w}{ }\PY{l+s}{\PYZdq{}}\PY{l+s}{all}\PY{l+s}{\PYZdq{}}\PY{p}{)}\PY{p}{;}
\PY{+w}{  }\PY{n}{STOREPAR}\PY{p}{(}\PY{l+s}{\PYZdq{}}\PY{l+s}{SFO1}\PY{l+s}{\PYZdq{}}\PY{p}{,}\PY{+w}{ }\PY{n}{sfo1}\PY{p}{)}\PY{p}{;}
\PY{+w}{  }\PY{n}{STOREPAR}\PY{p}{(}\PY{l+s}{\PYZdq{}}\PY{l+s}{O1}\PY{l+s}{\PYZdq{}}\PY{p}{,}\PY{+w}{ }\PY{n}{o1}\PY{p}{)}\PY{p}{;}
\PY{+w}{  }\PY{n}{FETCHPAR}\PY{p}{(}\PY{l+s}{\PYZdq{}}\PY{l+s}{SFO1}\PY{l+s}{\PYZdq{}}\PY{p}{,}\PY{+w}{ }\PY{o}{\PYZam{}}\PY{n}{sfo1}\PY{p}{)}\PY{p}{;}
\PY{+w}{  }\PY{n}{FETCHPAR}\PY{p}{(}\PY{l+s}{\PYZdq{}}\PY{l+s}{O1}\PY{l+s}{\PYZdq{}}\PY{p}{,}\PY{+w}{ }\PY{o}{\PYZam{}}\PY{n}{o1}\PY{p}{)}\PY{p}{;}
\PY{+w}{  }\PY{n}{fprintf}\PY{p}{(}\PY{n}{logfp}\PY{p}{,}\PY{+w}{ }\PY{l+s}{\PYZdq{}}\PY{l+s}{In IR expjeriment, I set SFO1 to \PYZpc{}f}\PY{l+s+se}{\PYZbs{}n}\PY{l+s}{\PYZdq{}}\PY{p}{,}
\PY{+w}{          }\PY{n}{sfo1}\PY{p}{)}\PY{p}{;}
\PY{+w}{  }\PY{n}{fprintf}\PY{p}{(}\PY{n}{logfp}\PY{p}{,}\PY{+w}{ }\PY{l+s}{\PYZdq{}}\PY{l+s}{In IR experiment, I set O1 to \PYZpc{}f}\PY{l+s+se}{\PYZbs{}n}\PY{l+s}{\PYZdq{}}\PY{p}{,}
\PY{+w}{          }\PY{n}{o1}\PY{p}{)}\PY{p}{;}
\PY{+w}{  }\PY{n}{FETCHPAR1}\PY{p}{(}\PY{l+s}{\PYZdq{}}\PY{l+s}{SFO1}\PY{l+s}{\PYZdq{}}\PY{p}{,}\PY{+w}{ }\PY{o}{\PYZam{}}\PY{n}{sfo1}\PY{p}{)}\PY{p}{;}
\PY{+w}{  }\PY{n}{FETCHPAR1}\PY{p}{(}\PY{l+s}{\PYZdq{}}\PY{l+s}{O1}\PY{l+s}{\PYZdq{}}\PY{p}{,}\PY{+w}{ }\PY{o}{\PYZam{}}\PY{n}{o1}\PY{p}{)}\PY{p}{;}
\PY{+w}{  }\PY{n}{fprintf}\PY{p}{(}
\PY{+w}{      }\PY{n}{logfp}\PY{p}{,}
\PY{+w}{      }\PY{l+s}{\PYZdq{}}\PY{l+s}{In IR experiment, I have SFO1 (indirect) as \PYZpc{}f}\PY{l+s+se}{\PYZbs{}n}\PY{l+s}{\PYZdq{}}\PY{p}{,}
\PY{+w}{      }\PY{n}{sfo1}\PY{p}{)}\PY{p}{;}
\PY{+w}{  }\PY{n}{fprintf}\PY{p}{(}
\PY{+w}{      }\PY{n}{logfp}\PY{p}{,}
\PY{+w}{      }\PY{l+s}{\PYZdq{}}\PY{l+s}{In IR experiment, I have O1 (indirect) as \PYZpc{}f}\PY{l+s+se}{\PYZbs{}n}\PY{l+s}{\PYZdq{}}\PY{p}{,}
\PY{+w}{      }\PY{n}{o1}\PY{p}{)}\PY{p}{;}
\PY{+w}{  }\PY{n}{STOREPAR}\PY{p}{(}\PY{l+s}{\PYZdq{}}\PY{l+s}{P1}\PY{l+s}{\PYZdq{}}\PY{p}{,}\PY{+w}{ }\PY{n}{p1}\PY{p}{)}\PY{p}{;}
\PY{+w}{  }\PY{n}{fprintf}\PY{p}{(}\PY{n}{logfp}\PY{p}{,}\PY{+w}{ }\PY{l+s}{\PYZdq{}}\PY{l+s}{In IR experiment, I set P1 to \PYZpc{}f}\PY{l+s+se}{\PYZbs{}n}\PY{l+s}{\PYZdq{}}\PY{p}{,}
\PY{+w}{          }\PY{n}{p1}\PY{p}{)}\PY{p}{;}
\PY{+w}{  }\PY{n}{FETCHPAR}\PY{p}{(}\PY{l+s}{\PYZdq{}}\PY{l+s}{P1}\PY{l+s}{\PYZdq{}}\PY{p}{,}\PY{+w}{ }\PY{o}{\PYZam{}}\PY{n}{p1}\PY{p}{)}\PY{p}{;}
\PY{+w}{  }\PY{n}{fprintf}\PY{p}{(}\PY{n}{logfp}\PY{p}{,}\PY{+w}{ }\PY{l+s}{\PYZdq{}}\PY{l+s}{In IR experiment, I get P1 of \PYZpc{}f}\PY{l+s+se}{\PYZbs{}n}\PY{l+s}{\PYZdq{}}\PY{p}{,}
\PY{+w}{          }\PY{n}{p1}\PY{p}{)}\PY{p}{;}
\PY{+w}{  }\PY{c+c1}{// pull + set r.d.}
\PY{+w}{  }\PY{n}{FETCHPAR}\PY{p}{(}\PY{l+s}{\PYZdq{}}\PY{l+s}{D1}\PY{l+s}{\PYZdq{}}\PY{p}{,}\PY{+w}{ }\PY{o}{\PYZam{}}\PY{n}{repDelay}\PY{p}{)}\PY{p}{;}
\PY{+w}{  }\PY{n}{fprintf}\PY{p}{(}\PY{n}{logfp}\PY{p}{,}\PY{+w}{ }\PY{l+s}{\PYZdq{}}\PY{l+s}{I got d1: \PYZpc{}f}\PY{l+s+se}{\PYZbs{}n}\PY{l+s}{\PYZdq{}}\PY{p}{,}\PY{+w}{ }\PY{n}{repDelay}\PY{p}{)}\PY{p}{;}
\PY{+w}{  }\PY{n}{STOREPAR}\PY{p}{(}\PY{l+s}{\PYZdq{}}\PY{l+s}{D1}\PY{l+s}{\PYZdq{}}\PY{p}{,}\PY{+w}{ }\PY{n}{my\PYZus{}d1}\PY{p}{)}\PY{p}{;}
\PY{+w}{  }\PY{n}{FETCHPAR}\PY{p}{(}\PY{l+s}{\PYZdq{}}\PY{l+s}{D1}\PY{l+s}{\PYZdq{}}\PY{p}{,}\PY{+w}{ }\PY{o}{\PYZam{}}\PY{n}{repDelay}\PY{p}{)}\PY{p}{;}
\PY{+w}{  }\PY{n}{fprintf}\PY{p}{(}\PY{n}{logfp}\PY{p}{,}\PY{+w}{ }\PY{l+s}{\PYZdq{}}\PY{l+s}{after change, I got d1: \PYZpc{}f}\PY{l+s+se}{\PYZbs{}n}\PY{l+s}{\PYZdq{}}\PY{p}{,}
\PY{+w}{          }\PY{n}{repDelay}\PY{p}{)}\PY{p}{;}
\PY{+w}{  }\PY{n}{ZG\PYZus{}OVERWRITE}\PY{p}{;}
\PY{+w}{  }\PY{c+c1}{//\PYZcb{}\PYZcb{}\PYZcb{}}
\PY{p}{\PYZcb{}}
\PY{c+c1}{// close down}

\PY{n}{fclose}\PY{p}{(}\PY{n}{logfp}\PY{p}{)}\PY{p}{;}
\PY{n}{QUIT}

\PY{c+c1}{// the above code relies on some standard C functions,}
\PY{c+c1}{// which rely on other (standard) headers: these are}
\PY{c+c1}{// defined here.}
\PY{c+c1}{//}
\PY{c+c1}{// Note that we do need to include the function}
\PY{c+c1}{// declarations at the top of the AU file, as well.}

\PY{c+cp}{\PYZsh{}}\PY{c+cp}{include}\PY{+w}{ }\PY{c+cpf}{\PYZlt{}math.h\PYZgt{}}
\PY{c+cp}{\PYZsh{}}\PY{c+cp}{include}\PY{+w}{ }\PY{c+cpf}{\PYZlt{}stdlib.h\PYZgt{}}

\PY{+w}{    }\PY{k+kr}{inline}\PY{+w}{ }\PY{k}{static}\PY{+w}{ }\PY{n}{REAL}
\PY{+w}{    }\PY{n}{sqr}\PY{p}{(}\PY{n}{REAL}\PY{+w}{ }\PY{n}{x}\PY{p}{)}\PY{+w}{ }\PY{p}{\PYZob{}}
\PY{+w}{  }\PY{k}{return}\PY{+w}{ }\PY{n}{x}\PY{+w}{ }\PY{o}{*}\PY{+w}{ }\PY{n}{x}\PY{p}{;}
\PY{p}{\PYZcb{}}
\PY{c+c1}{// linear regression function}
\PY{k+kt}{int}\PY{+w}{ }\PY{n}{linreg}\PY{p}{(}\PY{k+kt}{int}\PY{+w}{ }\PY{n}{n}\PY{p}{,}\PY{+w}{ }\PY{k}{const}\PY{+w}{ }\PY{n}{REAL}\PY{+w}{ }\PY{n}{x}\PY{p}{[}\PY{p}{]}\PY{p}{,}\PY{+w}{ }\PY{k}{const}\PY{+w}{ }\PY{n}{REAL}\PY{+w}{ }\PY{n}{y}\PY{p}{[}\PY{p}{]}\PY{p}{,}
\PY{+w}{           }\PY{n}{REAL}\PY{+w}{ }\PY{o}{*}\PY{n}{m}\PY{p}{,}\PY{+w}{ }\PY{n}{REAL}\PY{+w}{ }\PY{o}{*}\PY{n}{b}\PY{p}{,}\PY{+w}{ }\PY{n}{REAL}\PY{+w}{ }\PY{o}{*}\PY{n}{r}\PY{p}{)}\PY{+w}{ }\PY{p}{\PYZob{}}
\PY{+w}{  }\PY{n}{REAL}\PY{+w}{ }\PY{n}{sumx}\PY{+w}{ }\PY{o}{=}\PY{+w}{ }\PY{l+m+mf}{0.0}\PY{p}{;}\PY{+w}{  }\PY{c+cm}{/* sum of x     */}
\PY{+w}{  }\PY{n}{REAL}\PY{+w}{ }\PY{n}{sumx2}\PY{+w}{ }\PY{o}{=}\PY{+w}{ }\PY{l+m+mf}{0.0}\PY{p}{;}\PY{+w}{ }\PY{c+cm}{/* sum of x**2  */}
\PY{+w}{  }\PY{n}{REAL}\PY{+w}{ }\PY{n}{sumxy}\PY{+w}{ }\PY{o}{=}\PY{+w}{ }\PY{l+m+mf}{0.0}\PY{p}{;}\PY{+w}{ }\PY{c+cm}{/* sum of x*y   */}
\PY{+w}{  }\PY{n}{REAL}\PY{+w}{ }\PY{n}{sumy}\PY{+w}{ }\PY{o}{=}\PY{+w}{ }\PY{l+m+mf}{0.0}\PY{p}{;}\PY{+w}{  }\PY{c+cm}{/* sum of y     */}
\PY{+w}{  }\PY{n}{REAL}\PY{+w}{ }\PY{n}{sumy2}\PY{+w}{ }\PY{o}{=}\PY{+w}{ }\PY{l+m+mf}{0.0}\PY{p}{;}\PY{+w}{ }\PY{c+cm}{/* sum of y**2  */}
\PY{+w}{  }\PY{k+kt}{int}\PY{+w}{ }\PY{n}{i}\PY{p}{;}

\PY{+w}{  }\PY{k}{for}\PY{+w}{ }\PY{p}{(}\PY{n}{i}\PY{+w}{ }\PY{o}{=}\PY{+w}{ }\PY{l+m+mi}{0}\PY{p}{;}\PY{+w}{ }\PY{n}{i}\PY{+w}{ }\PY{o}{\PYZlt{}}\PY{+w}{ }\PY{n}{n}\PY{p}{;}\PY{+w}{ }\PY{n}{i}\PY{o}{+}\PY{o}{+}\PY{p}{)}\PY{+w}{ }\PY{p}{\PYZob{}}
\PY{+w}{    }\PY{n}{sumx}\PY{+w}{ }\PY{o}{+}\PY{o}{=}\PY{+w}{ }\PY{n}{x}\PY{p}{[}\PY{n}{i}\PY{p}{]}\PY{p}{;}
\PY{+w}{    }\PY{n}{sumx2}\PY{+w}{ }\PY{o}{+}\PY{o}{=}\PY{+w}{ }\PY{n}{sqr}\PY{p}{(}\PY{n}{x}\PY{p}{[}\PY{n}{i}\PY{p}{]}\PY{p}{)}\PY{p}{;}
\PY{+w}{    }\PY{n}{sumxy}\PY{+w}{ }\PY{o}{+}\PY{o}{=}\PY{+w}{ }\PY{n}{x}\PY{p}{[}\PY{n}{i}\PY{p}{]}\PY{+w}{ }\PY{o}{*}\PY{+w}{ }\PY{n}{y}\PY{p}{[}\PY{n}{i}\PY{p}{]}\PY{p}{;}
\PY{+w}{    }\PY{n}{sumy}\PY{+w}{ }\PY{o}{+}\PY{o}{=}\PY{+w}{ }\PY{n}{y}\PY{p}{[}\PY{n}{i}\PY{p}{]}\PY{p}{;}
\PY{+w}{    }\PY{n}{sumy2}\PY{+w}{ }\PY{o}{+}\PY{o}{=}\PY{+w}{ }\PY{n}{sqr}\PY{p}{(}\PY{n}{y}\PY{p}{[}\PY{n}{i}\PY{p}{]}\PY{p}{)}\PY{p}{;}
\PY{+w}{  }\PY{p}{\PYZcb{}}

\PY{+w}{  }\PY{n}{REAL}\PY{+w}{ }\PY{n}{denom}\PY{+w}{ }\PY{o}{=}\PY{+w}{ }\PY{p}{(}\PY{n}{n}\PY{+w}{ }\PY{o}{*}\PY{+w}{ }\PY{n}{sumx2}\PY{+w}{ }\PY{o}{\PYZhy{}}\PY{+w}{ }\PY{n}{sqr}\PY{p}{(}\PY{n}{sumx}\PY{p}{)}\PY{p}{)}\PY{p}{;}
\PY{+w}{  }\PY{k}{if}\PY{+w}{ }\PY{p}{(}\PY{n}{denom}\PY{+w}{ }\PY{o}{=}\PY{o}{=}\PY{+w}{ }\PY{l+m+mi}{0}\PY{p}{)}\PY{+w}{ }\PY{p}{\PYZob{}}
\PY{+w}{    }\PY{c+c1}{// singular matrix. can\PYZsq{}t solve the problem.}
\PY{+w}{    }\PY{o}{*}\PY{n}{m}\PY{+w}{ }\PY{o}{=}\PY{+w}{ }\PY{l+m+mi}{0}\PY{p}{;}
\PY{+w}{    }\PY{o}{*}\PY{n}{b}\PY{+w}{ }\PY{o}{=}\PY{+w}{ }\PY{l+m+mi}{0}\PY{p}{;}
\PY{+w}{    }\PY{k}{if}\PY{+w}{ }\PY{p}{(}\PY{n}{r}\PY{p}{)}
\PY{+w}{      }\PY{o}{*}\PY{n}{r}\PY{+w}{ }\PY{o}{=}\PY{+w}{ }\PY{l+m+mi}{0}\PY{p}{;}
\PY{+w}{    }\PY{k}{return}\PY{+w}{ }\PY{l+m+mi}{1}\PY{p}{;}
\PY{+w}{  }\PY{p}{\PYZcb{}}

\PY{+w}{  }\PY{o}{*}\PY{n}{m}\PY{+w}{ }\PY{o}{=}\PY{+w}{ }\PY{p}{(}\PY{n}{n}\PY{+w}{ }\PY{o}{*}\PY{+w}{ }\PY{n}{sumxy}\PY{+w}{ }\PY{o}{\PYZhy{}}\PY{+w}{ }\PY{n}{sumx}\PY{+w}{ }\PY{o}{*}\PY{+w}{ }\PY{n}{sumy}\PY{p}{)}\PY{+w}{ }\PY{o}{/}\PY{+w}{ }\PY{n}{denom}\PY{p}{;}
\PY{+w}{  }\PY{o}{*}\PY{n}{b}\PY{+w}{ }\PY{o}{=}\PY{+w}{ }\PY{p}{(}\PY{n}{sumy}\PY{+w}{ }\PY{o}{*}\PY{+w}{ }\PY{n}{sumx2}\PY{+w}{ }\PY{o}{\PYZhy{}}\PY{+w}{ }\PY{n}{sumx}\PY{+w}{ }\PY{o}{*}\PY{+w}{ }\PY{n}{sumxy}\PY{p}{)}\PY{+w}{ }\PY{o}{/}\PY{+w}{ }\PY{n}{denom}\PY{p}{;}
\PY{+w}{  }\PY{k}{if}\PY{+w}{ }\PY{p}{(}\PY{n}{r}\PY{+w}{ }\PY{o}{!}\PY{o}{=}\PY{+w}{ }\PY{n+nb}{NULL}\PY{p}{)}\PY{+w}{ }\PY{p}{\PYZob{}}
\PY{+w}{    }\PY{o}{*}\PY{n}{r}\PY{+w}{ }\PY{o}{=}
\PY{+w}{        }\PY{p}{(}\PY{n}{sumxy}\PY{+w}{ }\PY{o}{\PYZhy{}}\PY{+w}{ }\PY{n}{sumx}\PY{+w}{ }\PY{o}{*}\PY{+w}{ }\PY{n}{sumy}\PY{+w}{ }\PY{o}{/}
\PY{+w}{                     }\PY{n}{n}\PY{p}{)}\PY{+w}{ }\PY{o}{/}\PY{+w}{ }\PY{c+cm}{/* compute correlation coeff */}
\PY{+w}{        }\PY{n}{sqrt}\PY{p}{(}\PY{p}{(}\PY{n}{sumx2}\PY{+w}{ }\PY{o}{\PYZhy{}}\PY{+w}{ }\PY{n}{sqr}\PY{p}{(}\PY{n}{sumx}\PY{p}{)}\PY{+w}{ }\PY{o}{/}\PY{+w}{ }\PY{n}{n}\PY{p}{)}\PY{+w}{ }\PY{o}{*}
\PY{+w}{             }\PY{p}{(}\PY{n}{sumy2}\PY{+w}{ }\PY{o}{\PYZhy{}}\PY{+w}{ }\PY{n}{sqr}\PY{p}{(}\PY{n}{sumy}\PY{p}{)}\PY{+w}{ }\PY{o}{/}\PY{+w}{ }\PY{n}{n}\PY{p}{)}\PY{p}{)}\PY{p}{;}
\PY{+w}{    }\PY{k+kt}{double}\PY{+w}{ }\PY{n}{slope}\PY{+w}{ }\PY{o}{=}\PY{+w}{ }\PY{p}{(}\PY{p}{(}\PY{n}{n}\PY{+w}{ }\PY{o}{*}\PY{+w}{ }\PY{n}{sumxy}\PY{p}{)}\PY{+w}{ }\PY{o}{\PYZhy{}}\PY{+w}{ }\PY{p}{(}\PY{n}{sumx}\PY{+w}{ }\PY{o}{*}\PY{+w}{ }\PY{n}{sumy}\PY{p}{)}\PY{p}{)}\PY{+w}{ }\PY{o}{/}\PY{+w}{ }\PY{n}{denom}\PY{p}{;}
\PY{+w}{    }\PY{k+kt}{double}\PY{+w}{ }\PY{n}{intercept}\PY{+w}{ }\PY{o}{=}
\PY{+w}{        }\PY{p}{(}\PY{p}{(}\PY{n}{sumy}\PY{+w}{ }\PY{o}{*}\PY{+w}{ }\PY{n}{sumx2}\PY{p}{)}\PY{+w}{ }\PY{o}{\PYZhy{}}\PY{+w}{ }\PY{p}{(}\PY{n}{sumx}\PY{+w}{ }\PY{o}{*}\PY{+w}{ }\PY{n}{sumxy}\PY{p}{)}\PY{p}{)}\PY{+w}{ }\PY{o}{/}\PY{+w}{ }\PY{n}{denom}\PY{p}{;}
\PY{+w}{    }\PY{k+kt}{double}\PY{+w}{ }\PY{n}{term1}\PY{+w}{ }\PY{o}{=}\PY{+w}{ }\PY{p}{(}\PY{p}{(}\PY{n}{n}\PY{+w}{ }\PY{o}{*}\PY{+w}{ }\PY{n}{sumxy}\PY{p}{)}\PY{+w}{ }\PY{o}{\PYZhy{}}\PY{+w}{ }\PY{p}{(}\PY{n}{sumx}\PY{+w}{ }\PY{o}{*}\PY{+w}{ }\PY{n}{sumy}\PY{p}{)}\PY{p}{)}\PY{p}{;}
\PY{+w}{    }\PY{k+kt}{double}\PY{+w}{ }\PY{n}{term2}\PY{+w}{ }\PY{o}{=}\PY{+w}{ }\PY{p}{(}\PY{p}{(}\PY{n}{n}\PY{+w}{ }\PY{o}{*}\PY{+w}{ }\PY{n}{sumx2}\PY{p}{)}\PY{+w}{ }\PY{o}{\PYZhy{}}\PY{+w}{ }\PY{p}{(}\PY{n}{sumx}\PY{+w}{ }\PY{o}{*}\PY{+w}{ }\PY{n}{sumy}\PY{p}{)}\PY{p}{)}\PY{p}{;}
\PY{+w}{    }\PY{k+kt}{double}\PY{+w}{ }\PY{n}{term3}\PY{+w}{ }\PY{o}{=}\PY{+w}{ }\PY{p}{(}\PY{p}{(}\PY{n}{n}\PY{+w}{ }\PY{o}{*}\PY{+w}{ }\PY{n}{sumy2}\PY{p}{)}\PY{+w}{ }\PY{o}{\PYZhy{}}\PY{+w}{ }\PY{p}{(}\PY{n}{sumy}\PY{+w}{ }\PY{o}{*}\PY{+w}{ }\PY{n}{sumy}\PY{p}{)}\PY{p}{)}\PY{p}{;}
\PY{+w}{    }\PY{k+kt}{double}\PY{+w}{ }\PY{n}{term23}\PY{+w}{ }\PY{o}{=}\PY{+w}{ }\PY{p}{(}\PY{n}{term2}\PY{+w}{ }\PY{o}{*}\PY{+w}{ }\PY{n}{term3}\PY{p}{)}\PY{p}{;}
\PY{+w}{    }\PY{k+kt}{double}\PY{+w}{ }\PY{n}{r2}\PY{+w}{ }\PY{o}{=}\PY{+w}{ }\PY{l+m+mf}{1.0}\PY{p}{;}
\PY{+w}{    }\PY{k}{if}\PY{+w}{ }\PY{p}{(}\PY{n}{fabs}\PY{p}{(}\PY{n}{term23}\PY{p}{)}\PY{+w}{ }\PY{o}{\PYZgt{}}\PY{+w}{ }\PY{n}{MIN\PYZus{}DOUBLE}\PY{p}{)}
\PY{+w}{      }\PY{n}{r2}\PY{+w}{ }\PY{o}{=}\PY{+w}{ }\PY{p}{(}\PY{n}{term1}\PY{+w}{ }\PY{o}{*}\PY{+w}{ }\PY{n}{term1}\PY{p}{)}\PY{+w}{ }\PY{o}{/}\PY{+w}{ }\PY{n}{term23}\PY{p}{;}
\PY{+w}{  }\PY{p}{\PYZcb{}}
\PY{+w}{  }\PY{k}{return}\PY{+w}{ }\PY{l+m+mi}{0}\PY{p}{;}
\PY{p}{\PYZcb{}}

\PY{k+kt}{void}\PY{+w}{ }\PY{n}{settitle}\PY{p}{(}\PY{k+kt}{char}\PY{+w}{ }\PY{o}{*}\PY{n}{string}\PY{p}{)}\PY{+w}{ }\PY{p}{\PYZob{}}
\PY{+w}{  }\PY{k+kt}{char}\PY{+w}{ }\PY{n}{titlename}\PY{p}{[}\PY{l+m+mi}{1000}\PY{p}{]}\PY{p}{;}
\PY{+w}{  }\PY{k+kt}{FILE}\PY{+w}{ }\PY{o}{*}\PY{n}{titlefp}\PY{p}{;}
\PY{+w}{  }\PY{p}{(}\PY{k+kt}{void}\PY{p}{)}\PY{n}{sprintf}\PY{p}{(}\PY{n}{titlename}\PY{p}{,}\PY{+w}{ }\PY{l+s}{\PYZdq{}}\PY{l+s}{\PYZpc{}s/\PYZpc{}s/\PYZpc{}d/pdata/\PYZpc{}d/title}\PY{l+s}{\PYZdq{}}\PY{p}{,}
\PY{+w}{                }\PY{n}{disk}\PY{p}{,}\PY{+w}{ }\PY{n}{name}\PY{p}{,}\PY{+w}{ }\PY{n}{expno}\PY{p}{,}\PY{+w}{ }\PY{n}{procno}\PY{p}{)}\PY{p}{;}
\PY{+w}{  }\PY{k}{if}\PY{+w}{ }\PY{p}{(}\PY{p}{(}\PY{n}{titlefp}\PY{+w}{ }\PY{o}{=}\PY{+w}{ }\PY{n}{fopen}\PY{p}{(}\PY{n}{titlename}\PY{p}{,}\PY{+w}{ }\PY{l+s}{\PYZdq{}}\PY{l+s}{wt}\PY{l+s}{\PYZdq{}}\PY{p}{)}\PY{p}{)}\PY{+w}{ }\PY{o}{=}\PY{o}{=}\PY{+w}{ }\PY{n+nb}{NULL}\PY{p}{)}\PY{+w}{ }\PY{p}{\PYZob{}}
\PY{+w}{    }\PY{n}{Proc\PYZus{}err}\PY{p}{(}\PY{n}{DEF\PYZus{}ERR\PYZus{}OPT}\PY{p}{,}\PY{+w}{ }\PY{l+s}{\PYZdq{}}\PY{l+s}{Can\PYZsq{}t open \PYZpc{}s}\PY{l+s+se}{\PYZbs{}n}\PY{l+s}{\PYZpc{}s}\PY{l+s}{\PYZdq{}}\PY{p}{,}\PY{+w}{ }\PY{n}{titlename}\PY{p}{,}
\PY{+w}{             }\PY{n}{strerror}\PY{p}{(}\PY{n}{errno}\PY{p}{)}\PY{p}{)}\PY{p}{;}
\PY{+w}{    }\PY{k}{return}\PY{+w}{ }\PY{l+m+mi}{0}\PY{p}{;}
\PY{+w}{  }\PY{p}{\PYZcb{}}
\PY{+w}{  }\PY{n}{fprintf}\PY{p}{(}\PY{n}{titlefp}\PY{p}{,}\PY{+w}{ }\PY{n}{string}\PY{p}{)}\PY{p}{;}
\PY{+w}{  }\PY{n}{fclose}\PY{p}{(}\PY{n}{titlefp}\PY{p}{)}\PY{p}{;}
\PY{p}{\PYZcb{}}
\end{Verbatim}

%% file: brukerPPG.tex
\begin{Verbatim}[commandchars=\\\{\}]
\PY{c+c1}{;ab\PYZus{}IR2h (written aug2020)}
\PY{c+c1}{;based off egr\PYZus{}IR}
\PY{c+c1}{;avance\PYZhy{}version (12/01/11)}
\PY{c+c1}{;1D sequence}
\PY{c+c1}{;using 2H lockswitch unit or BSMS 2H\PYZhy{}TX board}
\PY{c+c1}{;}
\PY{c+c1}{;\PYZdl{}CLASS=HighRes}
\PY{c+c1}{;\PYZdl{}DIM=1D}
\PY{c+c1}{;\PYZdl{}TYPE=}
\PY{c+c1}{;\PYZdl{}SUBTYPE=}
\PY{c+c1}{;\PYZdl{}COMMENT=}

\PYZsh{}include\PYZlt{}Avance.incl\PYZgt{}
\PYZsh{}include\PYZlt{}Sysconf.incl\PYZgt{}

define loopcounter total\PYZus{}ph\PYZus{}steps

\PY{l+s+s2}{\PYZdq{}}\PY{n+nv}{p2}\PY{o}{=}\PY{n+nv}{p1}\PY{o}{*}\PY{l+m+mi}{2}\PY{l+s+s2}{\PYZdq{}}
\PY{l+s+s2}{\PYZdq{}}\PY{n+nv}{d11}\PY{o}{=}\PY{l+m+mf}{30m}\PY{l+s+s2}{\PYZdq{}}
\PY{l+s+s2}{\PYZdq{}}\PY{n+nv}{acqt0}\PY{o}{=}\PY{o}{\PYZhy{}}\PY{n+nv}{p1}\PY{o}{*}\PY{l+m+mi}{2}\PY{o}{/}\PY{l+m+mi}{3}\PY{err}{.}\PY{l+m+mi}{1416}\PY{l+s+s2}{\PYZdq{}}
\PY{l+s+s2}{\PYZdq{}}\PY{n+nv}{l20}\PY{o}{=}\PY{l+m+mi}{2}\PY{l+s+s2}{\PYZdq{}} \PY{c+c1}{;steps in ph1}
\PY{l+s+s2}{\PYZdq{}}\PY{n+nv}{l21}\PY{o}{=}\PY{l+m+mi}{4}\PY{l+s+s2}{\PYZdq{}} \PY{c+c1}{;steps in ph2}
\PY{l+s+s2}{\PYZdq{}}\PY{n+nv}{total\PYZus{}ph\PYZus{}steps} \PY{o}{=} \PY{n+nv}{l20}\PY{o}{*}\PY{n+nv}{l21}\PY{l+s+s2}{\PYZdq{}}
\PY{l+s+s2}{\PYZdq{}}\PY{n+nv}{l23}\PY{o}{=}\PY{n+nv}{td1}\PY{o}{/}\PY{n+nv}{total\PYZus{}ph\PYZus{}steps}\PY{l+s+s2}{\PYZdq{}}
\PY{c+c1}{;td1=number of phase steps * vdlist}
\PY{c+c1}{;td2=4096}

\PY{l+m+mf}{1} \PY{k}{ze}
  \PY{n+nv}{d11} LOCKDEC\PYZus{}ON
  \PY{n+nv}{d11} H2\PYZus{}PULSE

\PY{l+m+mf}{2} \PY{l+m+mf}{30m} \PY{k}{rpp1}
  \PY{l+m+mf}{1m} \PY{k}{rpp2}

\PY{l+m+mf}{3}       \PY{l+m+mf}{30m} H2\PYZus{}LOCK 
        \PY{n+nv}{d1}
        \PY{n+nv}{d11} H2\PYZus{}PULSE 
        \PY{n+nv}{p2}:D ph1
        vd
        \PY{n+nv}{p1}:D ph2
        goscnp ph31
        \PY{n+nv}{d11} \PY{k}{wr} \PYZsh{}\PY{l+m+mf}{0} \PY{k}{if} \PYZsh{}\PY{l+m+mf}{0}
        \PY{l+m+mf}{2u} \PY{k}{ipp2}
      \PY{k}{lo} \PY{k}{to} \PY{l+m+mf}{3} \PY{k}{times} \PY{n+nv}{l21}
    \PY{l+m+mf}{2u} \PY{k}{ipp1}
    \PY{k}{lo} \PY{k}{to} \PY{l+m+mf}{3} \PY{k}{times} \PY{n+nv}{l20}
  \PY{l+m+mf}{0.1u} \PY{k}{ivd}
  \PY{k}{lo} \PY{k}{to} \PY{l+m+mf}{2} \PY{k}{times} \PY{n+nv}{l23}
  \PY{n+nv}{d11} H2\PYZus{}LOCK
  \PY{n+nv}{d11} LOCKDEC\PYZus{}OFF
exit

ph1=\PY{l+m+mf}{0} \PY{l+m+mf}{2}
ph2=\PY{l+m+mf}{0} \PY{l+m+mf}{1} \PY{l+m+mf}{2} \PY{l+m+mf}{3}
ph31=\PY{l+m+mf}{0}

\PY{c+c1}{;pl1 : f1 channel \PYZhy{} power level for pulse (default)}
\PY{c+c1}{;p1 : f1 channel \PYZhy{}  90 degree high power pulse}
\PY{c+c1}{;d1 : relaxation delay; 1\PYZhy{}5 * T1}
\PY{c+c1}{;d11: delay for disk I/O                             [30 msec]}
\PY{c+c1}{;ns: 1 * n, total number of scans: NS * TD0}

\PY{c+c1}{;locnuc: off}

\PY{c+c1}{;\PYZdl{}Id: zg2h,v 1.14.8.1 2012/01/31 17:56:41 ber Exp \PYZdl{}}
\end{Verbatim}